%% file: main.tex
\pdfoutput=1
\documentclass[12pt]{article}
\usepackage{natbib}
\usepackage{amssymb, amsmath,natbib,bm,graphicx}
\usepackage{setspace}
\usepackage{mathrsfs,amsthm}
\usepackage[normalem]{ulem}
\usepackage{times,bm}
\usepackage[plain,noend]{algorithm2e}
\usepackage{multirow}
\usepackage{anyfontsize}
\usepackage{graphicx}
\usepackage{float}
\usepackage{tabularx}
\usepackage{url}

\def\RR{\mathbb{R}}

\newtheorem{theorem}{Theorem}
\newtheorem{lemma}{Lemma}
\newtheorem{prop}{Proposition}
\newenvironment{remark}{\begin{quote}{\bf Remark.\quad}\rm\small}{\end{quote}}
\newtheorem{corollary}{Corollary}

\newcommand{\I}{\ensuremath{{\text{I}}}}
\newcommand{\xx}{\ensuremath{{\bm{x}}}}

\newcommand{\lstar}{\ensuremath{L^{\star}}}
\newcommand{\rstar}{\ensuremath{R^{\star}}}

\newcommand{\PP}{\ensuremath{\tilde{P}}}
\newcommand{\sstar}{\ensuremath{s^{\star}}}
\newcommand{\ex}{\ensuremath{\mathbb{E}}}
\DeclareMathOperator*{\argmin}{arg\,min}
\DeclareMathOperator*{\argmax}{arg\,max}
\DeclareMathOperator*{\dmin}{\underline{m}}

\newtheorem{definition}{Definition}
\newcommand{\cS}{{\cal S}}
\newcommand{\cs}{\textsl{{\cal sp}}}

\usepackage{bigstrut}

\begin{document}

\title{Convex clustering via $\ell_1$ fusion penalization}
\date{}
\author{
{\sc Peter Radchenko and Gourab Mukherjee}
\thanks{University of Southern California}
}

\maketitle

\begin{abstract}
We study the large sample behavior of a convex clustering framework, which minimizes the sample within cluster sum of squares under an~$\ell_1$
fusion constraint on the cluster centroids.   This recently proposed approach has been gaining in popularity, however, its asymptotic properties have remained mostly unknown.   Our analysis is based on a novel representation of the sample clustering procedure as a sequence of cluster splits determined by a sequence of maximization problems.  We use this representation to provide a simple and intuitive formulation for the population clustering procedure.  We then demonstrate that the sample procedure consistently estimates its population analog, and derive the corresponding rates of convergence.  The proof conducts a careful simultaneous analysis of a collection of M-estimation problems, whose cardinality grows together with the sample size.  Based on the new perspectives gained from the asymptotic investigation, we propose a key post-processing modification of the original clustering framework.  We show, both theoretically and empirically, that the resulting approach can be successfully used to estimate the number of clusters in the population.   Using simulated data, we compare the proposed method with existing number of clusters and modality assessment approaches, and obtain encouraging results.  We also demonstrate the applicability of our clustering method for the detection of cellular subpopulations in a single-cell virology study.
\end{abstract}

\noindent {\small{\it Some key words}:
Convex Clustering;
Fusion Penalties;
Number of Clusters;
Rates of Convergence}

\section{Introduction}

Clustering is one of the most popular statistical techniques for unsupervised classification and taxonomy detection \citep{Hartigan-75,Kaufman-09}.  One serious limitation of the traditional methods, such as $k$-means, is the non-convexity of the corresponding optimization problems.  Recently, several convex clustering algorithms have been proposed \citep{Xu-04,Bach-08,Chi-13}.  Speed and scalability of these algorithms make them increasingly popular for cluster analysis of massive modern datasets.  These approaches use convex relaxations of the traditional non-convex clustering criteria, however, they do not naturally inherit the statistical properties associated with the original methods.  Here we study the large sample behavior of a popular convex clustering framework that is based on an~$\ell_1$ fusion penalty \citep{Hocking11}.

Consider the problem of clustering~$n$ observations, $\xx_1,\ldots,\xx_n$, which are sampled from a Euclidean space, $\RR^d$.  The well-studied $k$-means approach \citep{Macqueen-67, Hartigan-78, Pollard-81,Pollard-82,Jain-10} is based on minimizing the within cluster sum of squares, $\sum_{i=1}^n\Vert \xx_i - \bm{\alpha}_i \Vert_2^2$, with respect to the cluster centroids, $\bm{\alpha}_1,\ldots,\bm{\alpha}_n$, under the restriction that the number of distinct cluster centroids is at most~$k$.  This restriction can be viewed as an~$\ell_0$ constraint on the centroids.
Motivated by the Lasso and its variants \citep{Tibshirani-96,Tibshirani-05b}, which successfully use the~$\ell_1$ constraint as a surrogate for the NP-hard $\ell_0$ constraint, \cite{Hocking11} consider the following modification of the $k$-means clustering criterion:
\begin{align}\label{criterion-multivariate}
\min_{\bm{\alpha}_1,\ldots,\bm{\alpha}_n}  \sum_{i=1}^n\Vert \xx_i-\bm{\alpha}_i \Vert_2^2 \;\text{ subject to }\; {\sum_{1 \leq i< j \leq n} \big \| \bm{\alpha_i} - \bm{\alpha_j} \big \|_1} \leq t.
\end{align}
When $t = 0$, the $\ell_1$ penalty fuses all the cluster centroids together. Thus, all the observations are placed in the same cluster.  When~$t\ge\sum_{i<j} \|\xx_i-\xx_j\|_1$, we have $\bm{\alpha}_i=\xx_i$ for all~$i$, and, thus, each observation forms its own cluster. Varying $t$ between the two extremes creates a path of solutions to the regularized clustering problem.
Note that the Lagrangian form of the above criterion, $\min_{\bm{\alpha}_i} \sum_{i=1}^n\Vert \xx_i-\bm{\alpha}_i \Vert_2^2 +\lambda {\sum_{i< j} \big \| \bm{\alpha_i} - \bm{\alpha_j} \big \|_1}$, is separable across dimensions.  Consequently, the corresponding optimization problem reduces to independently minimizing~$d$ univariate convex clustering criteria.

Thus, to understand the large sample behaviour of the multivariate solution, it is sufficient to focus on the analysis of the univariate clustering criterion,
\begin{align}\label{clus.criterion}
\min_{\alpha_1,...,\alpha_n} \sum_{i=1}^n (x_i-\alpha_i)^2 +  \lambda \sum_{1 \leq i< j \leq n} \big| \alpha_i - \alpha_j \big|.
\end{align}
As the penalty parameter~$\lambda$ varies from $0$ to $\infty$, each corresponding solution determines a cluster partition.  We are interested in the asymptotics of the entire collection of such partitions, which we view as the outcome of the \textit{sample clustering procedure}.

{\bf Summary of the Main Contributions.}
We analyze the large sample behavior of the sample clustering procedure determined by the solution path for criterion~(\ref{clus.criterion}).  We develop a simple and intuitive formulation for the population clustering procedure, show that under some very mild regularity conditions the sample procedure consistently estimates its population analog, and derive the corresponding rates of convergence.

More specifically, we first demonstrate that the path of solutions to~(\ref{clus.criterion}) determines a clustering tree, which can be formed by either successive merges of clusters, in a bottom up fashion, or successive splits, in a top down approach.  We then study the asymptotic behavior of the full clustering tree by representing each split as a solution to a maximization problem.  We define the corresponding population clustering procedure in a similar fashion, but replace sample averages with the corresponding expected values.   The asymptotic analysis is significantly complicated by the fact that, unlike in the standard M-estimation setup (e.g. \citealt{vaartwellner96book, vaart98book}), the number of maximization problems at the sample level tends to infinity together with~$n$, and the number of the corresponding population problems is infinite.
We establish consistency and the rates of convergence of the sample clustering procedure through a careful analysis of the population procedure and the corresponding empirical process.

Motivated by the results of our large sample investigation, we introduce a key postprocessing modification to the sample clustering procedure.  We show, both theoretically and empirically, that the resulting approach can be successfully used to estimate the number of clusters in the population.  We also compare the new methodology with a wide variety of existing modality assessment and number of clusters approaches.  Our results provide strong support for the use of fusion penalization in clustering.

{\bf Connections to Related Work.}
\cite{Hocking11}, \cite{Chi-13} and \cite{tan2015statistical}  have studied modifications of optimization problem~(\ref{criterion-multivariate}).  These include using~$\ell_2$ or~$\ell_{\infty}$ regularization, as well as incorporating weights \citep{pelckmans2005convex,lindsten2011clustering,zhu2014convex}.  The large sample analysis in the papers listed above focusses on showing that if the distance between clusters grows at a sufficiently fast rate, then the corresponding method can separate the groups perfectly.  Here we consider a completely different perspective and investigate the  asymptotics of a clustering approach in the classical sense of \cite{Pollard-81}.  We study a clustering procedure that is applied to a random sample, and analyze its convergence to the outcome of the corresponding population procedure, which is based on the underlying probability distribution.  As we point out in Section~\ref{sec.disc}, the general framework of our theoretical analysis has the potential to handle the aforementioned modifications of the optimization problem.

The criterion in~(\ref{criterion-multivariate}) can be viewed \citep{Hocking11} as a convex relaxation of the hierarchical clustering criterion \citep{Hartigan-75}. However, as clustering is a very mature subject, approaches built on several other philosophies are also widely used in practice. A detailed review of clustering methods can be found in \citet{Kaufman-09}. One of the most popular methods is the k-means algorithm \citep{Macqueen-67}, which follows a partitioning approach for making clusters. Other popular partitioning methods, such as PAM \citep{kaufman1990partitioning} and CLARA \citep{kaufman1986clustering}, are based on the k-medoids algorithm.
 Density driven approaches, which include mixture model based methods, such as \cite{fraley2002model} and \cite{li2005clustering}, as well as non-parametric methods (see \citealt{li2007nonparametric} and the references therein), provide a flexible clustering framework, while spectral clustering methods, such as \citep{belkin2001laplacian,rohe2011spectral,shi2009data} perform efficient dimension reduction before segmenting the data.  In our empirical analysis we compare the performance of the proposed approach with most of the aforementioned clustering methods.

The~$\ell_1$ penalty, which is extensively used for variable selection \citep{Tibshirani-11}, also finds its use in trend filtering \citep{Tibshirani-13} and high-dimensional clustering problems \citep{Soltanolkotabi12,Witten-10}.  Another related approach, the fused Lasso \citep{Rinaldo-09,Tibshirani-05,Hoefling-10}, deals with applications having ordered features and checks for local constancy of their associated coefficients.  This approach penalizes the successive differences of the coefficients. \cite{shen-10,shen-12,ke-13,Bondell-08} have proposed methods based on fusion penalties, which apply to all the pairwise differences of coefficients.  These approaches can successfully recover the grouping structure of predictors in a high-dimensional regression setup.  However, the theory developed for these methods focusses on the  homogeneity of regression coefficients  and cannot be applied in the unsupervised clustering setup considered in this paper.

{\bf Organization of the Paper.}  In Section~\ref{sec.path.alg} we derive two equivalent algorithmic representations of the sample clustering procedure, which we use to formulate the  corresponding population procedure.  Section~\ref{sec.gen.results} contains our main results, in which we establish consistency and the rates of convergence.  Our asymptotic analysis reveals that an overwhelming majority of the sample clusters are in some sense negligible.  Motivated by this observation, we introduce a key post-processing modification to the clustering procedure.
In Section~\ref{sec.sim} we conduct a detailed empirical analysis of our approach.  More specifically, we use simulated data to show its strong performance relative to popular existing approaches for assessing modality and estimating the number of clusters.  We also illustrate the use of our method in analysis of single-cell virology datasets.  All the proofs, together with additional technical details, are relegated to the Supplementary Material.

\section{Sample and Population Clustering Procedures}\label{sec.path.alg}

In this section we derive two equivalent representations of the sample clustering procedure.  First, we develop a computationally efficient merging algorithm for producing a path of solutions to the clustering criterion~(\ref{clus.criterion}).  Then, in order to understand the large sample behavior of the solution path, we introduce an equivalent splitting procedure, which can recover all the corresponding cluster splits by solving a sequence of maximization problems.   We use the splitting representation to define the population clustering procedure, and describe its basic properties.

\subsection{Equivalent Representations for the Sample Solution Path}
\label{eqv.repr}

Note that a solution path for problem~(\ref{clus.criterion}) could be produced using the highly general fused lasso algorithm in \cite{Hoefling-10}.  Instead, we obtain a very simple and computationally efficient fitting procedure by analyzing our clustering criterion, (\ref{clus.criterion}), directly.  The path algorithm we describe here is a bottom up procedure, which starts at $\lambda=0$, with each observation forming its own cluster, and then gradually merges suitable clusters as~$\lambda$ increases.  Fix~$\lambda$, and suppose that~$C$ is one of the clusters identified by the solution to the optimization problem~\eqref{clus.criterion}. Write~$\alpha_C$ for the centroid of cluster~$C$, and denote the corresponding cluster average by~$\overline{X}_C$.  As pointed out in \cite{Hocking11}, the first order conditions for criterion~\eqref{clus.criterion} imply
\begin{align}\label{eq:clus.centroid}
\alpha_C=\overline{X}_C+ \lambda\sum_{j, \alpha_j\ne \alpha_C}\text{sign}(\alpha_j-\alpha_C).
\end{align}
Until the cluster partition or the ordering or the centroids are modified, parameter~$\lambda$ is the only component on the right-hand side of the equation that can change.  Thus, equation~\eqref{eq:clus.centroid} provides a simple way of tracking the piecewise linear paths of the centroids~$\alpha_i$. Another consequence of the first order conditions is that as~$\lambda$ increases, the only way the clusters get modified is some of them get merged together \citep{Hocking11}.  Hence, we can store the full cluster partition path by keeping track of the merges and the corresponding values of the tuning parameter~$\lambda$.  Algorithm~\ref{al1} makes this idea precise, and Theorem~\ref{merge.crit.eq} provides a rigorous justification.  Here we use~$|\cdot|$ to denote the cardinality of a set.

\begin{algorithm}[!h]
\caption{Merging Algorithm} \label{al1}
\vspace*{-12pt}
\begin{tabbing}
   \enspace INITIALIZE: \\[1ex]
    \qquad Sort data in ascending order and store them as $\xx_n=\{x_1,\ldots,x_n\}$.\\[1ex]
    \qquad Set $K$, the number of clusters, equal to~$n$.  For each~$i$ in $1,...,n$, set~$C_i=\{x_i\}$.\\[1ex]
   \enspace REPEAT: \\[1ex]
   \qquad Find the adjacent centroid distances standardized by cluster sizes:\\[1ex]
   \qquad \qquad  $d(j,j+1)  \leftarrow (\overline{X}_{C_{j+1}} - \overline{X}_{C_j}) \left/ (|C_j|+|C_{j+1}|)\right.$. \\[1ex]
   \qquad Find the clusters that minimize this distance: $ j^* \leftarrow \argmin_j d(j,j+1)$.\\[1ex]
  \qquad Merge the clusters that were found: $C_{j^*}\leftarrow C_{j^*}\cup C_{j^*+1}$.\\[1ex]
  \qquad Store the above merge and the corresponding $\lambda$ value: $\lambda=d(j^*,j^*+1)$.\\[1ex]
  \qquad Relabel the remaining clusters:  for $j>j^*$ set $C_j\leftarrow C_{j+1}$.\\[1ex]
  \qquad Reduce the total number of clusters:   $K\leftarrow K-1$.\\[1ex]
  \enspace UNTIL $K=1$. \\[1ex]
\enspace OUTPUT: Sequence of cluster merges and corresponding~$\lambda$ values.
\end{tabbing}
\end{algorithm}

The following result shows that Algorithm~\ref{al1} reproduces the sequence of cluster partitions and the corresponding $\lambda$ values from the optimization problem~(\ref{clus.criterion}).  In the proof, which is provided in the Supplementary Material, we also verify that the sequence of $\lambda$ values, corresponding to successive merges in Algorithm~\ref{al1}, is increasing.
\begin{prop}
\label{merge.crit.eq}
Suppose that the observations are generated from a continuous distribution.  Then, with probability one, the sequence of merges and~$\lambda$ values produced by the merging algorithm is the same as the sequence corresponding to the optimization criterion~$($\ref{clus.criterion}$)$.
\end{prop}

For the asymptotic analysis, it is helpful to recover the sequence of cluster partitions in a top down approach: we start with everything in one cluster and then split the clusters iteratively.  We call a representation of the cluster~$C$ as $C=C_1\cup C_2$ a split if $\max C_1 < \min C_2$.  The full collection of splits corresponding to the optimization problem~(\ref{clus.criterion}) is given by the splitting procedure, described in Algorithm~\ref{al2} below.  Proposition~\ref{split.merge.eq}, proved in the Supplementary Material, provides theoretical justification.  In particular, it shows that each of the cluster splits is chosen to maximize the distance between the two sub-cluster means.

\begin{algorithm}[!h]
\caption{Splitting Procedure} \label{al2}
\vspace*{-12pt}
\begin{tabbing}
   \enspace INITIALIZE: \\[1ex]
    \qquad Sort data in ascending order and store them as $\xx_n=\{x_1,\ldots,x_n\}$.\\[1ex]
    \qquad Set the current partition of~$\xx_n$ to~$\xx_n$.\\[1ex]
   \enspace REPEAT: \\[1ex]
   \qquad Select one cluster, $C$, with~$|C|>1$, from a current cluster partition of~$\xx_n$.\\[1ex]
   \qquad Find a split partition $C=C_1\cup C_2$, that maximizes the distance $\overline{X}_{C_2}-\overline{X}_{C_1}$.\\[1ex]
   \qquad Store the split $C=C_1\cup C_2$ and the corresponding value $\lambda=(\overline{X}_{C_2} - \overline{X}_{C_1}) \left/ |C|\right.$. \\[1ex]
  \qquad Replace $C$ with $C_1\cup C_2$ in the current partition of~$\xx_n$. \\[1ex]
  \enspace UNTIL: All the clusters in the current partition of~$\xx_n$ are of size one. \\[1ex]
\enspace OUTPUT: Collection of cluster splits and corresponding~$\lambda$ values.
\end{tabbing}
\end{algorithm}

\begin{prop}
\label{split.merge.eq}
Suppose that the observations are generated from a continuous distribution.  Then, with probability one, the collection of splits and corresponding~$\lambda$ values produced by the splitting procedure in Algorithm~\ref{al2}  exactly matches the sequence of  merges and corresponding~$\lambda$ values produced by the merging algorithm .
\end{prop}

Note that, unlike the merging algorithm, the splitting procedure does not provide a computationally efficient way for producing the clustering tree.  Instead, we use the splitting procedure to understand the large sample behavior of the sample clustering procedure.  It is reasonable to expect that, as $n$ tends to infinity, the collection of splits in the sample procedure should resemble the collection of splits in an analogous procedure defined on the population.   The population procedure can be defined by replacing the averages with the corresponding conditional means.  The formal definition is given in the next section.

\subsection{Population Clustering Procedure}
\label{pop.clust.proc}
For the remainder of the paper we assume that the underlying distribution has a finite first moment and a real valued density, $f$.  For concreteness, we focus on the case where the support of the distribution is of the form~$(L_0,R_0)$, where $-\infty\le L_0<R_0\le\infty$.  Thus, every open interval in~$(L_0,R_0)$ contains positive probability.
Given an interval $(l,r)\subseteq (L_0,R_0)$, we write $\mu_{l,r}$ for the population conditional mean on $(l,r)$,
\begin{equation}
\mu_{l,r}=\left(\int\limits_l^r f(x)dx\right)^{-1} \int\limits_l^r xf(x)dx.
\end{equation}
We set $\mu_{r,r}=r$, by continuity.
Given an interval $(L,R)\subseteq (L_0,R_0)$, we define
\begin{equation}
G_{L,R}(a)=\mu_{a,R}-\mu_{L,a},
\end{equation}
for $a\in[L,R]$.   Note that $G_{L,R}(L)=\mu_{L,R}-L$ and $G_{L,R}(R)=R-\mu_{L,R}$.

According to the results in Section~\ref{eqv.repr}, the sample clustering procedure determines the split partition of a cluster by maximizing the distance between the empirical sub-cluster means.
We define the \emph{population clustering procedure} by analogy.  Given a cluster $(L,R)$, the population procedure chooses the split that maximizes the distance between the population sub-cluster means.  In other words,
it finds a point~$s$ that maximizes $G_{L,R}$, then partitions $(L,R)$ into subintervals $(L,s)$ and $(s,R)$, on which the procedure is repeated.    If~$s$ is an interior point of $(L,R)$, we call it a \emph{split point}, and we call the corresponding partition a \emph{split}.  Otherwise, the population procedure essentially wants to split off an endpoint,  which forces the cluster to be truncated rather than split.  More formally, given a cluster $(L,R)$, we distinguish three types of \emph{truncation}, as specified below.
\begin{definition}
\label{def.trunc}
\begin{enumerate}
\item[(i)] if $\arg\max G_{l,R}=\{l\}$ for all $l\in [L,L^*)$, and $\arg\max G_{L^*,R}\ne\{L^*\}$, then the interval $(L,R)$ is truncated from the \textbf{left} to $(L^*,R^*)$, where $R^*=R$;
\item[(ii)] if $\arg\max G_{L,r}=\{r\}$ for all $r\in (R^*,R]$, and $\arg\max G_{L,R^*}\ne\{R^*\}$, then $(L,R)$ is truncated from the \textbf{right} to $(L^*,R^*)$, where $L^*=L$;
\item[(iii)] if there exists a continuous decreasing function $l\mapsto R_l$, satisfying $R_L=R$, for which $\arg\max G_{l,R_l}=\{l,R_l\}$ for all $l\in [L,L^*)$, and $\arg\max G_{L^*,R_{L^*}}\ne\{L^*, R_{L^*}\}$, then $(L,R)$ is truncated, in a \textbf{two-sided} fashion, to $(L^*,R^*)$, where $R^*=R_{L^*}$.
\end{enumerate}
\end{definition}
Note that we incorporated a continuity requirement into the definition of a two-sided truncation.  In the next subsection we give regularity conditions under which this requirement is satisfied.  We are now ready to formulate the full population clustering procedure.

\begin{algorithm}[!h]
\caption{Population Clustering Procedure} \label{al3}
\vspace*{-12pt}
\begin{tabbing}
   \enspace INITIALIZE: \\[1ex]
    \qquad Set the current cluster collection, $\Sigma$, equal to~$\{(L_0,R_0)\}$.\\[1ex]
   \enspace REPEAT: \\[1ex]
   \qquad Select one non-empty cluster, $(L,R)$, from the current cluster collection, $\Sigma$.\\[1ex]
   \qquad $-\,$ If the maximum value of $G_{L,R}$ is achieved at a point~$s$ in $(L,R)$, then
   store~$s$ \\[1ex]
   \qquad $\;\;\;\,$ as a split point and replace~$(L,R)$ in $\Sigma$ with $(L,s)$ and $(s,R)$. \\[1ex]
      \qquad $-\,$ Otherwise,
   replace~$(L,R)$  in $\Sigma$ with the interval $(L^*,R^*)$ from Definition~\ref{def.trunc}. \\[1ex]
  \enspace UNTIL: The current cluster collection, $\Sigma$, consists only of empty clusters. \\[1ex]
\enspace OUTPUT: Set of split points.
\end{tabbing}
\end{algorithm}

\begin{figure}[h]
\includegraphics[height=15pc,width=35pc]{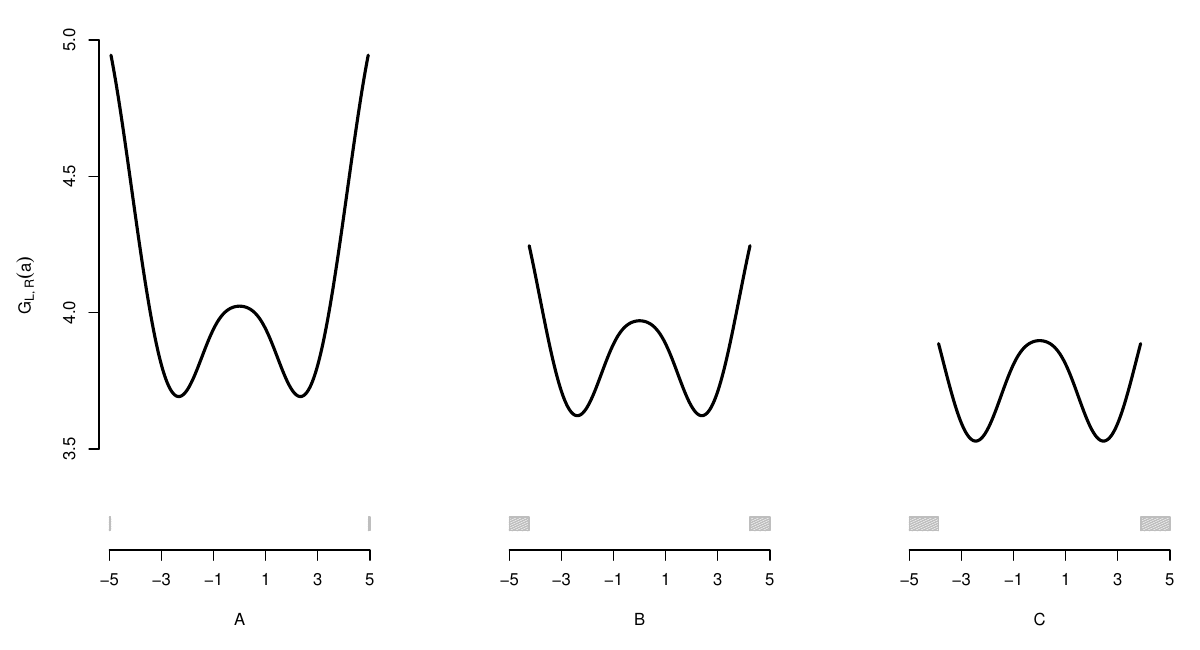}
\caption{Population criterion function, $G_{L,R}$, corresponding to the Gaussian mixture $0.5\,N(-2,1)+0.5\,N(2,1)$, is plotted for three choices of $(L,R)$, such that $R=-L$ and $G_{L,R}(L)=G_{L,R}(R)=\max G_{L,R}$.  The population clustering procedure continuously truncates the support of the distribution, symmetrically in a two-sided fashion, until $\max G_{L,R}$ can be achieved at an interior point (plot C).  Then, the procedure places a split point at zero.  Note that the resulting sub-clusters are then truncated down to empty sets according to Proposition~\ref{prop.unimod} in Section~\ref{sec.reg.cond}.}\label{fig1}
\end{figure}

The collection of population split points determines the corresponding clusters.  For example, consider the symmetric mixture of two Gaussian distributions examined in Figure~\ref{fig1}.  The population procedure identifies one split point, located at zero.  This specifies the population cluster partition: $(-\infty,0)\cup(0,\infty)$.

Given an underlying distribution, Algorithm~\ref{al3} determines the exact behaviour of the population clustering procedure. In Section~4 of the Supplementary Material we document the performance of the population procedure for a variety of Gaussian mixtures.    In Section~\ref{sec.reg.cond} we establish an important fact that the population procedure produces no splits for unimodal distributions.  We also provide conditions under which the population procedure is \emph{well defined}, by which we mean that it implements finitely many uniquely defined steps.

\subsection{Properties of the Population Procedure}
\label{sec.reg.cond}

The proofs of the results established in this section are provided in the Supplementary Material.
We first consider an important special case, where the underlying distribution is unimodal.  We suppose that the density~$f$ is either strictly monotone on its support, $(L_0,R_0)$, or there exists a point~$c$ for which~$f$ is strictly increasing on $(L_0,c)$ and strictly decreasing on $(c,R_0)$.  The following result shows that in this setting, under just a continuity assumption on~$f$, the population procedure is unique and does not reveal any clusters.
\begin{prop}
\label{prop.unimod}
If~$f$ is continuous and unimodal, then the population clustering procedure is uniquely defined and produces no splits.
\end{prop}

We now move to the general setting.  The following simple regularity condition ensures existence of a population clustering procedure with finitely many steps, as we demonstrate in the proof of Proposition~\ref{prop.gen} below.
\begin{enumerate}
\item[{\bf C1}.] Density~$f$ is nonzero and differentiable on $(L_0,R_0)$.  It has finitely many modes and, at each of its interior modes, admits a non-constant Taylor approximation.
\end{enumerate}
\begin{remark}
The last requirement means the following: for each interior mode~$c$, there exists a positive integer~$k$, such that $f$ is $k$ times differentiable at~$c$ with $f^{(k)}(c)\ne 0$.
\end{remark}
Note that the differentiability assumption can be slightly relaxed: for example, the results that follow hold for continuous piece-wise linear densities with no constant segments.  However, we prefer to keep this assumption, as it simplifies the presentation of the results.

To address the question of uniqueness, consider the following counterexample.  Suppose the underlying distribution is uniform on $(L,R)$.  Then, the criterion function $G_{L,R}$ is constant on its domain, and the population procedure applied to cluster $(L,R)$ may place a split point anywhere in its interior.   Thus, there are infinitely many versions of the population clustering procedure. The following regularity condition explicitly rules out such settings, by requiring that the interior of~$(L,R)$ contains at most one maximizer of~$G_{L,R}$.
\begin{enumerate}
\item[{\bf C2}.]  When the population procedure performs a split, the location of the split point is uniquely determined.
\end{enumerate}
\noindent Note that condition C2 holds for each distribution with a continuous unimodal density, as a direct consequence of Proposition~\ref{prop.unimod}.  In the proof of Proposition~\ref{prop.unimod} we also show that C2 holds for all bimodal densities, provided the smoothness condition, C1, is satisfied.  The following result establishes existence and uniqueness of the population clustering procedure in the general setting.
\begin{prop}
\label{prop.gen}
If regularity conditions C1 and C2 are satisfied, then the population clustering procedure is uniquely defined and implements finitely many steps.
\end{prop}
\noindent In Section~\ref{sec.gen.results} we show that under the same regularity conditions, C1 and C2, the sample procedure consistently estimates its population counterpart.

\section{Main Results}
\label{sec.gen.results}

In this section we show that the clustering tree produced by criterion~(\ref{clus.criterion}) consistently estimates the clustering tree produced by the population procedure defined in Section~\ref{pop.clust.proc}.  We also derive the corresponding rates of convergence and propose a novel post-processing modification of the sample clustering procedure.  Recall that we assume a finite first moment for the underlying distribution.

\subsection{Consistency}

We start with some useful notation.
In both the sample and the population, each split is characterized by a triple $(L,s,R)$, where interval $(L,R)$ is the cluster being split, and~$s$ is the split point, located inside~$(L,R)$.  We write~$P_{L,R}$ for the probability assigned to the interval~$(L,R)$ by the underlying distribution.
For a split $\cs=(L,s,R)$ we define its size as~$size(\cs)=\min\left\{P_{L,s}, P_{s,R}\right\}$.  When the probabilities in the above definition are replaced with the corresponding sample frequencies, we write $\widehat{siz}e(\cs)$ for the resulting quantity, and refer to it as the empirical size. 

The set of all the population splits, denoted by $\cS$, defines the population clustering tree.  Similarly, the set of all sample splits, $\widehat\cS$, defines the sample tree.  The cardinality of~$\widehat\cS$ tends to infinity as the sample size grows.  Alternatively, according to Proposition~\ref{prop.gen}, under mild regularity conditions the population procedure produces finitely many splits, together with some truncations.  To establish consistency, we divide the sample splits into ``big'' and ``small'', based on their empirical size, then show that the first group converges to the population splits, while the second is asymptotically negligible.   The formal definition is given below.  We write~$d_H$ for the Hausdorff distance between subsets of a Euclidean space.

\begin{definition}
Write~$\widehat\cS_{\alpha}$ for the set~$\{\cs\in\widehat\cS:\,\widehat {siz}e(\cs)>\alpha\}$ and let~$\alpha^*$ be the smallest split size in the population procedure.  We call the sample clustering procedure \textbf{strongly consistent} if, for each~$\alpha$ in~$(0,\alpha^*)$, the following statements hold almost surely,
\begin{eqnarray}
&& |\widehat \cS_{\alpha}|\rightarrow|\cS|\label{eq1.cons.def}\\
&&d_H(\widehat \cS_{\alpha},\cS)\rightarrow 0\qquad\text{and}\label{eq2.cons.def}\\
&&\max\{size(\cs):\,\cs\in\widehat \cS\setminus \widehat \cS_{\alpha}\}\rightarrow 0. \label{eq3.cons.def}
\end{eqnarray}
\end{definition}

\noindent If we replace almost sure convergence with convergence in probability, we have a weaker notion of consistency, which holds automatically when the sample procedure is strongly consistent.  In particular,
displays~(\ref{eq1.cons.def}) and~(\ref{eq2.cons.def}) imply that, except on a set of probability tending to zero, there is a one to one correspondence between~$\widehat\cS_{\alpha}$ and the set of all population splits, such that each split in~$\widehat\cS_{\alpha}$ converges to its population counterpart with respect to the usual Euclidean distance.  The next result, which is proved in the Supplementary Material, establishes consistency of the sample procedure.

\begin{theorem}
\label{gen.thm}
Suppose that regularity conditions C1 and C2, given in Section~\ref{sec.reg.cond}, are satisfied. Then, the sample clustering procedure is strongly consistent.
\end{theorem}

\begin{remark}
It follows from the proof that the result continues to hold if we replace $\widehat {siz}e$ with $size$ in the definition of $\widehat\cS_{\alpha}$ and/or replace $size$ with $\widehat {siz}e$ in display~(\ref{eq3.cons.def}).
\end{remark}

\noindent If condition C2 is violated, and the locations of the population splits are not uniquely determined, a modification of Theorem~\ref{gen.thm} continues to hold.  More specifically, suppose that the number of versions of the population procedure is finite.  Then, the sample procedure converges to the \emph{set} of population versions, rather than a specific one.  In other words, the number and the locations of the big sample splits approach the corresponding quantities for an appropriately chosen population version, where the choice depends on the sample.

Consider the important case of a unimodal underlying distribution.    Proposition~\ref{prop.unimod} in Section~\ref{sec.reg.cond} and the proof of Theorem~\ref{gen.thm} imply that in this case condition C1 is not required for consistency.  Note also that the population procedure produces no splits, by Proposition~\ref{prop.unimod}.  It follows that all of the sample splits are uniformly asymptotically negligible.
\begin{corollary}
\label{unimod.crl}
If~$f$ is continuous and unimodal, then the maximum size of all the splits in the sample clustering procedure goes to zero almost surely.
\end{corollary}

In the next section we extend the results in Theorem~\ref{gen.thm} by establishing the rates of convergence for the sample clustering procedure.

\subsection{Rates of Convergence}

To establish the rates of convergence for the sample splitting procedure, we need an additional regularity condition.  We use the term \emph{population cluster} to refer to all intervals that appear along the path of the population procedure.
\begin{enumerate}
\item[{\bf C3}.]  For each population cluster $(L,R)$ and each $t\in\arg\max_{[L,R]} G_{L,R}$, if $t\in(L,R)$, then $G''(t)\ne0$, otherwise $G_{L,R}'(t)\ne0$.
\end{enumerate}
The requirement on $G_{L,R}(t)''$, imposed for each population split $(L,t,R)$, is the standard M-estimation assumption that requires the second derivative of the population criterion function to be nonsingular at the population maximum (e.g. \citealt{vaartwellner96book,vaart98book}).  The requirement on $G'_{L,R}$ is the analog of the aforementioned M-estimation assumption in the case where the population criterion function, $G_{L,R}$, is maximized at an endpoint of $[L,R]$, rather than in the interior.  In this case,  the behaviour of $G_{L,R}$ near its maximum is characterized by the first derivative, rather than by the second.

Let $\widehat \cS(\tau)$  contain all the sample splits $\cs=(L,\widehat t ,R)$, for which the sample frequencies of $(L,R)$, $(-\infty,\widehat t )$ and $(\widehat t ,\infty)$ are greater than or equal to~$\tau$.  In Theorem~\ref{rates.thm} we restrict our attention to the sample splits in $\widehat \cS(\tau)$, for arbitrarily small but positive $\tau$.  Without this restriction, the rate of convergence in~(\ref{eq2.rates.thm}) would change.  In particular, split sizes larger than $O_p(\log n/n)$ are produced when the sample procedure is applied to intervals whose widths tend to zero.  Also, larger split sizes may appear near the boundary of the support of the distribution.  The general approach used in the proof of Theorem~\ref{rates.thm} would also establish these slower rates of convergence.  However, the exact form of the new rates depends on the behavior of the density near the boundary of the support and on the aforementioned intervals of negligible width.  Instead, we present a clean result, with just one rate of convergence for all the small sample splits, while only imposing some simple regularity conditions.
\begin{theorem}
\label{rates.thm}
Suppose that regularity conditions C1-C3 are satisfied. Let~$\alpha^*$ be the smallest split size in the population procedure.  Then, for each~$\alpha$ in~$(0,\alpha^*)$,
\begin{eqnarray}
&&d_H(\widehat \cS_{\alpha},\cS)=O_p\left(n^{-1/3}\right)\qquad\text{and}\label{eq1.rates.thm}\\
&&\max\{size(\cs):\,\cs\in(\widehat \cS\setminus \widehat \cS_{\alpha})\cap \widehat \cS(\tau)\}=O_p\left(\log n/n\right). \label{eq2.rates.thm}
\end{eqnarray}
\end{theorem}
\begin{remark}
As we point out in the proof, if the domain, $(L_0,R_0)$, of the underlying distribution is bounded, then we can remove the lower bounds on the sample frequencies of $(-\infty,\widehat t )$ and $(\widehat t ,\infty)$ from the definition of $\widehat \cS(\tau)$ by assuming, instead, that $f(L_0)$ and $f(R_0)$ are nonzero.  It also follows from the proof that the result continues to hold if we replace $size$ with $\widehat {siz}e$ in display~(\ref{eq2.rates.thm}).
\end{remark}
The proof of Theorem~\ref{rates.thm} is provided in the Supplementary Material.  The intuition for the presented rates of convergence can be described, informally, as follows.  We bound the distance between the sample split point, $\widehat t$, and its population counterpart, $t$, by characterizing the behaviour of the sample criterion function near $t$.  The sample criterion, $\widehat G_{L,R}(a)$, is the empirical analog of the population criterion, $G_{L,R}(a)$, and is defined as the difference between the averages of the observations in $[a,R]$ and $[L,a]$, respectively.  We examine the decrease in $G_{L,R}(t)$ that occurs when~$t$ is perturbed by a small amount, $\delta$.  Then, we contrast this decrease with the stochastic term given by the difference between the corresponding deviations in $\widehat G_{L,R}$ and $G_{L,R}$.  The order of this term is roughly $\sqrt{\delta/n}$.  When~$t$ is a population split point and, thus, lies in the interior of~$(L,R)$, the corresponding decrease in $G_{L,R}$ is quadratic in~$\delta$.  Balancing out the two terms yields the cube root asymptotic behaviour (cf. \citealt{kim1990cube}) for the sample split point.  When function~$G_{L,R}$ is maximized at an endpoint of the interval $[L,R]$, as in the case of truncation, the decrease in~$G_{L,R}$ is linear in~$\delta$.  Balancing this decrease with the stochastic term of order $\sqrt{\delta/n}$ suggests that the sample split point is a $O_p(n^{-1})$ amount away from the boundary of $(L,R)$. Uniformity of the rate over all the small sample splits requires an additional $\log n$ factor.

In the next section we take advantage of our asymptotic results to propose a key modification to the sample clustering procedure.

\subsection{Big Merge Tracker: Post-processing the Sample Procedure}

Theorem~\ref{gen.thm} demonstrates that the sample analog of the truncation operation is peeling a large number of tiny clusters off the ends of a large cluster.  It follows that when recording sample splits we should distinguish between those that correspond to splits in the population procedure and those that correspond to truncations.  Based on this observation, we propose to post-process the sample clustering procedure by only keeping the splits with significant empirical sizes.  More specifically, given a threshold~$\alpha$, if the cardinality of one of the sub-clusters is below~$\alpha n$, the corresponding split is removed from the final output.

\begin{figure}[h]
\includegraphics[height=10pc,width=35pc]{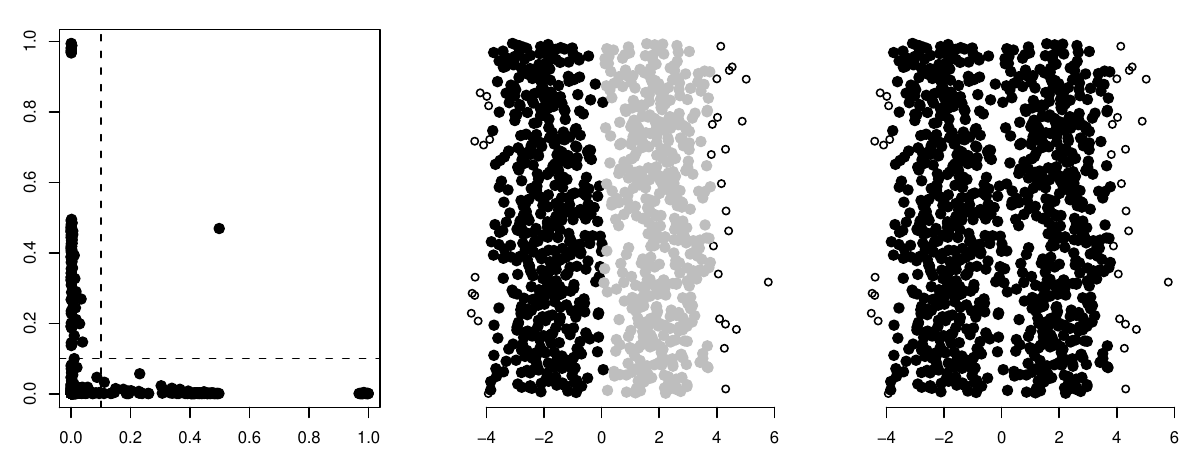}
\caption{The plots illustrate the path of Algorithm~\ref{al1} on a sample of~$1000$ observations from a symmetric Gaussian mixture, $0.5\,N(-2,1)+0.5\,N(2,1)$.  The scatter plot on the left displays the sample frequencies for each pair of clusters merged along the path.  The next two plots show the cluster memberships before and after the \emph{big merge}.  For the leftmost plot, the $x$ and~$y$ axes denote the proportions of the points in the two merging clusters. For the other two plots, only the $x$ axis, which marks the locations of the points, is informative.}\label{fig2}
\end{figure}
Figure~\ref{fig2} illustrates the path of Algorithm~\ref{al1} on a sample of~$1000$ observations, generated independently from the symmetric Gaussian mixture distribution used in Figure~\ref{fig1}.  The scatter plot on the left displays the sample frequencies for each pair of clusters merged along the path.  We found only one merge in which both clusters pass the $\alpha=0.1$ threshold.  The \emph{big merge} occurs at a point where the current number of clusters is~$32$.  The two rightmost plots display cluster memberships before and after the merge. The non-shaded points belong to clusters with non-appreciable size.

The equivalence between the splitting procedure and the merging algorithm implies that in the post-processing step of Algorithm~\ref{al1} we only keep the merges with the cardinality of each of the merging clusters above~$\alpha n$.  For any such merge, we place the split point midway between the two closest representatives of the two clusters being merged.   We also replace the stored merges with the corresponding split points.  The resulting sequence of split points can then be reinterpreted as a sequence of splits, or a sequence of merges, using the full sample.  For example, if the final output contains no split points, then all of the observations in the sample are placed in the same cluster.  We call this modified approach the \emph{Big Merge Tracker} (BMT) with threshold~$\alpha$.  As a direct consequence of Theorems~\ref{gen.thm} and~\ref{rates.thm}, under the respective regularity conditions, the BMT consistently estimates the number of population clusters, and its split points converge to their population counterparts at the $O_p(n^{-1/3})$ rate.  In the next section we analyze the empirical performance of the BMT approach.
%
%
%
%

%
%

\section{Simulation Study and Real Data Analysis}\label{sec.sim}

In this section we show strong performance of the proposed BMT approach relative to popular existing methods for assessing modality and estimating the number of clusters.  We also illustrate the use of our methodology in analysis of single-cell virology datasets.  In addition, in Section~5.3 of the Supplementary Material we apply BMT on very large simulated data sets and demonstrate its superior scalability properties.

When the separation between two clusters is very small, the population splitting procedure can still be successful at finding a split point by massively truncating the support. This \textit{zooming-in} effect may result in larger sizes of the \textit{small} sample splits, as we discussed, from a theoretical standpoint, in the paragraph above the statement of Theorem~\ref{rates.thm}.
To counteract this phenomenon, we propose an \textit{adjustment} to the Big Merge Tracker. If the sum of the sample frequencies for the two merging clusters in the last big merge is less than~50\%, we do not report any merges.  Preventing the corresponding splitting procedure from truncating more than~$50\%$ of the data, while searching for the first split, slightly reduces its efficiency, but makes it more robust to sampling fluctuations.  Throughout this section we use the adjustment described above and set the BMT threshold, $\alpha$, equal to~$0.1$ (Algorithm~1 in  Section~5 of the Supplementary Material contains the full pseudocode).  Note that a large number of additional simulation results, corresponding to a wider range of sample sizes, together with an analysis regarding the choice of~$\alpha$, are provided in Section~5.4 of the Supplementary Material.

\subsection{Modality Assessment} \label{sec.modal}
Testing for homogeneity of a population is an important statistical problem \citep{Aitkin-85,Muller-91,Roeder-94}. Here, we use the BMT to detect the presence of two or more dominant modes in the density.  In Table~\ref{table2} we compare our approach with two popular modality assessment procedures:
(i) kernel density estimate based test of \cite{Silverman-81} (ii) histogram based Diptest proposed by \cite{Hartigan-85}.
P-values of the Silverman test are calculated using the R-package referenced in \cite{Vollmer-13}.  R-packge of \cite{Diptest} is used for implementing the Dip test. Detailed descriptions of these procedures are given in Section~5.1 of the Supplementary Material.
\par
We consider~$5$ different simulation scenarios, in which~$100$ independent samples of size~$10000$ were generated and subjected to modality analysis. Table~\ref{table2} reports the percentage of cases in which multi-modality was detected.  P-values for the Dip and Silverman tests were computed based on~$1000$ MCMC simulations, and decision on the null hypothesis of unimodality was made at the~$5\%$ level of significance.  The mean and the standard deviation of the p-values from these tests are also reported.  In the two unimodal scenarios the BMT is on par with the Silverman and the Dip tests in confirming unimodality of the population distribution with high certainty. In the three non-unimodal cases,  which include normal and beta mixtures, the BMT shows better performance in detecting multi-modality.

\input{table-2}

\subsection{Estimating the Number of Clusters}
We  study the potency of the BMT in detecting the true number of clusters. We compare its performance with the following \textit{number of clusters} estimation methods:
(i) the CH index of \cite{Calinski-74} (ii) the KL index of  \cite{Krzanowski-88} (iii) the H measure of \cite{Hartigan-75} (iv) the Silhouette statistic based KR index of \cite{Kaufman-09}, (v) the Gap statistic of \cite{Tibshirani-01} (vi) the Jump statistic of \citet{Sugar-03} (vii) the clustering prediction strength criterion of \cite{Tibshirani-05}, and (viii) the bootstrap based cluster instability minimizing criterion of \cite{fang2012selection}, which is inspired by the stable clusters selection approach of \citet{wang2010consistent}. Detailed descriptions of these procedures are provided in Section~5.2 of the Supplementary Material.
We consider one multivariate and five univariate regimes. $100$ independent replications with the sample size of~$5000$ are used in each simulation setting, and the distribution of the number of clusters detected by each method is reported.  The eight competitor methods are implemented using a number of different clustering approaches via the \texttt{NbClust} R-package of \cite{NbClust} and the \texttt{fpc} package of \cite{fpc}.  More specifically, we use $k$-means clustering with the Euclidean distance metric (the corresponding results are reported in Table~\ref{table3}), as well as Ward's method \citep{ward1963hierarchical}, Centroid-based clustering \citep{Kaufman-09}, PAM \citep{kaufman1990partitioning}, CLARA \citep{kaufman1986clustering}, clustering by merging Gaussian mixture components \citep{hennig2010methods} and  hierarchical clustering initialized Gaussian mixture model based clustering method of \citet{fraley2002model}.  All the results for the non-$k$-means clustering approaches are reported in the tables provided in Section~5.2  
of the Supplementary Material.
\input{table-3}
\par
In our first univariate example, we consider a non-symmetric mixture of two normal densities. Each of them has unit variance and their means are fairly well-separated. We observe that the  CH, KL and Hartigan methods struggle to recover the bimodal structure, while the others successfully detect the two clusters (we note that in this setting, the CH index performs better when it uses the centroid based clustering algorithm).
The next three simulation scenarios correspond to non-symmetric tri-modal population densities that are mixtures of standard normals, non-central $t$-densities with one degree of freedom and double exponential densities with the unit rate parameter, respectively.
The medians of the mixing densities and the mixture weights match across the three settings, and the separation between the adjacent medians is not large.  In these three simulation settings, the Gap statistic approach, as well as the CH, KL and Hartigan measures, has difficulty detecting the true number of clusters.  The Silhouette method performs well for Gaussian mixtures but has difficulties in the other two cases.  Jump statistic, prediction strength and bootstrap stability approaches do well in the normal and the double exponential cases, however, they do not show good performance in the considerably thicker-tailed case of the mixture of $t$-densities.  For our fifth simulation setting we consider a bounded population density that is a mixture of three Beta densities. Here, only the BMT and the Silhouette do well in recovering the true number of clusters. In our last example we consider a~$5$ dimensional data set, which is generated from a product density.  The first dimension is generated  from a symmetric mixture of two Gaussians; the next two dimensions contain white noise, while the forth and fifth dimensions are generated from a central chi-square distribution with one degree of freedom.  We observe that,  together with the CH, KL and Hartigan measures, the Silhouette and the Jump approaches do not perform well in detecting the two clusters in this data.
\par
BMT does consistently well across each of the six simulation scenarios, outperforming all the other approaches overall.   The prediction strength and the cluster stability methods, which are modern state of the art approaches, deliver the best results among the competitors.  However, these two methods have significant trouble in the cases of the beta and the non-central~$t$ mixtures.  When implemented with various non-$k$-means clustering approaches (see Section~5.2 of the Supplementary Material), neither of the competitors considerably improves the performance reported in Table~\ref{table3}.
Figure~3 in Section~5.2 of the Supplementary Material provides plots of the densities used in the numerical experiments.

\subsection{Sub-population Analysis in Single Cell Virology}
We demonstrate an application of our clustering approach in a \textit{single-cell Mass Cytometry} \citep{Bendall11} based virology study.
We analyze the data reported in  \citet{Sen14}, where the effect of Varicella Zoster Virus (VZV) on human tonsil T cell is studied. VZV is a human herpesvirus and causes varicella and zoster \citep{Zerboni-14}. We study protein expressions  from five independent experiments, each containing an Uninfected (UN) and a Bystander (BY) populations. Bystanders are cells in the VZV infected population, which are not directly infected by the virus, but are influenced by neighboring virus infected cells. Protein expression values are studied on the $\text{arcsinh}$ scale. Non-expressed values are uniformly distributed between $[-1,1]$.  Cellular sub-populations are detected by clustering the populations based on the expressions of ``core-proteins", which are associated with T cell activation \citep{Newell12}.
Most of the samples have large sizes, usually on the order of $\sim 10^5$. Traditional clustering techniques fail to accommodate such large sample sizes and resort to sub-sampling based approaches \citep{Qui11,Linderman12}. The BMT, on the other hand, has the advantage of being scalable enough to conduct clustering analysis on  the entire sample.
\input{table-5}
\par
We treat three proteins, CD4, CD8 and CD45RA (naive), as core-proteins, as they are typically used to classify T cells.  For each of the $10$ samples (UN and BY from experiments I-V), based on the expressions of the above three proteins, we performed automated clustering by using BMT in the three dimensional space. Figure~4 and~5 in Section~5.5 of the Supplementary Material show that in all the cases, the BMT detects unimodality for CD4 and CD45RA  and bimodality for CD8 expression values. Using the bi-modality of  CD8 and the BMT detected splits, we classify cells as CD8-high and CD8-low. Also, considering the expression and non-expressions of the other two markers we simultaneously classify cells into the following clusters, or sub-populations:
(i) Dual positive: CD4 expressed and CD8 High
(ii) Dual-negative: CD4 non-expressed and CD8 low
(iii) CD4 Non-Naive: CD4 expressed and CD45RA non-expressed
(iv) CD4 Naive: CD4 expressed and CD45RA non-expressed
(v) CD8 Naive: CD8 high and CD45RA expressed.
\par
Table~\ref{table5} reports the sizes and proportional representations of these sub-populations, across the five experiments. BMT sub-populations resemble the T-cell biology based phenotypic  classification in \citet{Sen14}. They also revalidate that the sub-population distribution in the Bystander cells is not much different from that of the uninfected, though the UN sub-population distribution varies across experiments.
Using this BMT based categorization of the T cells, sub-population level cell-signaling patterns can be subsequently studied. Figure~6 in Section~5.5 of the Supplementary Material shows the heatmap of the protein expressions (core + signaling proteins) of the sub-populations from Experiment I.



\section{Discussion}
\label{sec.disc}

In this paper we focus on the analysis of a popular convex clustering approach that is based on the $\ell_1$ fusion regularization.  However, we believe that our general theoretical framework can be extended to handle other types of fusion penalties.  In particular, based on a preliminary analysis of the corresponding~$\ell_2$ approach, we conjecture that the asymptotic results in this paper also hold in the $\ell_2$ case, under appropriate regularity conditions.  The corresponding population procedure can be defined by analogy, as a collection of cluster splits and truncations, where each operation seeks to maximize the Euclidian distance between the corresponding sub-cluster means.  The proofs require a rigorous formulation and a thorough analysis of a multivariate analog of the truncation operation.

High computational efficiency of the proposed BMT approach allows it to be applied to massive high-dimensional data sets.  In particular, BMT can be used for high-dimensional feature screening, to rule out predictors that do not reveal any clusters in the data.  Moreover, we can take further advantage of BMT's computational efficiency and apply the screening procedure to several linear combinations for each pair of variables.  This way we can move beyond marginal screening, similarly to how regression models with interaction terms move beyond the simple additive structure.


\section{Supplementary Material}
\input{appendix-1.tex}

\input{appendix-2.tex}

\input{appendix-3.tex}

\section*{Acknowledgement}
Radchenko's research was partially supported by NSF Grant DMS-1209057.  Computing resources were provided by the USC Center for High Performance Computing.  The R code and the data sets used for producing the numerical results of Section~4 can be downloaded from \url{http://www-bcf.usc.edu/~gourab/code-bmt}.


\bibliographystyle{natbib}
{\setstretch{1.0}
\bibliography{ref-1,ref-2,append}}
\end{document}

%% file: table-2.tex
\begin{table}[h]
\centering
\scalebox{0.68}{
\begin{tabular}{|r|r|r|r||r|r|r||r||}
 \hline
 Population Density  & \multicolumn{3}{c||}{Dip Test P-value (D)} & \multicolumn{3}{|c||}{Silverman Test P-value (S)} &  BMT\\
 \cline{2-4} \cline{5-7}
 &  \small{Mean (D)} & \small{Std(D)} & \small{\textbf{\% multi-mode}} & \small{Mean (S)} & \small{Std(S)} & \small{\textbf{\% multi-mode}}  & \small{\textbf{\% multi-mode}}\\
  \hline & & & & & & & \\
\multirow{1}{*}{N(0,1)}  & 0.99 & 0.04 & 0.00 & 0.48 & 0.25 & 0.00 & 0.00\\[1ex] 
\hline
  \hline & & & & & & & \\
\multirow{1}{*}{Beta(2,4)}   & 0.98 & 0.04 & 0.00 & 0.54 & 0.28 & 2.00 & 0.00\\[1ex]     
\hline
  \hline & & & & & & & \\
\multirow{1}{*}{$ \big \{N(-1.1,1) + N(1.1,1)\big\}/2$}   & 0.81 & 0.22 & 0.00 & 0.22 & 0.21 & 29.00 & 69.00\\[1ex]    
\hline
  \hline & & & & & & & \\
\multirow{1}{*}{$\big \{Beta(4,6) + Beta(7,3)\big\}/2$} & 0.84 & 0.22 & 0.00 & 0.31 & 0.25 & 21.00 & 49.00\\[1ex]     
\hline
  \hline & & & & & & & \\
\multirow{1}{*}{$ \big\{N(-2.5,1) + N(0,1) + N(2.5,1)\big\}/3$} & 0.10 & 0.14 & 52.00 & 0.03 & 0.03 & 79.00 & 96.00\\[1ex]     
\hline
\hline
\end{tabular}
}
\caption{Simulation study to compare multi-modality detection methods}\label{table2}
\end{table}

%% file: table-3.tex
\begin{table}
\centering
\scalebox{0.75}{
\begin{tabular}{|c|r|r|r|r|r|r|r|r|r|r|r|}
\cline{1-12}
True Population Density & Methods &\multicolumn{10}{ |c| }{\;\;\;\;Number of Clusters} \\ \cline{3-12}
&   &  1 & 2 & 3 & 4 & 5 & 6 & 7 & 8 & 9 & 10+\\ \cline{1-12}
\multicolumn{1}{ |c  }{\multirow{9}{*}{$0.3\, N(-4,1) + 0.7 \,N(4,1) $} } &
\multicolumn{1}{ |c| }{CH} & 0 & \textbf{29} & 12 &   4  & 2&   2 &  4&  11&  14 &  22  \\ \cline{2-12}
\multicolumn{1}{ |c  }{}       &
\multicolumn{1}{ |c| }{KL} & 0 & \textbf{39} & 10 &  5 &  5  & 8 &  8 &  8 &  9 &  8  \\ \cline{2-12}
\multicolumn{1}{ |c  }{}       &
\multicolumn{1}{ |c| }{Hartigan} & 0 & \textbf{0} & 32 &16& 12 &10 &11&  5 & 7  &7  \\ \cline{2-12}
\multicolumn{1}{ |c  }{}       &
\multicolumn{1}{ |c| }{Silhouette} & 0 & \textbf{100} & 0 & 0 & 0 & 0 & 0 & 0 & 0 & 0  \\ \cline{2-12}
\multicolumn{1}{ |c  }{}       &
\multicolumn{1}{ |c| }{Gap} & 0 & \textbf{100} & 0 & 0 & 0 & 0 & 0 & 0 & 0 & 0    \\ \cline{2-12}
\multicolumn{1}{ |c  }{}       &
\multicolumn{1}{ |c| }{Jump} & 0 & \textbf{99} & 0 & 0 & 0 & 0 & 0 & 0 & 0 & 1  \\ \cline{2-12}
\multicolumn{1}{ |c  }{}       &
\multicolumn{1}{ |c| }{Pred Str.} & 0 & \textbf{100} & 0 & 0 & 0 & 0 & 0 & 0 & 0 & 0  \\ \cline{2-12}
\multicolumn{1}{ |c  }{}       &
\multicolumn{1}{ |c| }{Stability} & 0 & \textbf{100} & 0 & 0 & 0 & 0 & 0 & 0 & 0 & 0  \\ \cline{2-12}
\multicolumn{1}{ |c  }{}                        &
\multicolumn{1}{ |c| }{BMT} & 0 & \textbf{93} & 7 & 0 & 0 & 0 & 0 & 0 & 0 & 0      \\ \cline{2-12}
\hline
\hline
\multicolumn{1}{ |c  }{\multirow{9}{*}{$0.3\, N(-3,1) + 0.35 \,N(0,1) + 0.35 \,N(3,1)$} } &
\multicolumn{1}{ |c| }{CH}  & 0 & 0 & \textbf{0} & 0 & 1  & 0 &  6 & 13 & 28 & 52    \\ \cline{2-12}
\multicolumn{1}{ |c  }{}       &
\multicolumn{1}{ |c| }{KL} & 0 & 11 & \textbf{8} &11 & 9 & 6 &14& 12& 19& 10      \\ \cline{2-12}
\multicolumn{1}{ |c  }{}       &
\multicolumn{1}{ |c| }{Hartigan} & 0 & 0 & \textbf{69} &13  &7  &4 & 2 & 4  &1 & 0  \\ \cline{2-12}
\multicolumn{1}{ |c  }{}       &
\multicolumn{1}{ |c| }{Silhouette} & 0 & 8 & \textbf{92}  & 0 & 0 & 0 & 0 & 0 & 0 & 0  \\ \cline{2-12}
\multicolumn{1}{ |c  }{}       &
\multicolumn{1}{ |c| }{Gap} & 0 & 100 & \textbf{0}  & 0 & 0 & 0 & 0 & 0 & 0 & 0    \\ \cline{2-12}
\multicolumn{1}{ |c  }{}       &
\multicolumn{1}{ |c| }{Jump} & 0 & 0 & \textbf{100}  & 0 & 0 & 0 & 0 & 0 & 0 & 0  \\ \cline{2-12}
\multicolumn{1}{ |c  }{}       &
\multicolumn{1}{ |c| }{Pred Str.} & 0 & 0 &\textbf{100} & 0 & 0 & 0 & 0 & 0 & 0 & 0  \\ \cline{2-12}
\multicolumn{1}{ |c  }{}       &
\multicolumn{1}{ |c| }{Stability} & 0 & 11 & \textbf{89} & 0 & 0 & 0 & 0 & 0 & 0 & 0  \\ \cline{2-12}
\multicolumn{1}{ |c  }{}       &
\multicolumn{1}{ |c| }{BMT} & 0 & 5 & \textbf{95} & 0 & 0 & 0 & 0 & 0 & 0 & 0      \\ \cline{2-12}
\hline
\hline
\multicolumn{1}{ |c  }{\multirow{9}{*}{$0.3\, t_1(-3) + 0.35 \,t_1(0) + 0.35 \,t_1(3)$} } &
\multicolumn{1}{ |c| }{CH}    & 0 & 7  & \textbf{9} & 7 & 11 &  7 &  6 & 10 &14 & 29\\ \cline{2-12}
\multicolumn{1}{ |c  }{}       &
\multicolumn{1}{ |c| }{KL} & 0 &18& \textbf{23} & {15} &14 & 5  &4 & 9  &6 & 6    \\ \cline{2-12}
\multicolumn{1}{ |c  }{}       &
\multicolumn{1}{ |c| }{Hartigan} & 0 & 0 &\textbf{46} & 24 &11 & 5 & 4  &5  &1  &4  \\ \cline{2-12}
\multicolumn{1}{ |c  }{}       &
\multicolumn{1}{ |c| }{Silhouette} & 0 & 71 & \textbf{26}  & 3 & 0 & 0 & 0 & 0 & 0 & 0  \\ \cline{2-12}
\multicolumn{1}{ |c  }{}       &
\multicolumn{1}{ |c| }{Gap} & 0 & 36 & \textbf{64}  & 0 & 0 & 0 & 0 & 0 & 0 & 0    \\ \cline{2-12}
\multicolumn{1}{ |c  }{}       &
\multicolumn{1}{ |c| }{Jump} & 17 & 11 & \textbf{9}  & 13 & 12 & 5 & 11 & 9 & 7 & 6  \\ \cline{2-12}
\multicolumn{1}{ |c  }{}       &
\multicolumn{1}{ |c| }{Pred Str.} & 0 & 24 &\textbf{53} & 19 & 4 & 0 & 0 & 0 & 0 & 0  \\ \cline{2-12}
\multicolumn{1}{ |c  }{}       &
\multicolumn{1}{ |c| }{Stability} & 0 & 69 & \textbf{30} & 0 & 1 & 0 & 0 & 0 & 0 & 0  \\ \cline{2-12}
\multicolumn{1}{ |c  }{}       &
\multicolumn{1}{ |c| }{BMT} & 0 & 1 & \textbf{99} & 0 & 0 & 0 & 0 & 0 & 0 & 0      \\ \cline{2-12}
\hline
\hline
\multicolumn{1}{ |c  }{\multirow{9}{*}{$0.3\, \text{dexp}(-3) + 0.35 \,\text{dexp}(0) + 0.35 \,\text{dexp}(3)$ }} &
\multicolumn{1}{ |c| }{CH} & 0 & 0 & \textbf{0} & 0 & 1  & 0 &  6 & 14 & 23 & 56   \\ \cline{2-12}
\multicolumn{1}{ |c  }{}       &
\multicolumn{1}{ |c| }{KL} & 0 & 13 & \textbf{8} &18 & 9& 11 &7& 20& 10& 4  \\ \cline{2-12}
\multicolumn{1}{ |c  }{}       &
\multicolumn{1}{ |c| }{Hartigan} & 0 & 0 & \textbf{55} &19  & 9  & 10 & 3 & 1  &1 & 2  \\ \cline{2-12}
\multicolumn{1}{ |c  }{}       &
\multicolumn{1}{ |c| }{Silhouette} & 0 & 39 & \textbf{61}  & 0 & 0 & 0 & 0 & 0 & 0 & 0  \\ \cline{2-12}
\multicolumn{1}{ |c  }{}       &
\multicolumn{1}{ |c| }{Gap} & 0 & 100 & \textbf{0}  & 0 & 0 & 0 & 0 & 0 & 0 & 0    \\ \cline{2-12}
\multicolumn{1}{ |c  }{}       &
\multicolumn{1}{ |c| }{Jump} & 0 & 0 & \textbf{100}  & 0 & 0 & 0 & 0 & 0 & 0 & 0  \\ \cline{2-12}
\multicolumn{1}{ |c  }{}       &
\multicolumn{1}{ |c| }{Pred Str.} & 0 & 0 &\textbf{100} & 0 & 0 & 0 & 0 & 0 & 0 & 0  \\ \cline{2-12}
\multicolumn{1}{ |c  }{}       &
\multicolumn{1}{ |c| }{Stability} & 0 & 0 & \textbf{100} & 0 & 0 & 0 & 0 & 0 & 0 & 0  \\ \cline{2-12}
\multicolumn{1}{ |c  }{}       &
\multicolumn{1}{ |c| }{BMT} & 0 & 0 & \textbf{100} & 0 & 0 & 0 & 0 & 0 & 0 & 0      \\ \cline{2-12}
\hline
\hline
\multicolumn{1}{ |c  }{\multirow{9}{*}{$ \big \{ Beta(8,2)+Beta(5,5)+Beta(2,8) \big\}\big /3$ }} &
\multicolumn{1}{ |c| }{CH} & 0 & 0 & \textbf{2} & 0 & 3  & 3 &  6 & 16 & 22 & 48   \\ \cline{2-12}
\multicolumn{1}{ |c  }{}       &
\multicolumn{1}{ |c| }{KL} & 0 & 13 & \textbf{3} &11 & 9 & 10 &15& 16& 9& 14   \\ \cline{2-12}
\multicolumn{1}{ |c  }{}       &
\multicolumn{1}{ |c| }{Hartigan} & 0 & 0 & \textbf{57} &14  &9  &6 & 5 & 5 & 0 & 4  \\ \cline{2-12}
\multicolumn{1}{ |c  }{}       &
\multicolumn{1}{ |c| }{Silhouette} & 0 & 0 & \textbf{100}  & 0 & 0 & 0 & 0 & 0 & 0 & 0  \\ \cline{2-12}
\multicolumn{1}{ |c  }{}       &
\multicolumn{1}{ |c| }{Gap} & 0 & 100 & \textbf{0}  & 0 & 0 & 0 & 0 & 0 & 0 & 0    \\ \cline{2-12}
\multicolumn{1}{ |c  }{}       &
\multicolumn{1}{ |c| }{Jump} & 99 & 0 & \textbf{1}  & 0 & 0 & 0 & 0 & 0 & 0 & 0  \\ \cline{2-12}
\multicolumn{1}{ |c  }{}       &
\multicolumn{1}{ |c| }{Pred Str.} & 0 & 78 &\textbf{17} & 4 & 1 & 0 & 0 & 0 & 0 & 0  \\ \cline{2-12}
\multicolumn{1}{ |c  }{}       &
\multicolumn{1}{ |c| }{Stability} & 0 & 0 & \textbf{25} & 60 & 15 & 0 & 0 & 0 & 0 & 0  \\ \cline{2-12}
\multicolumn{1}{ |c  }{}       &
\multicolumn{1}{ |c| }{BMT} & 0 & 0 & \textbf{100} & 0 & 0 & 0 & 0 & 0 & 0 & 0      \\ \cline{2-12}
\hline
\hline
\multicolumn{1}{ |c  }{}       &
\multicolumn{1}{ |c| }{CH} & 0 & \textbf{0} & 1 & 0  & 4 &  5 & 11 & 10 & 23 & 46  \\ \cline{2-12}
\multicolumn{1}{ |c  }{$\{ 0.5\, N(-2,1) + 0.5\,N(2,1) \}$}       &
\multicolumn{1}{ |c| }{KL} & 0 &  \textbf{18} & 11 & 14 & 7 & 13 & 9 & 11 & 11 & 6   \\ \cline{2-12}
\multicolumn{1}{ |c  }{$\quad \otimes \; N(0,1) $}       &
\multicolumn{1}{ |c| }{Hartigan} & 0 &  \textbf{0} & 52 & 16  & 13  & 4 & 7 & 4  & 4 & 0  \\ \cline{2-12}
\multicolumn{1}{ |c  }{$ \quad \otimes \; N(0,1) $}       &
\multicolumn{1}{ |c| }{Silhouette} & 0 &  \textbf{0} & 87 & 12 & 0 & 1 & 0 & 0 & 0 & 0  \\ \cline{2-12}
\multicolumn{1}{ |c  }{$ \quad \otimes \; \chi^2_1 $}       &
\multicolumn{1}{ |c| }{Gap} & 0 & \textbf{100} & 0 & 0 & 0 & 0 & 0 & 0 & 0 & 0    \\ \cline{2-12}
\multicolumn{1}{ |c  }{$ \quad \otimes \; \chi^2_1 $}       &
\multicolumn{1}{ |c| }{Jump} & 0 & \textbf{14} & 0  & 62 & 0 & 0 & 0 & 0 & 8 & 16  \\ \cline{2-12}
\multicolumn{1}{ |c  }{}       &
\multicolumn{1}{ |c| }{Pred Str.} & 0 & \textbf{100} & 0 & 0 & 0 & 0 & 0 & 0 & 0 & 0  \\ \cline{2-12}
\multicolumn{1}{ |c  }{}       &
\multicolumn{1}{ |c| }{Stability} & 0 & \textbf{85} & 0 & 15 & 0 & 0 & 0 & 0 & 0 & 0  \\ \cline{2-12}
\multicolumn{1}{ |c  }{} &
\multicolumn{1}{ |c| }{BMT} & 0 & \textbf{96} & 4 & 0 & 0 & 0 & 0 & 0 & 0 & 0      \\ \cline{2-12}
\hline
\hline

\end{tabular}
}
\caption{Number of clusters detected in $100$ trials for six simulation scenarios}\label{table3}
\end{table}

%% file: table-5.tex
\begin{table}[h]
\centering
\scalebox{0.665}{
\begin{tabular}{||c||c|c||c|c||c|c||c|c||c|c||}
 \hline
 \hline
 & \multicolumn{2}{|c|}{Experiment I} & \multicolumn{2}{|c|}{Experiment II}  & \multicolumn{2}{|c|}{Experiment III} & \multicolumn{2}{|c|}{Experiment IV} & \multicolumn{2}{|c|}{Experiment V} \\[0.5ex]
  \cline{2-3}\cline{4-5}\cline{6-7}\cline{8-9}\cline{10-11}
 SUB-POPULATIONS & UN & BY & UN & BY & UN & BY & UN & BY& UN & BY\\[0.5ex]
\hline
  \hline & & & & &  & & & & & \\
 DUAL POSITIVE & 8411 & 6596 & 5253 & 4169 & 4971 & 2703 & 3795 & 1510 & 8047 & 5225\\
 & (8.8\%) & (7.3\%) & (5.8\%)& 5.7\%)& (6.0\%)& (5.4\%) & (5.0\%) & (4.6\%) & (8.5\%) & (8.0\%) \\[1ex]
  \hline & & & & &  & & & & & \\
 DUAL NEGATIVE & 2723 & 2973 & 3537 & 2631 & 4433 & 2935 & 4354 & 2196 & 5012 & 2881 \\
 & (2.8\%) & (3.3\%) & (3.9\%) & (3.6\%) & (5.3\%) & (5.9\%) & (5.8\%) & (6.7\%) & (5.3\%) & (4.4\%) \\[1ex] 
  \hline & & & & &  & & & & & \\
 CD4 NON-NAIVE & 7993 & 10636 & 15144 & 11556 & 21444 & 12429 & 22149 & 8508 & 30034 & 20469 \\ 
 & (8.4\%) & (11.8\%) & (16.7\%) & (15.9\%) & (25.9\%) & (25.0\%) & (29.7\%) & (26.1\%) & (31.9\%) & (31.3\%) \\[1ex] 
  \hline & & & & &  & & & & & \\
CD4 NAIVE & 69977 & 64119 & 57744 & 47374 & 45764 & 27987 & 35458 & 16390 & 40398 & 28524 \\ 
& (73.7\%)& (71.1\%) & (63.7\%) & (65.1\%) & (55.3\%) & (56.3\%) & (47.5\%) & (50.4\%) & (43.0\%) & (43.7\%) \\[1ex] 
  \hline & & & & &  & & & & & \\
CD8 NAIVE & 5654 & 5671 & 8599 & 6571 & 5798 & 3490 & 8271 & 3774 & 9869 & 7829 \\ 
& (6.0\%) & (6.3\%) & (9.5\%) & (9.0\%) & (7.0\%) & (7.0\%) & (11.1\%) & (11.6\%) & (10.5\%) & (12.0\%) \\[1ex] 
  \hline
  \hline & & & & &  & & & & & \\[1ex]
POPULATION SIZE & 94837 & 90157 & 90641 & 72699 & 82637 & 49672 & 74540 & 32497 & 93878 & 65244 \\[1ex]
   \hline
  \hline
\end{tabular}
}
\caption{Sizes and Proportions of dominant clusters detected by BMT across $5$ independent Virology experiments}\label{table5}
\end{table}

%% file: appendix-1.tex
\subsection{Proof of Theorem~\ref{gen.thm}}

\subsubsection*{Preliminaries}

Here we provide three lemmas that make important contributions to the proof.  The first allows us to focus on a compact support, the second established uniform convergence of the sample criterion functions on compact sets, and the third derives useful continuity properties of the population solution.   The proofs of the lemmas are provided below, after the main argument for Theorem~\ref{gen.thm}.   All of the results take advantage of regularity conditions C1 and C2, which are given in Section~\ref{sec.reg.cond} of the main paper.  In the proof of the lemmas we verify that these conditions can be removed for densities~$f$ that are unimodal and continuous.
First, we handle the case where the support of density is unbounded.  By analogy with the term population cluster (page~13 of the main paper), we use \textit{sample cluster} to refer to intervals that appear along the path of the sample clustering procedure.  We follow the traditional approach of using~$\omega$ to denote a general element of the sample space.

\begin{lemma}
\label{lem.comp}
For every positive~$\epsilon$, $\delta$ and~$\delta'$, and almost all~$\omega$, there exist a sample cluster $(\widehat L, \widehat R)$, a population cluster $(L,R)$, and a bounded interval~$B_\epsilon$, which depends only on~$\epsilon$, such that, for all sufficiently large~$n$,
\begin{enumerate}
\item[(a)] $P_{-\infty,\widehat L}\le\epsilon$ and $P_{\widehat R,\infty}\le\epsilon$
\item[(b)] $|\widehat R - R|\le \delta$ and $|\widehat L - L|\le \delta'$
\item[(c)] $(\widehat L,\widehat R)\subseteq B_\epsilon$.
\end{enumerate}
\end{lemma}
\begin{remark}
Note that $L$, $R$, $\widehat L$ and~$\widehat R$ depend on~$\omega$, while $B_\epsilon$ does not.
\end{remark}

To draw connections between the sample and the population clustering procedures, we define the sample criterion function, $\widehat G_{L,R}$, as the empirical analog of the population criterion, $G_{L,R}$. More formally, we write~$\widehat \mu_{l,r}$ for the the sample average on $[l,r]$ and let
\begin{equation}
\widehat G_{L,R}(a)=\widehat\mu_{a,R}-\widehat\mu_{L,a},
\end{equation}
for~$a\in[L,R]$, with the following important caveat.  If either of the intervals $[L,a]$ and $[a,R]$ contains no observations, we consider $\widehat G_{L,R}(a)$ to be \textit{undefined}.  We now summarize the sample clustering procedure using the newly defined criterion function.  Given a cluster $(L,R)$, the sample procedure finds a point~$s$ that maximizes $\widehat G_{L,R}$, then splits $(L,R)$ into sub-clusters $(L,s)$ and $(s,R)$.  Note that $s$ is guaranteed to be located strictly between two of the observations.

To understand the behavior of the sample clustering procedure on a bounded support, we establish uniform convergence of the of the sample criterion functions to their population counterparts.
\begin{lemma}
\label{lem.unif.appr}
Define $A_{M}=\{(L,R),\, -M\le L< R\le M\}$.  Then, as~$n\rightarrow\infty$,
\begin{equation}
\sup_{(L,R) \in A_{M}}\;\;\max_{a\in [L,R]} |\widehat G_{L,R}(a)-G_{L,R}(a)|\;\rightarrow \;0,
\end{equation}
almost surely, for each positive $M$.
\end{lemma}
\begin{remark}
In the expression $\max_{a\in[L,R]}\widehat G_{L,R}(a)$ the maximum is taken over the closed interval $[\,\min\{\xx_n\cap[L,R]\}\,,\,\max\{\xx_n\cap[L,R]\}\,]$.  By the classical Glivenko-Cantelli theorem, the endpoints of this interval converge to~$L$ and~$R$ almost surely, and the convergence is uniform over $(L,R)$ in $A_M$, for each fixed~$M$.
\end{remark}

The next result derives some important properties of the population procedure, which we then relate to the sample procedure with the help of Lemma~\ref{lem.unif.appr}.
To state the result, we need to modify~$R_L$, which is a function of~$L$ that appears in the definition of the two sided truncation.  Suppose that the population procedure truncates interval $(L_s,R_s)$ to interval $(L^*,R^*)$ in a two-sided fashion.  We will extend function~$R_L$ to be defined in open neighborhoods of~$L_s$ and $L^*$, using the equation $G_{L,R_L}(L)=G_{L,R_L}(R_L)$.  The proof of Proposition~\ref{prop.gen} shows that, provided the neighborhoods are small enough, the extension is uniquely determined if we require that function~$R_L$ remains continuous and decreasing.  To simplify the notation we write $\max G_{L,R}$ for $\max_{a\in[L,R]}G_{L,R}(a)$.
\begin{lemma}
\label{lem.pop.ineq}
Let $(L_s,R_s)$ be a bounded cluster, produced by the population procedure, such that $\argmax G_{L_s,R_s}=\{L_s,R_s\}$.
Suppose that the first split of the population procedure applied to $(L_s,R_s)$ is given by $(L^*,s^*,R^*)$.
For every sufficiently small positive~$\epsilon$ there exist positive $\epsilon'$, $\delta$ and $\delta'$, such that $\epsilon'\le\epsilon$, $\delta'\le\delta\le\epsilon/2$ and $|R_L-R_{L+\delta'}|\le\delta$ for all~$L$ in~$[L_s-\delta', L^*+\epsilon']$, and the following sets of inequalities hold as long as $|R-R_L|\le 2\delta$:
\begin{equation*}
\begin{array}{lcrcl}
\max G_{L,R}&>&\max\limits_{[L+\delta',R-\delta]}  G_{L,R}& \qquad \text{for}& L\in [L_s-\delta', L^*-\epsilon'],\\
\\
\max  G_{L,R}&>&\max\limits_{[L+\delta',R-\delta]\setminus(s^*-\epsilon,s^*+\epsilon)}  G_{L,R} &\qquad \text{for}& L\in[ L^*-\epsilon',L^*+\epsilon'/2],\\
\\
\max  G_{L,R}&>&\max\limits_{[L,R]\setminus(s^*-\epsilon,s^*+\epsilon)}  G_{L,R} &\qquad \text{for}& L\in[L^*+\epsilon'/2,L^*+\epsilon'].
\end{array}
\end{equation*}
\begin{equation*}
\begin{array}{lcrcl}
\max\limits_{[R-\delta,R]}  G_{L,R} &>& \max\limits_{[L,L+\delta']}  G_{L,R} & \;\text{for}& L\in [L_s-\delta', L^*+\epsilon'], \;\; R\in[R_L+\delta,R_L+2\delta],\\
\\
\max\limits_{[L,L+\delta']}  G_{L,R} &>& \max\limits_{[R-\delta,R]}  G_{L,R}&  \;\text{for}& L\in [L_s-\delta', L^*+\epsilon'], \;\; R\in[R_L-2\delta,R_L-\delta].\\
\end{array}
\end{equation*}
\end{lemma}
\begin{remark}
A minor modification is required when $L_s=L_0$ or $R_s=R_0$, where $(L_0,R_0)$ is the support of~$f$: each interval appearing in the statement needs to be replaced by its intersection with~$[L_0,R_0]$.
\end{remark}
We will now use Lemma~\ref{lem.unif.appr} to show that if we replace~$G$ with~$\widehat G$ in the five sets of inequalities above, then for each~$\omega$ in a set of probability one the resulting sample inequalities simultaneously hold for all sufficiently large~$n$.  For example, the first set becomes $\max \widehat G_{L,R}>\max\limits_{[L+\delta,R-\delta]}  \widehat G_{L,R}$.  We also allow the cluster $(L_s,R_s)$ to depend on~$\omega$, under an additional assumption that there exists a finite deterministic $M$, for which bound $\max\{|L_s-\delta'|,|R_s+\delta|\}\le M$ holds almost surely.  To establish the sample inequalities, we define $\Delta_n=\sup_{A_{M}}\max |\widehat G_{L,R}-G_{L,R}|$ and
\begin{equation*}
\tau\;=\;\min_{L\in [L_s-\delta', L^*-\epsilon'],\, |R-R_L|\le 2\delta}\,\left(\max G_{L,R}-\max_{[L+\delta',R-\delta]}  G_{L,R}\right).
\end{equation*}
Continuity of $G_{L,R}(a)$ with respect to~$a$, $L$ an~$R$, together with the first set of inequalities in Lemma~\ref{lem.pop.ineq}, implies $\tau>0$.
Taking into account Lemma~\ref{lem.unif.appr}, as well as the continuity of~$G_{L,R}(a)$, we derive the following inequalities, which hold for all~$\omega$ in a set of probability one, and all sufficiently large~$n$:
\begin{equation*}
\max \widehat G_{L,R}-\max\limits_{[L+\delta,R-\delta]}  \widehat G_{L,R}>\max G_{L,R}-\max\limits_{[L+\delta,R-\delta]}  G_{L,R}-2\Delta_n-o(1)\ge\tau-o(1)>0.
\end{equation*}
The argument for the rest of the inequalities is analogous.

\subsubsection*{Main Body of the Proof}

We restrict our attention to the set of probability one that remains after we cast out the negligible sets during the application of the almost sure results from Lemma~\ref{lem.comp} and Lemma~\ref{lem.unif.appr}.  We focus on a fixed~$\omega$, but suppress the dependence on~$\omega$ to simplify the notation.

Consider an arbitrary positive~$\epsilon$.
We will investigate the behavior of the sample procedure, starting with an appropriately bounded sample cluster, $(\widehat L_s, \widehat R_s)$, that is close to some population cluster, $(L_s,R_s)$.  More specifically, we assume $(\widehat L_s, \widehat R_s)\subseteq B_\epsilon$, where~$B_\epsilon$ is a bounded interval that does not depend on~$\omega$.  We also assume $|\widehat R_s - R_s|\le\delta$ and $|\widehat L_s - L_s|\le\delta'$, for all sufficiently large~$n$, where~$\delta$ and~$\delta'$ are to be specified later.    We will track the sample and population procedures up to the first population split and show, under only the above assumptions, that the two procedures are close in an appropriate sense.
We need to consider the following options for the population procedure applied to $(L_s,R_s)$.  First: two-sided truncation, then a split; second: one-sided truncation, then a two sided truncation, followed by a split; third: one-sided truncation, followed by a split.  There are three more possibilities, where in each of the initial options the procedure does not produce a split.  Finally, the cluster $(L_s,R_s)$ may be split right away.

We first focus on the important case of a two-sided truncation, followed by a split.  The rest of the cases can be handled with only minor modifications to the argument.   Let $sp=(L^*,s^*,R^*)$ be the first split of the population procedure applied to $(L_s,R_s)$.  The population procedure performs a two-sided truncation along $(L,R_L)$, where $R_L$ is a continuous decreasing function of~$L$, such that $R_{L_s}=R_s$ and $R_{L^*}=R^*$.  Given our values of~$\epsilon$, $L_s$, $R_s$ and $sp$,  we let~$\epsilon'$, $\delta$ and $\delta'$ be the quantities from Lemma~\ref{lem.pop.ineq}.  Recall that $\epsilon'\le\epsilon$ and $\delta'\le\delta\le\epsilon/2$.  As we demonstrate in the paragraph below the statement of Lemma~\ref{lem.pop.ineq}, the five sets of population inequalities given in the lemma have natural sample counterparts.  The first three sets of inequalities hold for all~$R$ with $|R-R_L|\le 2\delta$ and all sufficiently large~$n$:
\begin{equation}
\label{ineq.sam.3sets}
\begin{array}{lcrcl}
\max \widehat G_{L,R}&>&\max\limits_{[L+\delta',R-\delta]}  \widehat G_{L,R}& \qquad \text{for}& L\in [L_s-\delta', L^*-\epsilon']\\
\\
\max  \widehat G_{L,R}&>&\max\limits_{[L+\delta',R-\delta]\setminus(s^*-\epsilon,s^*+\epsilon)}  \widehat G_{L,R} &\qquad \text{for}& L\in[ L^*-\epsilon',L^*+\epsilon'/2]\\
\\
\max  \widehat G_{L,R}&>&\max\limits_{[L,R]\setminus(s^*-\epsilon,s^*+\epsilon)}  \widehat G_{L,R} &\qquad \text{for}& L\in[L^*+\epsilon'/2,L^*+\epsilon'].
\end{array}
\end{equation}
To simplify the notation, we write $(\widehat L, \widehat R)$ for the current cluster in the sample procedure.  We call a sample split \textit{small} if the split point is placed in $[\widehat L,\widehat L+\delta']$ or $[\widehat R -\delta,\widehat R]$ and \textit{big} if the split point is placed in $[s^*-\epsilon,s^*+\epsilon]$.   Inequalities~(\ref{ineq.sam.3sets}) directly imply that the following properties hold for all sufficiently large~$n$, provided $|\widehat R- R_{\widehat L}|\le2\delta$ is satisfied:
\begin{enumerate}
\item[(i)] if $\widehat L \le L^*-\epsilon'$, then the next split of the sample procedure is small;
\item[(ii)] if $\widehat L\in[ L^*-\epsilon',L^*+\epsilon'/2]$, then the next sample split is either small or big;
\item[(iii)] if $\widehat L\in[L^*+\epsilon'/2, L^*+\epsilon']$, then the next sample split is big.
\end{enumerate}
We refer to $|\widehat R- R_{\widehat L}|\le 2\delta$ as the \textit{requirement}.  Note that it is satisfied for the starting cluster, $(\widehat L_s, \widehat R_s)$, by assumption, and because $|R_L-R_{L+\delta'}|\le\delta$ for all~$L$ in $[L_s-\delta', L^*+\epsilon']$, by Lemma~\ref{lem.pop.ineq}.  We will verify that the requirement remains valid until the sample procedure makes a big split.  First, suppose $|\widehat R- R_{\widehat L}|\le \delta$.  By properties (i)-(ii), together with the bound $|R_{\widehat L}-R_{\widehat L+\delta'}|\le\delta$, either the next sample split is big or it reduces $(\widehat L, \widehat R)$ to a cluster that still satisfies the requirement.   Now consider the case $|\widehat R- R_{\widehat L}|\in[\delta, 2\delta]$.  Consider the last two inequalities in Lemma~\ref{lem.pop.ineq}, and, again, derive the sample sample counterparts,
\begin{equation*}
\begin{array}{lcrcl}
\max\limits_{[R-\delta,R]}  \widehat G_{L,R}&>& \max\limits_{[L,L+\delta']}  \widehat G_{L,R} & \;\text{for}& L\in [L_s-\delta', L^*+\epsilon'], \;\; R\in[R_L+\delta',R_L+2\delta]\\
\\
\max\limits_{[L,L+\delta']}  \widehat G_{L,R} &>& \max\limits_{[R-\delta,R]}  \widehat G_{L,R}&  \;\text{for}& L\in [L_s-\delta', L^*+\epsilon'], \;\; R\in[R_L-2\delta',R_L-\delta].\\
\end{array}
\end{equation*}
Taking inequality $|R_{\widehat L}-R_{\widehat L+\delta'}|\le \delta$ into account, we deduce that the next sample split, if small, reduces the distance $|\widehat R- R_{\widehat L}|$.  Thus, when the sample procedure is applied to $(\widehat L_s, \widehat R_s)$, the requirement remains satisfied until the big split, and properties  (i)-(iii) remain valid.
Consequently, we can use these properties repeatedly to establish that, for all sufficiently large~$n$, the sample procedure starting at $(\widehat L_s, \widehat R_s)$ makes a number of small splits, followed by a big split, $\widehat{sp}=(\widehat L^*, \widehat s^*, \widehat R^*)$, such that $||\widehat{sp} - sp||_{\infty}<\epsilon$.

Now consider the case where the population procedure truncates, in a two-sided fashion, the cluster $(L_s,R_s)$ all the way to an empty set.  To establish that the sample procedure, applied to~$(\widehat L_s,
\widehat R_s)$, makes only small splits, we can apply the same reasoning as in the previous case, but using only the first of the three inequality sets in display~(\ref{ineq.sam.3sets}) and only the first of the properties (i)-(iii).   The rest of the cases can be handled analogously, with only minor modifications to the original argument.  Thus, for all sufficiently large~$n$, the sample procedure starting at $(\widehat L_s, \widehat R_s)$ either makes small splits until the end or makes small splits until a split $\widehat{sp}$, such that $||\widehat{sp} - sp||_{\infty}<\epsilon$.   Note that the width of the smaller sub-cluster produced by a small split is less than~$\epsilon$.  Because~$\epsilon$ can be chosen arbitrarily small, we conclude that $\widehat{sp}$ converges to $sp$, while the maximum size of the splits prior to~$\widehat{sp}$ goes to zero.  Note that $size(\widehat{sp})$ converges to~$size(sp)=\alpha^*$, by continuity of the function $size$.  Then, by the classical Glivenko-Cantelli theorem, $\widehat{siz}e(\widehat{sp})$ converges to~$\alpha^*$ as well.

Because the number of population splits is finite, we can repeat all of the preceding arguments sequentially. For example, to analyze the clustering procedures between the first and, potentially, second population split, we first consider the left sub-cluster and set $L_s$, $R_s$, $\widehat L_s$ and $\widehat R_s$ to $L^*$, $s^*$, $\widehat L^*$ and $\widehat s^*$, respectively.  Note that $(L_s,R_s)$ is appropriately bounded, and, for every positive~$\delta$ and~$\delta'$,  inequalities $|\widehat R_s - R_s|\le\delta$ and $|\widehat L_s - L_s|\le\delta'$ hold for all sufficiently large~$n$.  This verifies the assumptions stated in the first paragraph of the proof, which are the only ones needed for the preceding argument.  The same is true for the right sub-cluster, on which we set $L_s$, $R_s$, $\widehat L_s$ and $\widehat R_s$ to $s^*$, $R^*$, $\widehat s^*$ and $\widehat R^*$, respectively.

To complete the proof of the theorem, it is only left to show that the maximum size of all the sample splits leading to the very first starting cluster, $(\widehat L_s, \widehat R_s)$, converges to zero.  Recall that the existence of $(\widehat L_s, \widehat R_s)$, with the required properties, is guaranteed by Lemma~\ref{lem.comp}.  The same result also tells us that $P_{-\infty,\widehat L_s}$ and $P_{\widehat R_s,\infty}$ are bounded above by~$\epsilon$.  As~$\epsilon$ can be chosen arbitrarily small, we have established the desired convergence.

\subsubsection*{Proof of Lemma~\ref{lem.comp}}

The proof takes advantage of several applications of Lemma~\ref{lem.unif.appr} and the strong law of large numbers.  We restrict our attention to the set of probability one that remains after casting out the negligible sets associated with the aforementioned convergence results.  We conduct a pointwise argument, at a fixed~$\omega$, but suppress the dependence on~$\omega$ to simplify the notation.

When $(L_0,R_0)$, the support of the underlying distribution, is bounded, the conclusion of the lemma holds for $L=L_0$ and $R=R_0$.
Consider an unbounded case of $R_0=\infty$ and~$|L_0|<\infty$. Note that the population procedure starts with a right-sided truncation.  Let~$T^*$ denote the right endpoint of the cluster at which the population procedure either transitions to a two-sided truncation or makes a split.  Fix a~$K_1$, such that $P_{K_1,\infty}\le\epsilon$, with a further requirement that $K_1>T^*$, if $T^*$ exists.   Take $K_2$ large enough to satisfy $K_2>\mu_{L_0,\infty}+\mu_{K_1,\infty}-L_0$.  By the law of large numbers, $K_2-\widehat\mu_{L_0,\infty}>\widehat\mu_{K_1,\infty}-L_0$, for all sufficiently large~$n$.  Suppose that~$n$ is large enough for the above lower bound to be satisfied.  Then, the following inequalities hold for all $R> K_2$ and $r\in[L_0,K_1]$:
\begin{equation*}
\widehat G_{L_0,R}(R)=R-\widehat\mu_{L_0,R}>K_2-\widehat\mu_{L_0,\infty}>\widehat\mu_{K_1,\infty}-L_0>\widehat\mu_{r,R}-\widehat\mu_{L_0,r}=\widehat G_{L_0,R}(r).
\end{equation*}
This guarantees that for every sample cluster $(L_0,\widehat R)$, with $\widehat R>K_2$, the next sample split point is placed to the right of~$K_1$, which implies that the sample procedure will eventually produce a cluster $(L_0,\widehat R)$ with $\widehat R\in[K_1,K_2]$.  Because the population procedure reduces $(L_0,\infty)$ to $(L_0,K_1)$ using the right-sided truncation, interval $(L_0,R)$ is a bounded population cluster for each~$R$ in $[K_1,K_2]$.  This completes the proof for the case $R_0=\infty$, $|L_0|<\infty$.  The case $L_0=-\infty$, $R<\infty$ can be handled analogously.

Now consider the case $(L_0,R_0)=(-\infty,\infty)$.
Let $K^-_1$ and $K^+_1$ satisfy $P_{-\infty,K^-_1}\le \epsilon$ and $P_{K^+_1,\infty}\le\epsilon$. If the population procedure makes a split, and the first one is applied to the cluster $(L^*,R^*)$, we further require $(K^-_1,K^+_1)\supset[L^*,R^*]$.  Additional conditions are placed on~$K^-_1$ and~$K^+_1$ below.   Take $K^+_2$ sufficiently large to ensure $K^+_2>\mu_{K^-_1,\infty}+\mu_{K^+_1,\infty}-\mu_{-\infty,K^-_1}$.  By the law of large numbers, $K_2-\widehat\mu_{K^-_1,\infty}>\widehat\mu_{K_1^+,\infty}-\widehat\mu_{-\infty,K^-_1}$ for all sufficiently large~$n$.  Consequently, for all $r\in[K^-_1,K_1^+]$, $R>K_2^+$, $L\le K^-_1$ we have:
\begin{equation*}
\widehat G_{L,R}(R)=R-\widehat\mu_{L,R}>K^+_2-\widehat\mu_{K^-_1,\infty}>\widehat\mu_{K^+_1,\infty}-\widehat\mu_{-\infty,K^-_1}\ge\widehat\mu_{r,R}-\widehat\mu_{L,r}=\widehat G_{L,R}(r).
\end{equation*}
This guarantees that for every sample cluster $(\widehat L,\widehat R)$, with $\widehat L\le K^-_1$ and $\widehat R>K^+_2$, the next sample split point is placed outside of $[K^-_1,K^+_1]$.

If $K^-_2$ lies below $\mu_{-\infty,K^+_1}+\mu_{-\infty,K^-_1}-\mu_{K^+_1,\infty}$, then, by similar arguments,  $\widehat G_{L,R}(L)>\widehat G_{L,R}(r)$ for all $r\in[K^-_1,K^+_1]$, $L< K^-_2$ and $R\ge K^+_1$. This guarantees that for every sample cluster $(\widehat L,\widehat R)$, with $\widehat L<K^-_2$ and $\widehat R\ge K^+_1$, the next sample split point is placed outside of $[K^-_1,K_1^+]$.

Consequently, the sample procedure produces a cluster $(\widehat L, \widehat R)$ with $\widehat L\in[K^-_2,K^-_1]$ and $\widehat R\in[K^+_1,K^+_2]$.  Because $K^-_1<L^*$,  interval $(\widehat L, R_{\widehat L})$ is a population cluster.  It is only left to show that we can find an~$\widehat R$ that is located within~$\delta$ of~$R_{\widehat L}$. We will use the following result, which is proved further below.
\begin{lemma}
\label{lem.pop.ineq2}
Let $(L,R_L)$ be a population cluster, achieved via a two-sided truncation of the real line.  If $K_1^-$ is sufficiently small, then for each positive~$\delta$ we have
\begin{equation*}
\max_{[R_L+\delta,R]} G_{L,R} > \max_{[L,R_L]}  G_{L,R} \qquad \text{for all\;} L\in[K^-_2,K^-_1] \;\text{and}\;  R\in[R_L+\delta, K_2^+].
\end{equation*}
\end{lemma}

Applying the argument in the paragraph below the statement of Lemma~\ref{lem.pop.ineq}, and replacing~$L$ with~$\widehat L$, we deduce that $\max_{[R_{\widehat L}+\delta,R]} \widehat G_{{\widehat L},R} > \max_{[{\widehat L},R_{\widehat L}]} \widehat G_{{\widehat L},R}$ for all $R\in[R_{\widehat L}+\delta, K_2^+]$, as long as~$n$ is sufficiently large.  Hence, if $\widehat R>R_{\widehat L}+\delta$, the sample procedure sequentially moves the right endpoint of cluster $(\widehat L,\widehat R)$ further left, until it falls in $[R_{\widehat L},R_{\widehat L}+\delta]$.  The case $\widehat R<R_{\widehat L}-\delta$ can be handled using analogous arguments.

\subsubsection*{Proof of Lemma~\ref{lem.unif.appr}}

Let~$P_n$ denote the empirical measure associated with observations $x_1,...,x_n$, and let~$P$ be the corresponding population distribution.  To simplify the presentation, we replace expressions $\int h(x) dP_n(x)$ and $\int h(x)dP(x)$ by $P_n h$ and $P h$, respectively.

The classical Glivenko-Cantelli theorem gives the following uniform convergence,
\begin{equation}
\label{emp.fact1}
\sup_{l<r}\left|P_n(l,r)-P(l,r)\right|\rightarrow 0,
\end{equation}
almost surely, as~$n$ goes to infinity.    Note that the collection of functions $h_{l,r}(x)=x1_{\{l<x<r\}}$, defined for all real~$l$ and~$r$, forms a VC class with an integrable envelope, $H(x)=|x|$.  Consequently, by a functional generalization of the Glivenko-Cantelli theorem (e.g. Theorem 19.13 in \citealt{vaart98book}), we have
\begin{equation}
\label{emp.fact2}
\sup_{l,r}\left|P_n h_{l,r}-P h_{l,r}\right|\rightarrow 0,
\end{equation}
almost surely, as~$n$ goes to infinity.  For the rest of the proof we cast out the negligible sets on which~(\ref{emp.fact1}) and~(\ref{emp.fact2}) break down, and conduct a pointwise argument on the remaining set of probability one.  We  suppress the dependence on~$\omega$ to simplify the notation.

Fix an arbitrarily small positive~$\epsilon$.  Define $A_M(\epsilon)=A_M\cap\{(l,r):\;r-l\ge\epsilon\}$.  Let $c_{\epsilon}=\inf_{A_M(\epsilon)}P(l,r)$, and note that~$c_{\epsilon}$ is positive.  Taking advantage of the convergence in~(\ref{emp.fact1}) and~(\ref{emp.fact2}), we can bound $\sup_{A_M(\epsilon)}|\widehat\mu_{l,r}-\mu_{l,r}|$ above by
\begin{eqnarray*}
&& \sup_{A_M(\epsilon)}\left|\frac{P_nh_{l,r}-Ph_{l,r}}{P(l,r)}\right|+\sup_{A_M(\epsilon)}|P_nh_{l,r}|\left|\frac{1}{P_n(l,r)}-\frac{1}{P(l,r)}\right|\\
&\le& c_\epsilon^{-1}\sup_{l,r}\left|P_nh_{l,r}-Ph_{l,r}\right|+c_\epsilon^{-1}\frac{\sup_{l,r} P_n|h_{l,r}|}{\inf_{A_M(\epsilon)}P_n(l,r)}\sup_{l<r}\left|{P_n(l,r)}-P(l,r)\right|\\
&=& c_\epsilon^{-1}o(1)+c_\epsilon^{-1}\frac{PH+o(1)}{c_\epsilon+o(1)}o(1)=o(1),
\end{eqnarray*}
as~$n$ tends to infinity.  Consequently,
\begin{equation*}
\sup_{(L,R) \in A_{M}}\;\;\max_{a\in [L,R]} |\widehat G_{L,R}(a)-G_{L,R}(a)|\le 2\epsilon+2\sup_{A_M(\epsilon)}\left|\widehat\mu_{l,r}-\mu_{l,r} \right|=2\epsilon+o(1).
\end{equation*}
This completes the proof, because~$\epsilon$ can be chosen arbitrarily small.

\subsubsection*{Proof of Lemma~\ref{lem.pop.ineq}}

For concreteness we assume~$L_s>L_0$ throughout the proof.  The case $L_s=L_0$ follows with only minor modifications.  Suppose that positive~$\epsilon'$ and~$\delta'$ are small enough for~$R_L$ to be defined on ${[L_s-\delta', L^*+\epsilon']}$.  Then, $\max_{L\in[L_s-\delta', L^*+\epsilon']}|R_L-R_{L+\delta'}|\le\delta$ for every positive~$\delta$, as long as~$\delta'$ is sufficiently small, by uniform continuity of~$R_L$ on a compact set.

Differentiating $G_{L,R}(s)$ with respect to~$s$, we derive the following formulas:
\begin{equation}
\label{prf.lem3.dervs}
G'_{L,R}(L)=\frac{f(L)(\mu_{L,R}-L)}{P_{L,R}} -\frac12 \quad\text{and} \quad G'_{L,R}=\frac12-\frac{f(R)(R-\mu_{L,R})}{P_{L,R}}.
\end{equation}
Note that for $L\in[L_s,L^*]$, function $G_{L,R}$ is maximized at the endpoints.  Consequently, $G'_{L,R_L}(L)\le0$ and $G'_{L,R_L}(R_L)\ge0$, when $L\in[L_s,L^*]$.   For the clarity of the exposition we first focus on the \textit{regular} case, where both inequalities are strict for $L\in[L_s,L^*]$.  By continuity of $G'_{L,R}$, there exist positive~$\delta_1$ and~$\epsilon_1$, such that $G'_{L,R}(L)<0$ and $G'_{L,R}(R)>0$ for all $L\in[L_s-\delta_1,L^*+\epsilon_1]$ and $|R-R_L|\le\delta_1$.  Using uniform continuity of $G'_{L,R}$ on compact sets, we conclude that there exists a positive~$\delta_3$, such that
\begin{equation}
\label{prf.lem3.in1}
G_{L,R}(L)\vee G_{L,R}(R) >G_{L,R}(s)  \qquad \text{for all}\quad s\in (L,L+\delta_3]\cup[R-\delta_3,R),
\end{equation}
when $L\in[L_s-\delta_1,L^*+\epsilon_1]$ and $|R-R_L|\le\delta_1$.
Because $\arg\max G_{L,R_L}=\{L,R_L\}$ for $L\in[L_s,L^*)$, continuity of $G_{L,R}$ implies that, given an arbitrarily small positive~$\epsilon'$,  we can find a positive~$\delta_2$, such that
\begin{equation}
\label{prf.lem3.in2}
G_{L,R}(L)\vee G_{L,R}(R) > \max_{[L+\delta_3,R-\delta_3]} G_{L,R},
\end{equation}
when $L\in[L_s-\delta_2,L^*-\epsilon']$ and $|R-R_L|\le\delta_2$.
Inequalities~(\ref{prf.lem3.in1}) and~(\ref{prf.lem3.in2}) yield
\begin{equation*}
G_{L,R}(L)\vee G_{L,R}(R) >G_{L,R}(s)  \qquad \text{for all}\quad s\in (L,R),
\end{equation*}
when $L\in[L_s-\delta_1\wedge\delta_2,L^*-\epsilon']$ and $|R-R_L|\le \delta_1\wedge \delta_2$.  This implies the \textit{first set of inequalities} in Lemma~\ref{lem.pop.ineq}.

By a similar argument, for every positive~$\epsilon$ there exist positive~$\epsilon_2$ and~$\delta_4$, such that
\begin{equation}
\label{prf.eq.setminus}
G_{L,R}(L)\vee G_{L,R}(R) > \max_{[L+\delta_3,R-\delta_3]\setminus(s^*-\epsilon,s^*+\epsilon)} G_{L,R},
\end{equation}
when $L\in[L_s-\delta_4,L^*+\epsilon_2]$ and $|R-R_L|\le\delta_4$.  Combining this bound with inequality~(\ref{prf.lem3.in1}), we derive the \textit{second set of inequalities} in Lemma~\ref{lem.pop.ineq}.

In the proof of Proposition~\ref{prop.unimod} we establish inequalities  $\mu_{L^*,s^*}\le(L^*+s^*)/2$ and {$\mu_{s^*,R^*}\ge(s^*+R^*)/2$}.  In our case they are equalities, because $G_{L^*,R^*}(s^*)=G_{L^*,R^*}(L^*)=G_{L^*,R^*}(R^*)$.  Taking this fact into account when solving $G'_{L^*,R^*}(s^*)=0$ we derive  a useful equality, $(R^*-s^*)/P_{s^*,R^*}=(s^*-L^*)/P_{L^*,s^*}$.  It follows that
\begin{equation}
\label{prf.lem.ineq.2ineq}
f(L^*)(R^*-s^*)/P_{s^*,R^*}<1 \qquad \text{and} \qquad  f(L^*)(s^*-L^*)/P_{L^*,s^*}<1,
\end{equation}
because otherwise $f(L^*)(R^*-L^*)/P_{L^*,R^*}\ge1$, contradicting the fact that $G'_{L,R_L}(L)<0$ for $L\in[L_s,L^*]$, which holds in the regular setting that we now focus on.

Consider the function $h(L)=G_{L,R_L}(s^*)-G_{L,R_L}(L)$, and note that $h(L^*)=0$.  Inequalities~(\ref{prf.lem.ineq.2ineq}) guarantee that $h'(L^*)>0$.  Hence, there exists a positive~$\epsilon_3$, such that for every~$\epsilon'\le\epsilon_3$,
\begin{equation*}
G_{L,R_L}(s^*)>G_{L,R_L}(L)=G_{L,R_L}(R_L),
\end{equation*}
when $L\in[L^*+\epsilon'/2,L^*+\epsilon']$.  Continuity of~$G_{L,R}(s)$, together with compactness of the intervals involved, implies that for every~$\epsilon'\le\epsilon_3$ there exists a positive~$\delta_5$, such that
\begin{equation}
\label{prf.eq.spl1}
G_{L,R}(s^*)>G_{L,R}(L)\vee G_{L,R}(R),
\end{equation}
when $L\in[L^*+\epsilon'/2,L^*+\epsilon']$ and $|R-R_L|\le\delta_5$.
Because $\arg\max G_{L^*,R_{L^*}}=\{L^*,R_{L^*},s^*\}$, we can also find positive $\epsilon_4$ and~$\delta_{6}$, for which
\begin{equation}
\label{prf.eq.spl2}
G_{L,R}(s^*)>\max_{[L+\delta_3,R-\delta_3]\setminus(s^*-\epsilon,s^*+\epsilon)} G_{L,R},
\end{equation}
when $L\in[L^*,L^*+\epsilon_4]$ and $|R-R_L|\le\delta_{6}$.  Combining bounds~(\ref{prf.lem3.in1}), (\ref{prf.eq.spl1}) and~(\ref{prf.eq.spl2}), we establish the \textit{third set of inequalities} in Lemma~\ref{lem.pop.ineq}.

Consider function $h_L(R)= G_{L,R}(R)-G_{L,R}(L)$.  Differentiation gives $h'_L(R_L)=1-2f(R_L)(R_L-\mu_{L,R_L})/P_{L,R_L}$, which is strictly positive for $L\in[L_s,L^*]$ in the regular case that we now focus on.  Taking advantage of equality $h_L(R_L)=0$, continuity of~$h'_L$ and compactness of the intervals involved, we deduce that, provided~$\epsilon'$, $\delta$ and~$\delta'$ are  sufficiently small, inequalities $G_{L,R}(R)>\max_{[L,L+\delta']}G_{L,R}$ are satisfied for all $L\in [L_s-\delta', L^*+\epsilon']$ and $R\in[R_L+\delta,R_L+2\delta]$.  This establishes the \textit{fourth set of inequalities} in Lemma~\ref{lem.pop.ineq}.  The \textit{fifth set} can be derived using  analogous arguments.

We now examine the irregular setting, where $G'_{L,R_L}(L)=0$ or $G'_{L,R_L}(R_L)=0$ for some~$L$ in $[L_s,L^*]$.  For concreteness, we examine the case $G'_{L^i,R^i}(R^i)=0\ne G'_{L^i,R^i}(L^i)$, where $L^i\in [L_s,L^*]$ and $R^i=R_{L^i}$.  Suppose that $R^i<R_s$; the case $R^i=R_0$ can be handled with minor modifications. Differentiation gives $G''_{L^i,R^i}(R^i)=-f'(R^i)/[6f(R^i)]$, which, together with $\arg\max G_{L^i,R^i}\supseteq\{L^i,R^i\}$, implies $f'(R^i)\ge0$.  However, a strict inequality, $f'(R^i)>0$ implies that $2f(R_L)(R_L-\mu_{L,R_L})>P_{L,R_L}$ for all~$L$ that are sufficiently close to~$L^i$ and satisfy $L<L^i$.  Because this contradicts inequality $G'_{L,R_L}(R_L)\ge0$, we conclude that $f'(R^i)=0$, and thus, $G'_{L^i,R^i}(R^i)=G''_{L^i,R^i}(R^i)=0$.  Further differentiation yields $G'''_{L^i,R^i}(R^i)=-f''(R^i)/[2f(R^i)]$, which implies~$f''(R^i)\le 0$.  This stepwise argument can be continued if $f''(R^i)=0$, however we will focus on the case $f''(R^i)<0$, for concreteness.  Note that~$R^i$ is an interior mode of~$f$.

Let $l=L-L^i$, $t=s-R^i$ and $r=R-R^i$.  After deriving the third order Taylor series expansion for the function $(L,s,R)\mapsto G_{L,R}(s)$ at $(L^i,R^i,R^i)$, we establish the following approximation, which holds for $(l,t,r)$ near zero:
\begin{equation*}
G_{L^i+l,R^i+r}(R^i+t)-G_{L^i+l,R^i+r}(R^i)=\frac{-f''(R^i)}{6f(R^i)}(3t^3-t^2r+3tr^2)+o(|l|^3+|t|^3+|r|^3).
\end{equation*}
Analysis of the above expression reveals that, because $f''(R^i)<0$, the approximating cubic function is increasing in~$t$ for every fixed~$r$ and~$l$.

We now revisit the first set of inequalities in Lemma~\ref{lem.pop.ineq}; the rest of the sets can be established using similar arguments.  For the corresponding proof given in the regular setting to still go through, it is sufficient to establish that for every small enough positive~$\delta_1$,
\begin{equation*}
G_{L,R}(L)\vee G_{L,R}(R) >G_{L,R}(s)  \qquad \text{for all}\quad s\in (L,R-\delta_1/4],
\end{equation*}
when $|L-L^i|\le\delta_1$ and $|R-R^i|\le\delta_1$.

By the established monotonicity of the cubic approximation, we can find a positive~$\delta_3$, such that for each positive~$\delta<4\delta_3$,
\begin{equation}
\label{lem.prf.irr1}
G_{L,R}(R) > G_{L,R}(s)  \qquad \text{for all}\quad s\in [R-\delta_3,R-\delta_1/4],
\end{equation}
when $|L-L^i|\le\delta_1$ and $|R-R^i|\le\delta_1$.  Because $\arg\max G_{L^i,R^i}=\{L^i,R^i\}$, we can also find a positive $\delta_2$, such that
\begin{equation}
\label{lem.prf.irr2}
G_{L,R}(R)\vee G_{L,R}(L) > \max_{[L+\delta_3,R-\delta_3]}G_{L,R},
\end{equation}
when $|L-L^i|\le\delta_2$ and $|R-R^i|\le\delta_2$.   Combining inequalities~(\ref{lem.prf.irr1}) and~(\ref{lem.prf.irr2}) establishes
\begin{equation*}
G_{L,R}(R)\vee G_{L,R}(L) >\max_{[L+\delta_3,R-\delta_1/4]}G_{L,R},
\end{equation*}
when $|L-L^i|\le\delta_1$, $|R-R^i|\le\delta_1$, and~$\delta_1$ is sufficiently small.  Using the regular case argument in the beginning of the proof we also establish the left side bound,
\begin{equation*}
G_{L,R}(L) >G_{L,R}(s)  \qquad \text{for all}\quad s\in (L,L+\delta_3],
\end{equation*}
when $|L-L^i|\le\delta_2$, $|R-R^i|\le\delta_2$, and~$\delta_2$ is sufficiently small, which completes the proof.

\subsubsection*{Proof of Lemma~\ref{lem.pop.ineq2}}

Let~$R^m$ be the maximum of $R^*$ and the right-most mode of~$f$.  Let~$L^m$ be the minimum of $L^*$ and the left-most mode of~$f$.  As we point out in the proof of Proposition~\ref{prop.unimod}, any interior maximizer of $G_{L,R}$ must lie in $(L^m,R^m)$.  Thus, it is sufficient to verify
\begin{enumerate}
\item[(i)] $G_{L,R}(R)>\max_{[L^m,R^m]}  G_{L,R}$ \qquad\text{and}
\item[(ii)] $G_{L,R}(R)>G_{L,R}(L)$,
\end{enumerate}
for all $L\in[K^-_2,K^-_1]$ and $R\in[R_L+\delta, K_2^+]$.  If we choose~$K^-_1$ sufficiently small, so that $K^-_1<L^m$, $R_{K^-_1}>R^m$ and $R_{K^-_1}>\mu_{-\infty,R_m}+\mu_{R^m,\infty}-\mu_{-\infty,L^m}$, then~(i) follows from
\begin{equation*}
G_{L,R}(R)>R_{K^-_1}-\mu_{-\infty,R_m}>\mu_{R^m,\infty}-\mu_{-\infty,L^m}>\max_{[L^m,R^m]}  G_{L,R}.
\end{equation*}

The existence argument in the proof of Proposition~\ref{prop.gen} implies that that the derivative of the function $h_L(R)= G_{L,R}(R)-G_{L,R}(L)$ is strictly positive on $[R_L,\infty)$, for every $L<L^m$. Statement~(ii) then follows from $h_L(R_L)=0$.

\subsection{Proof of Theorem~\ref{rates.thm}}

Note that for the purposes of this proof we can replace ``sample frequency'' with ``underlying probability'' in the definition of the set $\widehat \cS(\tau)$.  By the classical Glivenko-Cantelli theorem, the corresponding modified statement of the theorem implies the original one.  Write $\cal{J}({\tau})$ for the collection of all intervals $(L,R)$, such that such that $P(L,R)\ge\tau$, $P(-\infty,L)\ge\tau/2$ and $P(R,\infty)\ge\tau/2$.  By Theorem~\ref{gen.thm}, all intervals $(L,R)$ corresponding to the splits in the redefined collection $\widehat \cS(\tau)$ lie in the set $\cal{J}({\tau})$, with probability tending to one.  We will only use the last two inequalities in the definition of $\cal{J}({\tau})$ to ensure that, with probability tending to one, all $(L,R)$ are contained in a fixed bounded interval, on which~$f$ is bounded away from zero.

\subsubsection*{Main Body of the Proof}

The following two lemmas give us appropriate control over the difference between small perturbations in $\widehat G_{L,R}$ and $G_{L,R}$.
\begin{lemma}
\label{rates.lem1}
For every positive $\epsilon$ and $\tau$, there exists a $O_p(1)$ random sequence $M_n$, such that
\begin{eqnarray*}
\left|(\widehat G_{L,R}(R) - \widehat G_{L,R}(a)) - ( G_{L,R}(R) -  G_{L,R}(a) ) \right|
&\le& \epsilon (R-a)+(\log n / n) M_n\\
\\
\left|(\widehat G_{L,R}(L) - \widehat G_{L,R}(a)) - ( G_{L,R}(L) -  G_{L,R}(a) ) \right|
&\le& \epsilon (a-L)+(\log n / n) M_n,
\end{eqnarray*}
for all $(L,R)\in\cal{J}({\tau})$  and $a\in(L,R)$, such that $\widehat G_{L,R}(a)$ is well-defined.
\end{lemma}
The proof of Lemma~\ref{rates.lem1} is given further below.  The next result can be established using a similar type of argument, however, it is a direct corollary of Lemma~4.1 in \cite{kim1990cube}.  Note that the~$\log n$ factor, which appears in Lemma~\ref{rates.lem1} but not in Lemma~\ref{rates.lem2}, is due to the fact that points~$L$ and~$R$, near which function $\widehat G_{L,R}$ is approximated in Lemma~\ref{rates.lem1}, are allowed to vary, while $s$, the corresponding point in Lemma~\ref{rates.lem2} is fixed.
\begin{lemma}
\label{rates.lem2}
Given positive~$\tau$ and $\epsilon$, and a point $s\in (L_0,R_0)$, there exists a $O_p(1)$ sequence $M_n$, such that
\begin{equation*}
\left|(\widehat G_{L,R}(s) - \widehat G_{L,R}(a)) - ( G_{L,R}(s) -  G_{L,R}(a) ) \right|
\le \epsilon |s-a|^2+n^{-2/3} M_n,
\end{equation*}
for all $L$,$a$ and $R$, such that $(L,s)\in\cal{J}({\tau})$, $(s,R)\in\cal{J}({\tau})$, $a\in(L,R)$ and $\widehat G_{L,R}(a)$ is well-defined.
\end{lemma}

We start by deriving the uniform rate of convergence for the \textit{small} sample splits.
As in the proof of consistency, we focus on the case of the population procedure performing a two-sided truncation, followed by a split.  The rest of the cases can be handled using analogous arguments.   Let $(L^*,s^*,R^*)$ be the first population split.  The population procedure truncates $(L_0,R_0)$, along $(L,R_L)$, down to $(L^*,R^*)$.  As we showed in the consistency proof, the sample procedure reduces the cluster consisting of all the observations down to, approximately, $(L^*,R^*)$, by repeatedly splitting off small clusters near the boundary.  The split point for the big split is placed near~$s^*$.  Given a positive~$\delta$, we define ${\cal{A}}(\delta)=\{(L,R): L\le L^*+\delta, |R-R_L|\le\delta\}$.  In the consistency proof we showed that the following property holds with probability tending to one for the starting cluster, $(L,R)$, of each small sample split: $(L,R)\in{\cal A}(\delta)$ and $\arg\max \widehat G_{L,R}\subset [L,L+\delta]\cup[R-\delta,R]$.  Let $\widehat a_{L,R} = \arg\max_{[L,L+\delta]\cup[R-\delta,R]} \widehat G_{L,R}$.  Then, to establish bound~(\ref{eq2.rates.thm}) in the statement of Theorem~\ref{rates.thm} it is sufficient to prove that for every positive~$\tau$ there exists a positive $\delta$ and a $O_p(\log n/n)$ sequence $B_n$, such that inequality $(R-\widehat a_{L,R})\wedge(\widehat a_{L,R}-L)\le B_n$ holds for all $(L,R)\in{\cal A}(\delta)\cap{\cal J}({\tau})$.

To characterize small perturbations in the population criterion function, $G_{L,R}$, we will use the derivations in the proof of Lemma~\ref{lem.pop.ineq} (we focus only on the \textit{regular} part of the proof, because of the new regularity condition, C3).  Those derivations imply that there exist positive~$c_0$ and~$\delta$, such that, for all $(L,R)\in{\cal A}(\delta)\cap{\cal J}({\tau})$,
\begin{eqnarray}
\label{G.boundary1}
G_{L,R}(R)-G_{L,R}(a)&\ge& c_0(R-a), \qquad \text{for}\;a\in[R-\delta,R], \;\text{and}\\
\nonumber\\
\label{G.boundary2}
G_{L,R}(L)-G_{L,R}(a)&\ge& c_0(a-L), \qquad \text{for}\;a\in[L,L+\delta].
\end{eqnarray}
Define
\begin{eqnarray*}
\Delta_n(L,a,R)&=&\left|(\widehat G_{L,R}(R) - \widehat G_{L,R}(a)) - ( G_{L,R}(R) -  G_{L,R}(a) ) \right|+\\
\\
&&\left|(\widehat G_{L,R}(L) - \widehat G_{L,R}(a)) - ( G_{L,R}(L) -  G_{L,R}(a) ) \right|.
\end{eqnarray*}
Combining Lemma~\ref{rates.lem1}, applied for $\epsilon=c_0/2$, with inequalities~(\ref{G.boundary1}) and~(\ref{G.boundary2}), we deduce
\begin{eqnarray*}
0&\ge& \widehat G_{L,R}(R) - \widehat G_{L,R}(\widehat a_{L,R})\\
\\
&\ge& c_0 (R-\widehat a_{L,R})\wedge(\widehat a_{L,R}-L) - \Delta_n(L,\widehat a_{L,R},R)\\
\\
&\ge& (c_0/2)(R-\widehat a_{L,R})\wedge(\widehat a_{L,R}-L) - (\log n / n) M_n.
\end{eqnarray*}
Hence, $(R-\widehat a_{L,R})\wedge(\widehat a_{L,R}-L)\le(c_0/2)^{-1}(\log n / n) M_n$ for all $(L,R)\in{\cal A}(\delta)\cap{\cal J}({\tau})$, which is what we needed to prove.  Thus, up to the first big split, the sizes of all the small sample splits are uniformly $O_p(\log n / n)$.    In the consistency theorem we showed that, with probability tending to one, the number of the big sample splits equals the (finite) number of the population splits.  Consequently, the behavior of the sample procedure after the first big split can be handled by repeating the argument given above.  This establishes bound~(\ref{eq2.rates.thm}) in the statement of Theorem~\ref{rates.thm}.

We now focus on deriving the rate of convergence in~(\ref{eq1.rates.thm}).  Recall that $(L^*,s^*,R^*)$ and $(\widehat L^*,\widehat s^*,\widehat R^*)$ denote the first population split and the first big sample split, respectively.  Given a positive~$\delta$, define ${\cal{B}}(\delta)=\{(L,s,R): |L- L^*|\le\delta, |s-s^*|\le\delta, |R-R^*|\le\delta\}$.  Note that $(\widehat L^*,\widehat s^*, \widehat R^*)\in{\cal B}(\delta)$ with probability tending to one.  Derivations in the proof of Lemma~\ref{lem.pop.ineq}, together with the assumption $G_{L^*,R^*}''(s^*)\ne0$, imply that there exist positive constants $\delta$, $c_1$ and $c_2$, such that
\begin{equation*}
G_{L,R}(s^*)-G_{L,R}(a)\ge c_1|a-s^*|^2-c_2|a-s^*|\left(|L-L^*|+|R-R^*|\right),
\end{equation*}
for all $(L,s,R)\in{\cal B}(\delta)$.  Define $\widehat s=\arg\max_{[s^*-\delta,s^*+\delta]} \widehat G_{\widehat L^*,\widehat R^*}$ and let
\begin{equation*}
D_n=\left|\widehat G_{\widehat L^*,\widehat R^*}(s^*)-\widehat G_{\widehat L^*,\widehat R^*}(\widehat s)-
G_{\widehat L^*,\widehat R^*}(s^*)-G_{\widehat L^*,\widehat R^*}(\widehat s)\right|.
\end{equation*}
We can handle~$D_n$ by applying Lemma~\ref{rates.lem2}, with  sufficiently small $\epsilon$ and $\tau$. It follows that, with probability tending to one,
\begin{eqnarray*}
0&\ge &\widehat G_{\widehat L^*,\widehat R^*}(s^*)-\widehat G_{\widehat L^*,\widehat R^*}(\widehat s)\\
\\
&\ge&G_{\widehat L^*,\widehat R^*}(s^*)-G_{\widehat L^*,\widehat R^*}(\widehat s) - D_n\\
\\
&\ge& c_1|\widehat s-s^*|^2-c_2|\widehat s-s^*|\left(|\widehat L^*-L^*|+|\widehat R^*-R^*|\right)-n^{-2/3}M_n.
\end{eqnarray*}
Consequently,
\begin{equation*}
|\widehat s-s^*|=O_p\left(n^{-1/3}+|\widehat L^*-L^*|+|\widehat R^*-R^*|\right).
\end{equation*}
Thus, to establish the rate in~(\ref{eq1.rates.thm}) for $(\widehat L^*,\widehat s^*, \widehat R^*)$ it is only left to show $|\widehat L^*-L^*|=O_p(n^{-1/3})$ and $|\widehat R^*-R^*|=O_p(n^{-1/3})$.  In the remainder of the proof we establish the last two stochastic bounds, and then extend the rate of convergence derived for $(\widehat L^*,\widehat s^*,\widehat R^*)$ to the subsequent big sample splits.

We again use the derivations in the proof of Lemma~\ref{lem.pop.ineq}, from which it follows that, given an arbitrarily small positive~$\tau$,  there exist positive constants $\delta$, $c_1$ and $c_2$, such that
\begin{equation*}
G_{L,R}(R) - G_{L,R}(a)\ge c_1 (R-R_L)-c_2(a-L),
\end{equation*}
for all $(L,R)\in{\cal A}(\delta)\cap{\cal J}({\tau})$ and $a\in[L,L+\delta]$.  Taking advantage of a maximal inequality for the empirical process indexed by a VC class of functions with a square integrable envelope (e.g. Lemma 19.38 in \citealt{vaart98book}), Lemma~\ref{rates.lem1}, and the already established rate of convergence for the small sample splits, we deduce that there exists a $O_p(n^{-1/2})$ sequence $C_n$, such that
\begin{equation*}
\widehat G_{L,R}(R) - \max_{a\in [L,L+\delta]} \widehat G_{L,R}(a)\ge c_1 (R-R_L) - C_n,
\end{equation*}
for all $(L,R)\in{\cal A}(\delta)\cap{\cal J}({\tau})$.  Consequently, when $R-R_L>c_1^{-1}C_n$, the maximizer of $\widehat G_{L,R}$ is at least a $\delta$ away from~$L$.  Because the split points of the small sample splits are uniformly within $O_p(\log n/n)$ of the cluster boundary, it follows that $R-R_L$ is bounded above by a $2c_1^{-1}C_n$ for all the sample clusters, $(L,R)$, produced up to the first big sample split.  A similar argument that focuses on~$L$ instead of~$R$ derives a lower bound for $R-R_L$.  Thus, $|R-R_L|=O_p(n^{-1/2})$, and it is only left to establish $|\widehat L^*-L^*|=O_p(n^{-1/3})$.

It follows from the derivations in the proof of Lemma~\ref{lem.pop.ineq} that there exist positive constants $\delta$, $c_1$, $c_2$ and $c_3$, such that
\begin{equation}
\label{bnd77}
G_{L,R}(s^*) - G_{L,R}(a)\ge c_1(L-L^*) - c_2(R-R_L)-c_3[(a-L)\wedge(R-a)],
\end{equation}
provided $L\in [L^*,L^*+\delta]$, $|R-R^*|\le\delta$ and $|a-L|\wedge|a-R|\le\delta$.  Handling the stochastic term the same way we did in the paragraph above, we conclude that there exists a $O_p(n^{-1/2})$ sequence~$T_n$, such that inequality
\begin{equation*}
\widehat G_{L,R}(s^*)\le \max_{a\in [L,L+\delta]\cup[R-\delta,R]} \widehat G_{L,R}(a)
\end{equation*}
implies $L-L^*\le T_n$. Note that the same $T_n$ is chosen for all intervals $(L,R)$ satisfying $L\in [L^*,L^*+\delta]$ and $|R-R^*|\le\delta$.  The companion lower bound on $L-L^*$ follows from an analogous argument, which replaces the lower bound in display~(\ref{bnd77}) with an upper bound.  The relationship $|\widehat L^*-L^*|=O_p(n^{-1/2})$ then follows directly after taking into consideration the rate of convergence already established for the small sample splits.

This completes the derivation of the rate of convergence for the first big sample split.  For the subsequent big splits the same argument can be repeated with the following negligible modification.  Due to the fact that endpoints of the clusters resulting from the first big sample split are within $O_p(n^{-1/3})$ of their population counterparts, all the $O_p(n^{-1/2})$ expressions should be replaced with $O_p(n^{-1/3})$.

\subsubsection*{Proof of Lemma~\ref{rates.lem1}}
We will only establish the first inequality. The second can be proved by using the same argument with the appropriate adjustment of the notation.  For points~$a$ that are bounded away from~$R$, the first inequality holds by the functional generalization of the Glivenko-Cantelli theorem (e.g. Theorem 19.13 in \citealt{vaart98book}).  Hence, from here on we only focus on $a\in[(L+R)/2,R]$.  We will use ``$\lesssim$'' to mean that inequality ``$\le$'' holds when the right hand side is multiplied by a positive constant, which is chosen independently from the involved parameters, such as $n$, $L$ and~$R$.
 Note that the statement of the lemma restricts $L$, $a$ and $R$ to a bounded interval, on which the infumum of~$f$ is positive, while the supremum is finite.  In particular, we have $R-a\lesssim P(a,R)$ and $P(a,R)\lesssim R-a$. Define $d_{l,r}(x)=(r-x)(r-l)^{-1}1_{\{l<x<r\}}$ and note that
\begin{eqnarray*}
\left|\widehat \mu_{a,R}-\mu_{a,R}\right|&=&\left| \frac{P_nd_{a,R}(R-a)}{P_n(a,R)}- \frac{Pd_{a,R}(R-a)}{P(a,R)} \right|\\
\\
&\le& \left|P_nd_{a,R}-Pd_{a,R}\right|\frac{(R-a)}{P(a,R)}+\left|P_n(a,R)-P(a,R)\right|\frac{(R-\widehat \mu_{a,R})}{P(a,R)}\\
\\
&\lesssim& \left|P_nd_{a,R}-Pd_{a,R}\right|+\left|P_n(a,R)-P(a,R)\right|\;=: \;  E_1+E_2.
\end{eqnarray*}
As before, we let $h_{l,r}(x)=x1_{l<x<r}$. Observe that $|P_nh_{a,R}-Ph_{a,R}|\lesssim E_2$, and stochastic bound $\sup_{l<r}(|P_nh_{l,r}-Ph_{l,r}|)+(|P_n(l,r)-P(l,r)|)=o_p(1)$ holds by the functional generalization Glivenko-Cantelli theorem. It follows that
\begin{eqnarray}
&&\left|(\mu_{L,R}-\mu_{L,a})-(\widehat \mu_{L,R}-\widehat \mu_{L,a})\right|\nonumber\\
\nonumber\\
&&\quad=\;\left|  \frac{Ph_{L,a}P(a,R)+Ph_{a,R}P(L,a)}{P(L,R)P(L,a)}-\frac{P_nh_{L,a}P_n(a,R)+P_nh_{a,R}P_n(L,a)}{P_n(L,R)P_n(L,a)} \right|\nonumber\\
\nonumber\\
&& \quad\lesssim\;E_2+\left[(R-a)+E_2\right]o_p(1),
\end{eqnarray}
where the $o_p(1)$ term comes from the functional generalization of the Glivenko-Cantelli theorem and is chosen uniformly for all $L$, $R$ and $a$ in consideration.

It is only left to show that the bound in the statement of the lemma holds for $E_1$ and $E_2$.  We start with $E_2$.  We need to show that there exists a $O_p(1)$ sequence of random variables $M_n$, such that
\begin{equation}
\left|P_n(l,r)-P(l,r)\right| \le \epsilon (r-l)+(\log n / n) M_n,
\end{equation}
for all $(l,r)$ contained within a given bounded set.  Define $M_n$ as the infimum of those values for which the above inequality holds.
Write $A_{j,n}$ for the set of intervals $(l,r)$ that satisfy $[2^{j-1}-1]\log n / n < P(l,r) \le 2^j\log n / n$.   Recall that we have restricted our attention to a bounded interval, on which $\|f\|_{\infty}$ is finite.  In the argument that follows the bounded interval is not explicitly present, however we do use the fact that $c_1:=\|f\|_{\infty}^{-1}$ is positive.  Observe that
\begin{eqnarray*}
P(M_n>t)&\le& \sum_{j=1}^{\infty} P(\{\exists (l,r)\in A_{j,n} \;\text{s.t.}\;  |P_n(l,r)-P(l,r)|>\epsilon(r-l)+t\log n / n\} \\
\\
&\le&\sum_{j=1}^{\infty} P(\{  \sup_{A_{j,n}}|P_n(l,r)-P(l,r)|>(\log n/n)[c_1\epsilon(2^{j-1}-1)+t]\}\\
\\
&\le& \sum_{j=1}^{\infty} \frac{E\sup_{A_{j,n}}|P_n(l,r)-P(l,r)|}
{(\log n/n)[c_1\epsilon(2^{j-1}-1)+t]}  := \sum_{j=1}^{\infty} S_j.
\end{eqnarray*}
We will bound the summands in the above expression using a maximal inequality from the empirical process theory.  Note that the class of indicator functions for all intervals has polynomial bracketing numbers with respect to~$L_2(P)$, and its envelope function is identically equal to one.  Consequently, an application of Lemma 19.36 in \cite{vaart98book} gives us the following bound:
\begin{equation}
\label{VdV.bnd}
E\sup_{P(l,r)<\theta^2} |P_n(l,r)-P(l,r)|\lesssim n^{-1/2}\theta[1+\sqrt{-\log(\theta\wedge 1)}]-\log(\theta\wedge 1)/n.
\end{equation}
We can bound each $S_j$ by applying inequality~(\ref{VdV.bnd}) with $\theta^2=2^j\log n / n$:
\begin{equation*}
S_j\lesssim\frac{2^{j/2}\log n / n}{(\log n/n)[\epsilon(2^{j-1}-1)+t]}=\frac{2^{j/2}}{(2^{j-1}-1)\epsilon+t}.
\end{equation*}
It follows that the sum bounding $P(M_n>t)$ can be made arbitrarily small, uniformly over all~$n$, by increasing~$t$.  Consequently, $M_n=O_p(1)$.

The proof of the bound for $E_1$ is essentially identical to the one above.  The class of functions $d_{l,r}$, for $l$ and $r$ in a fixed bounded interval, also has polynomial bracketing numbers with respect to~$L_2(P)$, and its envelope function is also identically equal to one.  Hence, the argument above works for $E_1$ after each $(l,r)$ is replaced with $d_{l,r}$.

\subsection{Proof of Propositions \ref{merge.crit.eq}-\ref{prop.gen}}\label{appendix}

\subsubsection*{Proposition~\ref{merge.crit.eq}}
Let~$\lambda_0=0$.  Suppose that the first merge happens at $\lambda=\lambda_1$, the second at $\lambda=\lambda_2$, and so on.  We will first show that, with probability one, values~$\lambda_k$ form an increasing sequence.  Consider two merges, $C_3=C_1\cup C_2$ and $C=C_3\cup C_4$.  For concreteness, we will focus on the case where cluster $C_4$ exists at the time of the first merge, and establish
\begin{equation}
\frac{\overline{X}_{C_2}-\overline{X}_{C_1}}{|C_3|}\le\frac{\overline{X}_{C_4}-\overline{X}_{C_3}}{|C|}.
\end{equation}
This will complete the proof because of the continuity of the underlying distribution.  The complementary case, where $C_4$ is formed after the first merge can be analyzed analogously.
Suppose that the above inequality does not hold.  Then, taking into account representation $|C_3|\overline{X}_{C_3}=|C_1|\overline{X}_{C_1}+|C_2|\overline{X}_{C_2}$, we can derive
\begin{align*}
\overline{X}_{C_4}-\overline{X}_{C_2}&=\overline{X}_{C_4}-\overline{X}_{C_3}-|C_1| \cdot {|C_2|}^{-1}\cdot(\overline{X}_{C_3}-\overline{X}_{C_1})\\
& =\overline{X}_{C_4}-\overline{X}_{C_3}-{|C_1|}\cdot{|C_3|}^{-1}\cdot(\overline{X}_{C_2}-\overline{X}_{C_1})\\
&<(\overline{X}_{C_2}-\overline{X}_{C_1})\bigg(\frac{|C|}{|C_3|}-\frac{|C_1|}{|C_3|}\bigg)=\frac{|C_2|+|C_4|}{|C_3|\,}(\overline{X}_{C_2}-\overline{X}_{C_1}).
\end{align*}
The resulting inequality contradicts the merge $C_3=C_1\cup C_2$.

We will now verify that the KKT conditions for optimization problem~(\ref{clus.criterion}) hold for the solutions produced by Algorithm~\ref{al1}.  The KKT conditions are satisfied if there exist $\beta_{ij}$ with $|\beta_{ij}|\le1$ and $\beta_{ij}=-\beta_{ji}$, such that for every~$i$:
\begin{equation}
\label{KKT.prel}
\alpha_i-x_i+\lambda\sum_{j\ne i, \alpha_j\ne \alpha_i}\text{sign}(\alpha_i-\alpha_j)+\lambda\sum_{j\ne i, \alpha_j=\alpha_i}\beta_{ij}=0.
\end{equation}
Write~$C(i)$ for the current cluster containing~$x_i$.  Taking into account equations~(\ref{eq:clus.centroid}), the KKT conditions can be rewritten as follows,
\begin{equation}
\label{KKT}
\overline{X}_{C(i)}-x_i+\lambda\sum_{j\ne i, j\in C(i)}\beta_{ij}=0.
\end{equation}
We will argue by induction over the number of merges and establish the KKT conditions, (\ref{KKT}), for all~$i$ and for all $\lambda=\lambda_k$.  If (\ref{KKT}) holds for a particular tuning parameter value~$\lambda_k$, it also holds for $\lambda>\lambda_k$, provided the cluster~$C(i)$ is not modified, if we shrink all the corresponding~$\beta_{ij}$ by a factor of~$\lambda_k/\lambda$.

For $\lambda=\lambda_0$, conditions~(\ref{KKT}) hold trivially, with each observation forming its own cluster.    Suppose we are able to verify the KKT conditions up to the merge~$k-1$.   Suppose that the $k$-th merge, at $\lambda=\lambda_k$, is $C=C_1\cup C_2$.  By the discussion above, conditions~(\ref{KKT}) hold at $\lambda=\lambda_k$ for all $i\notin C$, and there exist $\beta_{ij}$ with $|\beta_{ij}|\le1$ and $\beta_{ij}=-\beta_{ji}$, such that
\begin{eqnarray}
\label{KKT.C1}
\overline{X}_{C_1}-x_i+\lambda_k\sum_{j\ne i, j\in C_1}\beta_{ij}&=&0 \quad\text{for all}\; i\in C_1\\
\label{KKT.C2}
\overline{X}_{C_2}-x_i+\lambda_k\sum_{j\ne i, j\in C_2}\beta_{ij}&=&0 \quad\text{for all}\; i\in C_2.
\end{eqnarray}
We will set $\beta_{ij}=-1$ for each $i\in C_1$ and $j\in C_2$, and we will keep the remaining $\beta_{ij}$ values intact. We need to show that
\begin{equation}
\label{KKT.C}
\overline{X}_{C}-x_i+\lambda_k\sum_{j\ne i, j\in C}\beta_{ij}=0\quad\text{for all}\; i\in C.
\end{equation}
Consider an~$i\in C_1$.  Equations~(\ref{KKT.C1}) imply $\lambda_k\sum_{j\ne i, j\in C_1}\beta_{ij}=x_i-\overline{X}_{C_1}$.  Recall that $\lambda_k=(\overline{X}_{C_2}-\overline{X}_{C_1})/|C|$.    It follows that
\begin{equation}
\overline{X}_{C}-x_i+\lambda_k\sum_{j\ne i, j\in C}\beta_{ij}=\overline{X}_C-\overline{X}_{C_1}+(\overline{X}_{C_1}-\overline{X}_{C_2})|C_2|/|C|=0,
\end{equation}
as required.  The argument for $i\in C_2$ is analogous, but uses equations~(\ref{KKT.C2}) instead of~(\ref{KKT.C1}).

\subsubsection*{Proposition~\ref{split.merge.eq}}

We will prove the result by induction over the number of merges.  For each merge we will establish the following claim: the splitting procedure applied to the last formed cluster matches the merging procedure that formed that cluster.  It follows that the sequence of clusters formed by the splitting procedure matches the sequence of clusters formed by the merging algorithm.

The claim for merge number one is trivial: the only possible split exactly matches the first merge.
Suppose that the claim has been established for the first~$k$ merges.  Let~$C$ be the cluster formed by the merge~$k+1$, which combines clusters~$C_1$ and~$C_2$, with~$C_1$ being the left one.  It is only left to show that the first splitting procedure applied to~$C$ produces clusters~$C_1$ and~$C_2$.  Consider a possible alternative split: $C=C_3\cup C_4$, with $C_3$ being the left cluster that is different from~$C_1$.  To verify the claim, we need to establish $\overline{X}_{C_2}-\overline{X}_{C_1}>\overline{X}_{C_4}-\overline{X}_{C_3}$.  For concreteness, we will focus on the case $C_3\subset C_1$.  The case $C_1\subset C_3$ can be handled analogously, taking advantage of the inclusion $C_4\subset C_2$.

Define $C_5=C_1\setminus C_3$.  Using representations $\overline{X}_{C_1}=\overline{X}_{C_3}|C_3|/|C_1|+\overline{X}_{C_5}|C_5|/|C_1|$ and $\overline{X}_{C_4}=\overline{X}_{C_5}|C_5|/|C_4|+\overline{X}_{C_2}|C_2|/|C_4|$,  we can rewrite the desired inequality as
\begin{equation}
\label{spl.prf.eq2}
\frac{\overline{X}_{C_2}-\overline{X}_{C_5}}{|C_4|}>\frac{\overline{X}_{C_5}-\overline{X}_{C_3}}{|C_1|}.
\end{equation}
Suppose that the cluster~$C_1$ was formed by the merge $C_{11}\cup C_{12}$.  The induction claim for merges one through~$k$ implies that $\overline{X}_{C_{12}}-\overline{X}_{C_{11}}$ maximizes the corresponding difference of the averages over all partitions of~$C_1$.  By the monotonicity of the~$\lambda$ values in the merging algorithm, we have $(\overline{X}_{C_2}-\overline{X}_{C_1})/|C|>(\overline{X}_{C_{12}}-\overline{X}_{C_{11}})/|C_1|$, which yields
\begin{equation}
\label{spl.prf.eq3}
\frac{\overline{X}_{C_2}-\overline{X}_{C_1}}{|C|}>\frac{\overline{X}_{C_5}-\overline{X}_{C_3}}{|C_1|}.
\end{equation}
Consequently, if we can establish that
\begin{equation}
\label{spl.prf.eq4}
\frac{\overline{X}_{C_2}-\overline{X}_{C_5}}{|C_4|}>\frac{\overline{X}_{C_2}-\overline{X}_{C_1}}{|C|},
\end{equation}
then the required inequality~(\ref{spl.prf.eq2}) is satisfied.  Using representation $\overline{X}_{C_1}=\overline{X}_{C_3}|C_3|/|C_1|+\overline{X}_{C_5}|C_5|/|C_1|$ we can rewrite~(\ref{spl.prf.eq4}) as~$(\overline{X}_{C_2}-\overline{X}_{C_1})/|C|>(\overline{X}_{C_5}-\overline{X}_{C_3})/|C_1|$.  The last inequality is true by~(\ref{spl.prf.eq3}), which completes the proof.

\subsubsection*{Proposition~\ref{prop.unimod}}

\begin{figure}[h]
\begin{center}
\includegraphics[width=25pc]{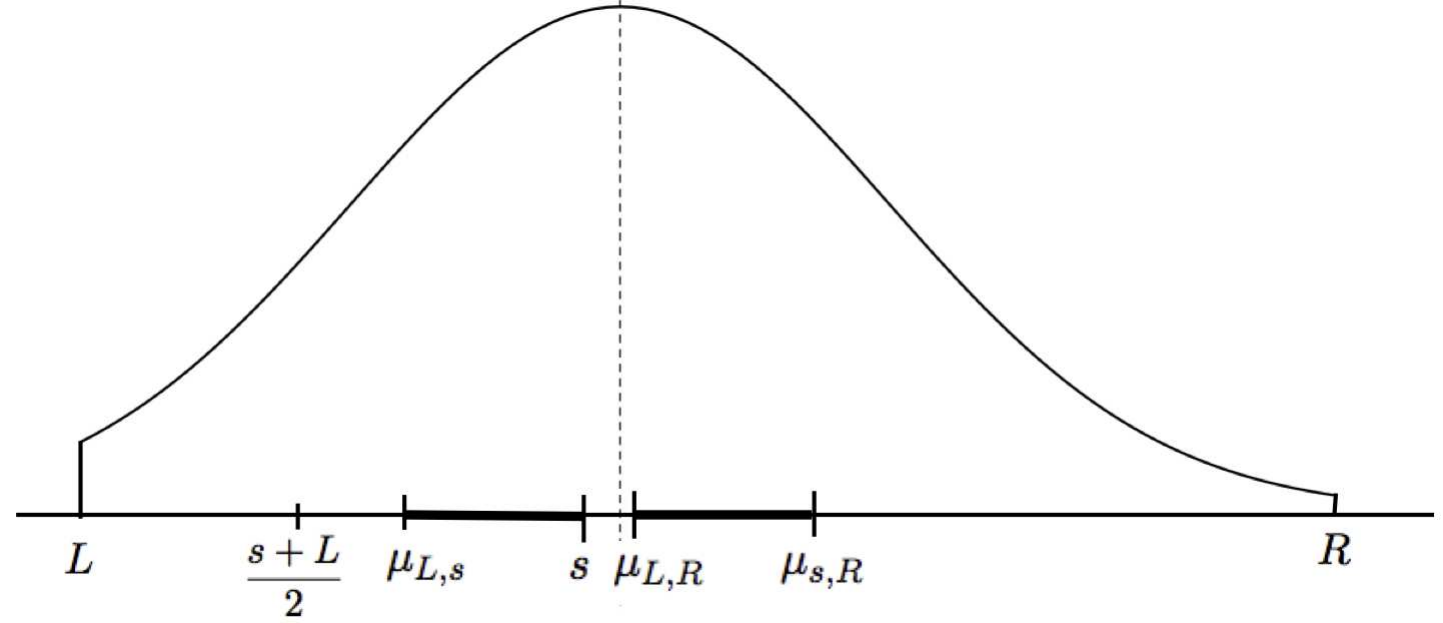}
\caption{Illustration of the proof of Proposition~\ref{prop.unimod} on a truncated normal density.}\label{fig-unimodal-split1}
\end{center}
\end{figure}

We will first establish that the population procedure produces no splits.
More specifically, we will show that for every~$s$ that is an extremum of~$G_{L,R}(\cdot)$ on the interval~$(L,R)$, we have $G_{L,R}(s)<G_{L,R}(L)\vee G_{L,R}(R)$.
First, consider the case where~$s$ is less than or equal to the mode of the density, $f$, restricted to $(L,R)$.  Differentiating function~$G_{L,R}$ gives
\begin{equation*}
G_{L,R}'(s) = \frac{f(s)P_{L,R}}{P_{L,s}P_{s,R}}[\mu_{L,s}+\mu_{s,R}-\mu_{L,R}-s].
\end{equation*}
Thus, if~$s$ is an extremum, then $\mu_{s,R}-\mu_{L,R}=s-\mu_{L,s}$ (see Figure~\ref{fig-unimodal-split1}).

\noindent Consequently,
\begin{eqnarray*}
G_{L,R}(s)\,=\,\mu_{s,R}-\mu_{L,s}&=&2(s-\mu_{L,s})+ \mu_{L,R}-s\\
&<& s-L+\mu_{L,R}-s=G_{L,R}(L).
\end{eqnarray*}
The last inequality follows from the fact that~$f$ is strictly increasing on~$(L,s)$, which implies $s-\mu_{L,s}<(s-L)/2$.  Hence, $G_{L,R}(s)<G_{L,R}(L)$.
In the case where~$s$ is greater than the mode of~$f$ on~$(L,R)$, an analogous argument establishes $G_{L,R}(s)<G_{L,R}(R)$.

We will now prove that the population clustering procedure is uniquely defined.  As we have established in the first part of the proof, the population procedure has the following options: (i) reduce the support to an empty set using a one-sided truncation; (ii) do the same using two-sided truncation; and (iii) start with a one-sided truncation, followed by the two-sided truncation, which reduces the support to an empty set.  Because one-sided truncation is always well defined, we can focus on the two-sided truncation exclusively.   Consider an interval $(L_1,R_1)$, such that $\argmax G_{L_1,R_1}=\{L_1,R_1\}$.  We need to show that there exists, one and only one, continuous and decreasing function $l\mapsto R_l$, defined on $[L_1,L_2]$, for some~$L_2>L_1$, such that $R_1=R_{L_1}$ and $L_2=R_{L_2}$.  Recalling the more general existence and uniqueness argument in the proof of Proposition~\ref{prop.gen}, we note that it is enough to verify that
\begin{equation}
\label{prf.prop.unimod.ineq}
[f(L)\vee f(R)](R-L)<P_{L,R},
\end{equation}
for all $L<R$ with $L\ge L_1$, $R\le R_1$ and $\argmax G_{L,R}=\{L,R\}$.  Suppose that inequality~(\ref{prf.prop.unimod.ineq}) is violated.  Then, unimodality of~$f$ implies $f(L)\ne f(R)$. Another consequence of the violation is that $\mu_{L,R}<(L+R)/2$ if $f(L)>f(R)$ and $\mu_{L,R}>(L+R)/2$, otherwise.  This contradicts the equality $\mu_{L,R}=(L+R)/2$, implied by $G_{L,R}(L)=G_{L,R}(R)$.  We conclude that $[f(L)\vee f(R)](R-L)<P_{L,R}$.  This completes the proof.

Note that in the first part of the proof we establish the following \textit{general property}, which holds for any continuous~$f$: if the population procedure splits $(L^*,R^*)$ at point~$s^*$, then $\mu_{L^*,s^*}\le(L^*+s^*)/2$ and $\mu_{s^*,R^*}\ge(s^*+R^*)/2$.  One consequence of this is that a population split point must lie to the right of left-most mode of~$f$ and to the left of the right-most mode.

This completes the proof of Proposition~\ref{prop.unimod}.  We will now show that C2 holds for all bimodal densities, provided the smoothness condition, C1, is satisfied.

Let~$\dmin$ denote the interior local minimum of density~$f$.  Consider an interval $(L,R)$ that is a subset of the support of~$f$.   Suppose there exist points~$s_1<s_2$, such that $\max G_{L,R}=G_{L,R}(s_1)=G_{L,R}(s_2)$.    We will establish that for every $s$ in $(s_1,s_2)$ equality $G'_{L,R}(s)=0$ implies $G''_{L,R}(s)<0$. Consequently, on the interval $(s_1,s_2)$ function $G_{L,R}$ is not flat and has no local minima, which results in a contradiction.

Define $g_R(s)=f(s)(\mu_{s,R}-s)/P_{s,R}$.  Note that
\begin{equation}
\label{pro.prop.2dG}
G''_{L,R}(s)=f(s)[2g_R(s)-1]\left(\frac1{P_{s,R}}+\frac1{P_{L,s}}\right) \quad  \text{when}\quad G'_{L,R}(s)=0.
\end{equation}
Because $G'(s_1)=0$ and $G''_{L,R}(s_1)\le0$, we have $2g_R(s_1)-1\le 0$.

In the proof of Proposition~\ref{prop.unimod} we showed that each interior maximizer of~$G_{L,R}$ must lie in between the two modes of~$f$.  Suppose that $s_1$ is located to the left of $\dmin$.   It follows that $f$ is decreasing on $(s_1,\dmin)$. Define $\kappa(s)=P_{s,R}(2g_R(s)-1)$ and recall that  $\kappa(s_1)\le0$.  Because $\kappa'(s)=2f'(s)(\mu_{s,R}-s)+f(s)(2g_R(s)-1)$, we have $\kappa'(s_1)<0$, and, consequently,  $\kappa'(s)<0$ and $\kappa(s)<0$ for all~$s$ in $(s_1,\dmin]$.  Thus, it follows from formula~(\ref{pro.prop.2dG}) that, for $s$ in $(s_1,\dmin]$, equality $G'_{L,R}(s)=0$ implies $G''_{L,R}(s)<0$.

Define $g_L(s)=f(s)(s-\mu_{L,s})/P_{L,s}$, and note that $g_R(s)=g_L(s)$ when $G'_{L,R}(s)=0$.  An argument analogous to the one in the previous paragraph, but with~$s_2$ and~$g_L$ used instead of~$s_1$ and~$g_R$, establishes that, for $s$ in $(\dmin,s_2)$, equality $G'_{L,R}(s)=0$ implies $G''_{L,R}(s)<0$.  This completes the proof.

\subsubsection*{Proposition~\ref{prop.gen}}

We will first focus on the case where the two-sided truncation starts at some bounded interval $(L_1,R_1)$.  Thus, $R_{L_1}=R_1$.  We will show that there exists a point~$L^*>L_1$ and a small $\epsilon>0$, such that the function~$R_L$, with the required properties, is uniquely defined either for~$L\in (L_1-\epsilon,L^*]$, where $L^*=R_{L^*}$, or for~$L\in (L_1-\epsilon,L^*+\epsilon)$, where the population procedure splits cluster $(L^*,R_{L^*})$.
Recall that the required properties of $R_L$ are that it is continuous and decreasing, $\arg\max G_{L,R_L}=\{L,R_L\}$ for $L\in[L_1,L^*)$, $\arg\max G_{L^*,R_{L^*}}\supseteq\{L^*,R_{L^*}\}$, and $G_{L,R_L}(L)=G_{L,R_L}(R_L)$ for $L\in(L_1-\epsilon,L^*+\epsilon)$.

Write $F(L,R)$ for $G_{L,R}(L)/2-G_{L,R}(R)/2$, and note that $F(L,R)=\mu_{L,R}-(R+L)/2$.  Observe that $\arg\max G_{L,R}=\{L,R\}$ implies $F(L,R)=0$.
To simplify the notation we write $F_1(L,R)$ for the derivative $\partial F(l,r)/\partial l$, evaluated at $(L,R)$, and write $F_2(L,R)$ for the corresponding value of $\partial F(l,r)/\partial r$.  Differentiation of~$F$ gives us
\begin{equation}
F_1(L,R)=\frac{f(L)(\mu_{L,R}-L)}{P_{L,R}}-\frac12  \quad\text{and}\quad  F_2(L,R)=\frac{f(R)(R-\mu_{L,R})}{P_{L,R}}-\frac12.
\end{equation}
According to display~(\ref{prf.lem3.dervs}) in the proof of Lemma~\ref{lem.pop.ineq}, we have $F_1(L,R)=G'_{L,R}(L)$ and $F_2(L,R)=-G'_{L,R}(R)$.  Because $\arg\max G_{L_1,R_1}=\{L_1,R_1\}$, we then also have inequalities $F_1(L_1,R_1)\le0$ and $F_2(L_1,R_1)\le0$.
Suppose, first, that both of the above inequalities are strict.  Then, both $F_1(L_1,R_1)$ and $F_2(L_1,R_1)$ are negative.  By the classical implicit function theorem, function $L\mapsto R_L$, for which $F(L,R_L)=0$, and hence $G_{L,R_L}(L)=G_{L,R_L}(R_L)$, is uniquely defined on $(L_1-\epsilon,L_1+\epsilon)$, for some small positive $\epsilon$.  By the same theorem $R_L$ is also continuous and decreasing.

As~$L$ increases from the value~$L_1$, we define $R_L$ as the largest~$R$ below~$R_1$, for which $F(L,R)=0$.  Note that the smallest $R_L$ can be is~$L$.  We continuously decrease~$L$ until $F_1(L,R_L)=0$, or $F_2(L,R_L)=0$, or $L=R_L$, and let~$L_2$ be the value of~$L$ corresponding to the stopping point.  Note that, by the implicit function theorem,  function~$R_L$ is continuous and decreasing on $(L_1-\epsilon,L_2)$, as well as uniquely defined.  Because $R_L$ is continuous, decreasing and bounded below by~$L_2$, we can define it at~$L_2$ by continuity, also preserving the monotonicity and the lower bound.  We write~$R_2$ for~$R_{L_2}$.

If $L_2=R_2$ or $L^*<L_2$, then the proof of existence and uniqueness of the two-sided truncation is complete.  Otherwise, we need to handle the case where $F_1(L_2,R_2)=0$ or $F_1(L_2,R_2)=0$.  Note that $G_{L_2,R_2}(L_2)=G_{L_2,R_2}(R_2)=\max G_{L_2,R_2}$.  We will now focus on the case $F_1(L_2,R_2)=F_2(L_2,R_2)=0$.  The remaining two cases follow from analogous arguments.

The second order Taylor series approximation to $F(L,R)$ at $(L_2,R_2)$ is
\begin{equation*}
F(L,R)-F(L_2,R_2)=\frac{(R_2-L_2)}{2P_{L_2,R_2}}\left[f'(L_2)(\Delta L)^2+f'(R_2)(\Delta R)^2\right]+o\left(|\Delta L|^2+|\Delta R|^2\right),
\end{equation*}
where $\Delta R = R-R_2$ and $\Delta L = L-L_2$.
As we point out in the proof of Lemma~\ref{lem.pop.ineq}, equality $G'_{L_2,R_2}(R_2)=F_2(L_2,R_2)=0$ implies $f'(R_2)\ge0$.  Similarly, we have $f'(L_2)\le0$.  Suppose, first, that both inequalities are strict; the remaining case can be handled using analogous arguments that use higher order Taylor expansions.  Then, given an arbitrarily small neighborhood~$\cal{N}$ of $(L_2,R_2)$ we can find $(L,R)\in\cal{N}$, such that $L>L_2$ and $R>R_2$ and $F(L,R)=0$.  Also note that $F_1(L,R)$ and $F_2(L,R)$ are both negative.  Consequently, by the implicit function theorem, the map $l\mapsto R_l$ is uniquely defined in a neighborhood of~$L$, such that $R=R_L$, and the function $R_l$ is continuous and decreasing.  Thus, the map is uniquely defined on $(L_2,L_2+\epsilon)$, for some positive~$\epsilon$, it is continuous and decreasing, and $R_l\uparrow R_2$ as $l\downarrow L_2$.  Thus, the continuity and monotonicity of the map is preserved when we glue together the functions defined on $(L_1,L_2]$ and $(L_2,L_2+\epsilon)$.  The above Taylor approximation implies that when~$\epsilon$ is sufficiently small, derivatives $F_1(L,R)$ and $F_2(L,R)$ are both negative for $L\in(L_2,L_2+\epsilon)$.  Thus, we are back in the regular setting of the previous paragraph.  Applying the above arguments sequentially establishes that the function $R_L$ is uniquely defined, in a continuous and decreasing fashion, either for~$L\in (L_1-\epsilon,L^*]$, where $L^*=R_{L^*}$, or for~$L\in (L_1-\epsilon,L^*+\epsilon)$, where the population procedure splits cluster $(L^*,R_{L^*})$.  The remaining case $(L_1,R_1)=(-\infty,\infty)$ can be handled using an additional, but similar, argument, where $L$ is decreased, rather than increased from the value~$L_1$.


Thus, we have shown that the two-sided truncation of the population procedure is uniquely defined.  Note that the one sided truncation exists and is unique directly by definition, while the uniqueness of the split is ensured by the regularity condition~$C2$.  It is only left to show that the population procedure will implement finitely many steps, for which we just need to check that the procedure will make finitely many splits.

In the proof of Proposition~\ref{prop.unimod} we establish the following general property: a population split point must lie to the right of left-most mode of the truncated version of~$f$, and to the left of the corresponding right-most mode, where the truncation is done to the cluster that is being split.  We will assume that the procedure makes infinitely many splits and conduct an argument to reach a contradiction.  The assumption implies that there exists an interval $(L^i,R^i)$, such that for every positive~$\epsilon$ there exists an infinitely large nested sequence of  distinct population clusters, of the form~$(L,R)$, sandwiched between $(L^i-\epsilon,R^i+\epsilon)$ and $[L^i,R^i]$, such that either the left or right endpoint of each cluster is a population split point for the previous cluster in the sequence.  Without loss of generality, we focus on the case where each right endpoint of the considered cluster sequence is a split point for the previous cluster.  By continuity of $G_{L,R}(s)$ we must then have $G_{L^i,R^i}(R^i)\ge G_{L^i,R^i}(L^i)$.  Below we examine the case where the last inequality is strict; the case of equality can be handled using an analogous argument.   It follows that when~$\epsilon$ is sufficiently small, the population procedure cannot place a split point in $(L,L^i]$ or do a left sided truncation of the cluster $(L,R)$.  Consequently, we have $L=L^i$ for each cluster in the considered sequence.

Recall that $G'_{L,R}(R)=1/2-f(R)(R-\mu_{L,R})/P_{L,R}$.  If $G'(L^i,R^i)(R^i)$ is negative, then, for a sufficiently small~$\epsilon$, $G'_{L,R}(R)$ is negative and bounded away from zero.  Consequently, when $R$ gets sufficiently close to~$R^i$, the corresponding split point must lie below $R^i$, which is a contradiction.  Similarly, we reach a contradiction when $G'(L^i,R^i)(R^i)$ is positive.  For the rest of the proof we focus on the remaining case $G'(L^i,R^i)(R^i)=0$.

The general property given in the beginning of the paragraph before the previous one implies that $f'(R)>0$ for $R\in(R^i,R^i+\epsilon)$ and $f'(R^i)\ge0$.  We first consider the case where the last inequality is strict.
Let $t=s-R^i$ and $r=R-R^i$.  After deriving the second order Taylor series expansion for the function $(s,R)\mapsto G_{L^i,R}(s)$ at $(R^i,R^i)$, we establish approximation
\begin{equation*}
G_{L^i,R^i+r}(R^i+t)-G_{L^i,R^i+r}(R^i)=\frac{-f'(R^i)}{6f(R^i)}(t^2+tr)+o(|t|^2+|r|^2),
\end{equation*}
which holds for $(t,r)$ near zero.  Because we now focus on the case $f'(R^i)>0$, we must have $G(L^i,R^i+r)(R^i-r/2)>\max_{s\in[R^i,R^i+r]} G(L^i,R^i+r)(s)$, for every sufficiently small positive $r$.  Consequently, when~$\epsilon$ is sufficiently small, the population procedure applied to an interval $(L^i,R)$ satisfying $R\in(R^i,R^i+\epsilon)$ must place the split point below~$R^i$.  This is again a contradiction of our assumption.

Now consider the remaining case $f'(R^i)=0$.  Let $t=s-R^i$ and $r=R-R^i$.  After deriving the third order Taylor series expansion for the function $(s,R)\mapsto G_{L^i,R}(s)$ at $(R^i,R^i)$, we establish approximation
\begin{equation*}
G_{L^i,R^i+r}(R^i+t)-G_{L^i,R^i+r}(R^i)=\frac{-f''(R^i)}{6f(R^i)}(3t^3-t^2r+3tr^2)+o(|t|^3+|r|^3),
\end{equation*}
for $(t,r)$ near zero.  We will focus on the case $f''(R^i)\ne0$, as the case $f''(R^i)=0$ can be handled using analogous arguments that use higher order Taylor approximation.  Because~$R^i$ is a local minimum of~$f$, we must have $f''(R^i)>0$.  Analysis of the above Taylor expansion reveals that $G(L^i,R^i+r)(R^i-r)>\max_{s\in[R^i,R^i+r]} G(L^i,R^i+r)(s)$, for every sufficiently small positive $r$.  Consequently, when~$\epsilon$ is sufficiently small, the population procedure applied to an interval $(L^i,R)$ satisfying $R\in(R^i,R^i+\epsilon)$ must place the split point below~$R^i$.  This contradiction completes the proof.

%% file: appendix-2.tex
\subsection{Analysis of the Population Procedure for Gaussian Mixtures}\label{appendix:normals}

Consider the case where~$f$ is a mixture of two Gaussian densities on the real line.    Note that regularity condition C1 is satisfied.   Proposition~\ref{prop.unimod} imply that condition C2 is satisfied as well.   Thus, the population procedure is well defined, and, according to the results in Section~\ref{sec.gen.results}, the sample clustering procedure is consistent.
In Table~\ref{table1} we document the behaviour of the population clustering procedure for various mixtures of two Gaussian distributions on the real line.  For~$7$ different levels of separation between the two normal means we consider~$9$ different mixing proportions, from the symmetric case of $50:50$ mixing to the highly non-symmetric $10:90$ mixing.  The behavior of the population splitting procedure in other cases can be interpolated from the table using continuity arguments.  We present, where applicable, the location of the split point, $s^*$ (``NO'' denotes that no splits was detected), as well as the endpoints, $L^*$ and~$R^*$, of the corresponding cluster.  The local minimum of the density, $\dmin$, and the split point minimizing the expected misclassification error, $s_{\text{MC}}$, are also provided.  We also report, under \textit{Excess MCE},  how much the misclassification error of the population clustering procedure exceeds that of the $s_{\text{MC}}$ based oracle rule.  Even though the population procedure focusses on cluster separation, rather than classification, its \textit{Excess MCE} is well controlled when the number of splits is detected correctly.
In all cases where~$s^*$ is not reported, we checked that the two-sided truncation takes the left endpoint of the cluster all the way to~$\dmin$, and thus reduces the density to a unimodal one.   Proposition~\ref{prop.unimod} then implies that the population procedure does not produce any splits.  Similarly, in the cases where~$s^*$ is provided, we verified that the bimodal sub-cluster is truncated down to a unimodal one, which means that the procedure does not produce a second split.

\input{table-1}

\subsubsection{Additional Technical Details}

Note that the population procedure for a distribution supported by the real line starts with a two-sided truncation of the support.  The truncation proceeds along the interval $(L,R_L)$, as~$L$ is increased, and stops when either the support is reduced all the way to an empty set or the maximum of $G_{L,R_L}$ is achieved at an interior point.  In the latter case, define $L^*$ as the smallest~$L$ for which there exists an~$s$ in $(L,R_L)$, such that $\mu_{L,s}=(L+s)/2$ and $\mu_{s,R_L}=(s+R_L)/2$.  Write~$s^*$ for the corresponding point~$s$, and let~$R^*$ stand for~$R_{L^*}$.  The population procedure truncates the support to the interval~$(L^*,R^*)$, which is then split at~$s^*$.  Note that $L^*$ must be smaller than~$\dmin$, the local minimum of the density, by Proposition~\ref{prop.unimod}.

Because analytical solutions are not available, we find~$L^*$, $R^*$ and~$s^*$ numerically.  For each~$L$ on a dense grid we locate~$R_L$ using equation $\mu_{L,R_L}=(L+R_L)/2$.   The symmetry of Gaussian distributions implies that in the search for~$L^*$ we only need to consider the values of~$L$ that satisfy  inequality $L\ge 2(\mu_1 - \mu_2)+ R_L$.    If the above inequality holds, then we focus on~$s$, the local maximum of~$G_{L,R_L}$, and compute
$\delta_1= \mu_{L,s}-(L+s)/2$ and $\delta_2=\mu_{s,R_L}-(s+R_L)/2$.  The proof of Proposition~\ref{prop.gen} shows that at $L=L^*$ we have $\delta_1=\delta_2=0$, and, as $L$ crosses~$L^*$, inequalities $\delta_1<0$ and $\delta_2>0$ are satisfied for the first time.   Thus, as~$L$ is increased, $L^*$ can be taken as the first point where the above inequalities hold.  To check for the second split we apply the same approach on the bimodal sub-cluster.  In all the reported cases, we verify that the two-sided truncation takes an endpoint of the sub-cluster all the way to~$\dmin$, and thus reduces the density to a unimodal one.   Proposition~\ref{prop.unimod} then implies that the population procedure does not produce a second split.

\begin{figure}
\begin{center}
\includegraphics[width=\textwidth]{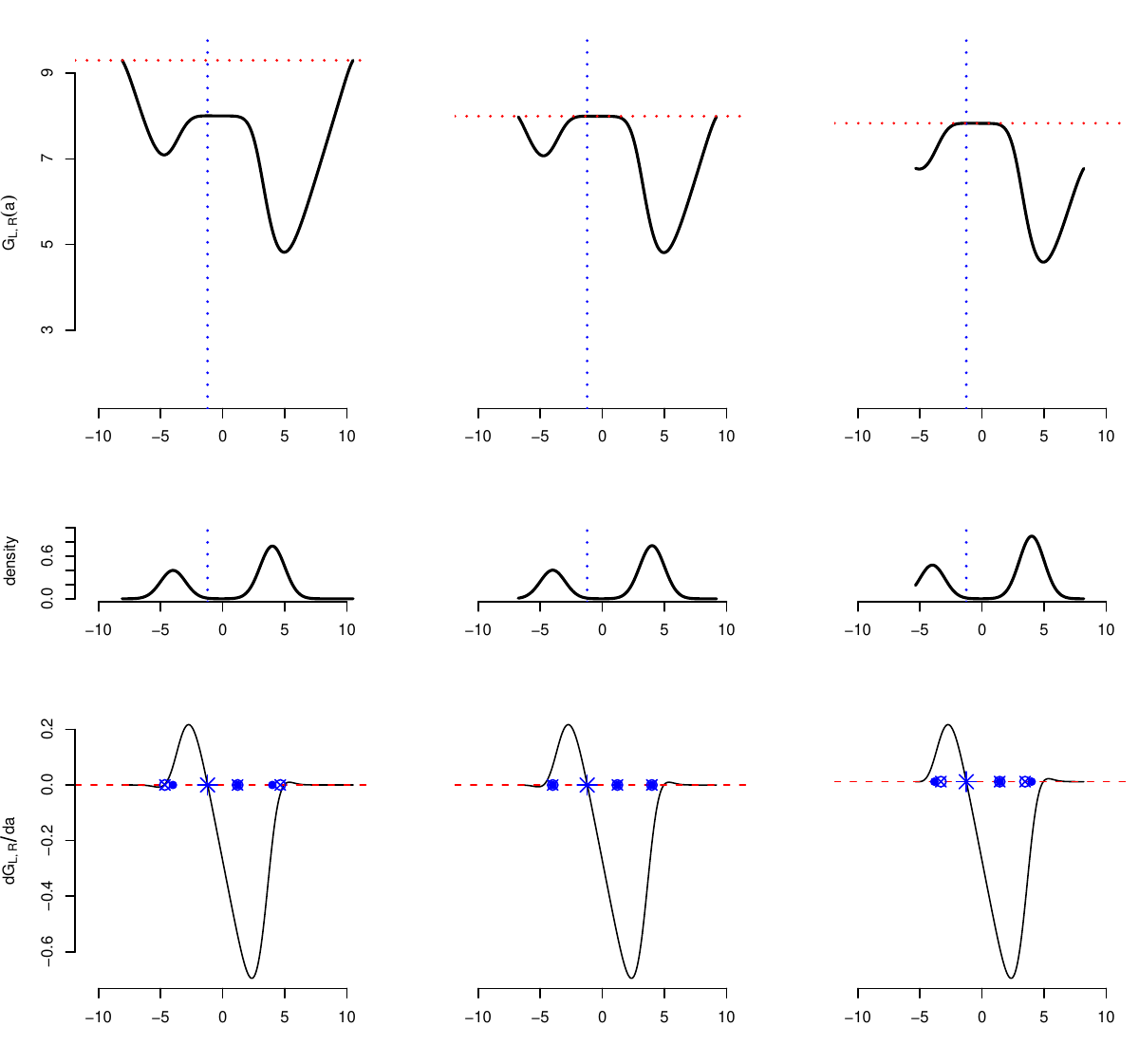}
\caption{Across rows we have the plots of $G_{L,R}(a)$, the truncated density and the normalized $G'_{L,R}(a)$, as $L$ varies across columns over the 3 cases: (i) $L < \lstar$    (ii) $L=\lstar$ (iii) $L>\lstar$.  The population density used here is $0.35\, N(-4,1)+ 0.65 N \, (4,1)$. The dotted blue line and the stars denote the position of the zero of $G'$. In the last row, the conditional means, $\mu_{Ls}$, $\mu_{LR}$ and $\mu_{sR}$, are denoted by circles and the corresponding mid-points, $(L+s)/2$, $(L+R_L)/2$ and $(s+R_L)/2$, by squares. The signs differences~$\delta_1$ and~$\delta_2$, defined in Section~\ref{sec.gen.results}, vary as follows  (i) $(\delta_1>0,\delta_2<0)$ (ii) $(\delta_1 = 0,\delta_2 = 0)$ (iii) $(\delta_1<0,\delta_2>0)$. }
\label{fig3}
\end{center}
\end{figure}

We now provide closed form expressions for~$\mu_{L,R}$, $G_{L,R}$ and $G_{L,R}'$.  Figure~\ref{fig3}  displays $G_{L,R}(a)$, $f$ and  $G_{L,R}'$ (normalized), corresponding to the $0.35\,N(-4,1)+0.65\,N(4,1)$ distribution, for three values of~$L$: (a) $L<L^*$, (b) $L=L^*$, (c) $L>L^*$.  Locations of $\mu_{L,s}, (L+s)/2, \mu_{s,R_L}$ and $(s+R_L)/2$ are also provided.


If the population density $f$ is a mixture of $k$ normals: $f (x):=\sum_{i=1}^k w_i \phi(x-\mu_i)$ then using the following inequality,
$$\int x \, \phi(x-\mu) \, dx = \mu\, \Phi(x-\mu) - \phi(x-\mu),$$
we get closed from expressions of $G_{L,R}$ which can be subsequently optimized:
$$G_{L,R}(a)=P^{-1}_{a,R} \, \int_{a}^R x f(x) \, dx - P^{-1}_{L,a} \, \int_{L}^a x f(x), \, dx \quad \text{ where}$$
\begin{align*}
&\int_{a}^R x f(x) \, dx = \sum_{i=1}^k w_i  \mu_i \{\Phi(R-\mu_i) - \Phi(a-\mu_i)\} - f(R)+f(a),\\
& \int_{L}^a x f(x) \, dx = \sum_{i=1}^k w_i \mu_i \{\Phi(a-\mu_i) - \Phi(L-\mu_i)\} - f(a)+f(L), \text{ and }\\
& P_{a,R}=\sum_{i=1}^k w_i \big\{\Phi(R-\mu_i) - \Phi(a-\mu_i)\big\},
\quad
P_{L,a}=\sum_{i=1}^k w_i \big\{\Phi(a-\mu_i) - \Phi(L-\mu_i)\big\}.
\end{align*}
The following alternative expression for $G_{L,R}(a)$ in terms of the conditional means and density
$f_{L,R}(x) = P^{-1}_{L,R} \, f (x) \, \I\{L,R\}$,  where $\PP_{a_1,a_2}= P^{-1}_{L,R} \, P_{a_1,a_2}$, is also useful:
\begin{align*}
G_{L,R}(a)&= \PP^{-1}_{L,a} \int_{a}^R x\, f_{L,R}(x) \, dx  - \PP^{-1}_{a,R} {\int_{L}^a x \, f_{L,R}(x) \, dx}\\
&=\{\PP_{L,a} \, \PP_{a,R}\}^{-1} \bigg\{ \PP_{L,a} \, \mu_{L,R} - \int_{\infty}^a x\, f_{L,R}(x) \, dx \bigg \}\\
&=f_{L,R}(a) \{\PP_{L,a} \, \PP_{a,R}\}^{-2} \bigg[\mu_{L,R} \PP_{a,L}^2+(1-2 \,\PP_{a,L}) \int_{-\infty}^a  x\, f_{L,R}(x) \,dx -  \PP_{a,L} \,{\PP}_{a,R}\bigg].
\end{align*}
Differentiating the above with respect to $a$, we arrive at:
$G'_{L,R}(a) = \kappa_{L,R}(a) \times H_{L,R}(a)$,
where
\begin{align*}
\kappa_{L,R}(a) & = f(a) \, P^{-2}_{L,a} \, P^{-2}_{a,R} \, P_{L,R}^{-1}, \;\;\text{ and}\\
H_{L,R}(a)& =\mu_{L,R} P_{L,a}^2 + \{P_{a,R}-P_{L,a}\} \int_{L}^a x f(x) \, dx - a\, P_{a,L} \,{P}_{a,R} .
\end{align*}
Note that $\kappa_{L,R}(a) > 0$ for all $a \in(L,R)$. Hence, to track the the extremas of $G_{L,R}$, it is enough to search for the zeros of $H_{L,R}(a)$.
We call $H_{L,R}$ the normalized $G'_{L,R}$. Figure~\ref{fig3} shows the plots of $G_{L,R}(a)$, the truncated density and  $H_{L,R}(a)$, when $f = 0.35\, N(-4,1)+ 0.65 N \, (4,1)$. The plot of $G_{L,R}$ appears flat in the neighborhood of the split. But, the plot of $H_{L,R}$ clearly shows only one zero-crossing and demonstrates uniqueness of the maximum of $G_{L,R}$ in the case of interest.

%% file: table-1.tex
\begin{table}
\centering
\scalebox{0.68}{
\begin{tabular}{|c|rrrr|r|rrr|c|cc|}
 \hline
 CASE & $p_1$ & $p_2$ & $\mu_1$ & $\mu_2$ &  $\dmin$ & $\sstar$ & $\lstar$ & $\rstar$ & 2nd split & $s_{\text{MC}}$ & \;\; Excess MCE \\ 
  \hline
  \hline
\multirow{9}{*}{$\bm{|\mu_2-\mu_1| = 9}$} 
  & 0.50 & 0.50 & -4.50 & 4.50 &  0.00 & 0.00 & -8.99 & 8.99 & NO & 0.00 & $0.000$\\ 
  & 0.45 & 0.55 & -4.50 & 4.50 &  -0.02 & -0.45 & -8.53 & 9.43 & NO & -0.02 & $0.000$\\ 
  & 0.40 & 0.60 & -4.50 & 4.50 &  -0.05 & -0.90 & -8.08 & 9.88 & NO & -0.04 & $0.000$\\ 
  & 0.35 & 0.65 & -4.50 & 4.50 &  -0.07 & -1.36 & -7.62 & 10.33 & NO & -0.07 & $0.000$ \\ 
  & 0.30 & 0.70 & -4.50 & 4.50 &  -0.10 & -1.82 & -7.17 & 10.79 & NO &  -0.09 & $0.001$ \\ 
  & 0.25 & 0.75 & -4.50 & 4.50 &  -0.13 & -2.31 & -6.67 & 11.24 & NO & -0.12 & $0.004$\\ 
  & 0.20 & 0.80 & -4.50 & 4.50 &  -0.16 & -2.90 & -6.09 & 11.70 & NO & -0.15 & $0.011$\\ 
  & 0.15 & 0.85 & -4.50 & 4.50 &  -0.20 & -3.82 & -5.09 & 12.16 & NO & -0.19 & $0.037$\\ 
  & 0.10 & 0.90 & -4.50 & 4.50 &  -0.26 & NO &  &  &   & -0.24 & $0.100$\\ 
   \hline
   \hline
   \multirow{9}{*}{$\bm{|\mu_2-\mu_1| = 8}$}
   & 0.50 & 0.50 & -4.00 & 4.00 & 0.00 & 0.00 & -7.99 & 7.99 & NO & 0.00 & $0.000$ \\ 
   & 0.45 & 0.55 & -4.00 & 4.00 & -0.03 & -0.40 & -7.58 & 8.38 & NO & -0.03 & $0.000$\\ 
   & 0.40 & 0.60 & -4.00 & 4.00  & -0.05 & -0.80 & -7.17 & 8.78 & NO & -0.05 & $0.000$\\ 
   & 0.35 & 0.65 & -4.00 & 4.00  & -0.08 & -1.22 & -6.77 & 9.19 & NO & -0.08 & $0.001$\\ 
   & 0.30 & 0.70 & -4.00 & 4.00  & -0.11 & -1.64 & -6.34 & 9.59 & NO & -0.11 & $0.003$\\ 
   & 0.25 & 0.75 & -4.00 & 4.00  & -0.15 & -2.12 & -5.86 & 9.99 & NO & -0.14 & $0.008$\\ 
   & 0.20 & 0.80 & -4.00 & 4.00 &  -0.18 & -2.72 & -5.25 & 10.40 & NO & -0.17 & $0.020$\\ 
   & 0.15 & 0.85 & -4.00 & 4.00 &  -0.23 & NO & & & & -0.22 & $0.150$\\ 
   & 0.10 & 0.90 & -4.00 & 4.00 & -0.29 & NO & & &  & -0.28 & $0.100$ \\ 
   \hline
   \hline
\multirow{9}{*}{$\bm{|\mu_2-\mu_1| = 7}$}
& 0.50 & 0.50 & -3.50 & 3.50 & 0.00 & 0.00 & -6.98 & 6.98 & NO & 0.00 & $0.000$ \\ 
 & 0.45 & 0.55 & -3.50 & 3.50  & -0.03 & -0.35 & -6.63 & 7.34 & NO & -0.03 & $0.000$\\ 
  & 0.40 & 0.60 & -3.50 & 3.50 & -0.06 & -0.71 & -6.27 & 7.69 & NO & -0.06 & $0.001$\\ 
 & 0.35 & 0.65 & -3.50 & 3.50  & -0.10 & -1.09 & -5.90 & 8.04 & NO & -0.09 & $0.003$\\ 
  & 0.30 & 0.70 & -3.50 & 3.50  & -0.13 & -1.49 & -5.49 & 8.39 & NO & -0.12 & $0.007$ \\ 
  & 0.25 & 0.75 & -3.50 & 3.50  & -0.17 & -1.97 & -5.01 & 8.75 & NO & -0.16 & $0.016$\\ 
  & 0.20 & 0.80 & -3.50 & 3.50 & -0.22 & -2.66 & -4.32 & 9.12 & NO & -0.20 & $0.040$\\ 
  & 0.15 & 0.85 & -3.50 & 3.50  & -0.27 & NO & & &  &-0.25 & $0.150$\\ 
  & 0.10 & 0.90 & -3.50 & 3.50  & -0.34 & NO & & &  &-0.31 & $0.100$\\ 
   \hline
   \hline
\multirow{9}{*}{$\bm{|\mu_2-\mu_1| = 6}$}
& 0.50 & 0.50 & -3.00 & 3.00  & 0.00 & 0.00 & -5.99 & 5.99 & NO & 0.00 & 0.000 \\ 
 & 0.45 & 0.55 & -3.00 & 3.00  & -0.04 & -0.32 & -5.66 & 6.28 & NO & -0.03 &0.001\\ 
  & 0.40 & 0.60 & -3.00 & 3.00  & -0.08 & -0.64 & -5.34 & 6.59 & NO &-0.07 & 0.004\\ 
  & 0.35 & 0.65 & -3.00 & 3.00  & -0.12 & -0.99 & -4.99 & 6.89 & NO &-0.10 & 0.008\\ 
   & 0.30 & 0.70 & -3.00 & 3.00 & -0.16 & -1.39 & -4.59 & 7.20 & NO & -0.14 &0.016\\ 
  & 0.25 & 0.75 & -3.00 & 3.00  & -0.21 & -1.91 & -4.07 & 7.52 & NO &-0.18 & 0.034\\ 
   & 0.20 & 0.80 & -3.00 & 3.00  & -0.26 & NO & & & &-0.23 & 0.200\\ 
   & 0.15 & 0.85 & -3.00 & 3.00  & -0.33 & NO & & & &-0.29 & 0.150\\ 
  & 0.10 & 0.90 & -3.00 & 3.00 & -0.41 & NO & & & &-0.37 & 0.100\\ 
   \hline
   \hline
\multirow{9}{*}{$\bm{|\mu_2-\mu_1| = 5}$} 
  & 0.50 & 0.50 & -2.50 & 2.50 &  0.00 & 0.00 & -4.97 & 4.97 & NO & 0.00 & 0.000 \\ 
  & 0.45 & 0.55 & -2.50 & 2.50 & -0.05 & -0.30 & -4.68 & 5.23 & NO &-0.04 & 0.005\\ 
  & 0.40 & 0.60 & -2.50 & 2.50 &  -0.10 & -0.61 & -4.37 & 5.49 & NO &-0.08 & 0.011\\ 
  & 0.35 & 0.65 & -2.50 & 2.50 & -0.15 & -0.96 & -4.01 & 5.75 & NO &-0.12 & 0.021\\ 
  & 0.30 & 0.70 & -2.50 & 2.50 &  -0.20 & -1.41 & -3.56 & 6.02 & NO &-0.17 & 0.041\\ 
  & 0.25 & 0.75 & -2.50 & 2.50 & -0.26 & NO & & & &-0.22 & 0.250\\ 
  & 0.20 & 0.80 & -2.50 & 2.50 &  -0.33 & NO & & & &-0.28 & 0.200\\ 
  & 0.15 & 0.85 & -2.50 & 2.50 & -0.41 & NO & & & &-0.35 & 0.149\\ 
  & 0.10 & 0.90 & -2.50 & 2.50 & -0.53 & NO & & & &-0.44 & 0.100\\ 
   \hline
   \hline
\multirow{9}{*}{$\bm{|\mu_2-\mu_1| = 4}$} 
  & 0.50 & 0.50 & -2.00 & 2.00 & 0.00 & 0.00 & -3.89 & 3.89 & NO & 0.00 & 0.000 \\ 
  & 0.45 & 0.55 & -2.00 & 2.00 & -0.07 & -0.32 & -3.62 & 4.15 & NO & -0.05 &  0.015\\ 
  & 0.40 & 0.60 & -2.00 & 2.00 & -0.14 & -0.67 & -3.28 & 4.38 & NO & -0.10 &0.034\\ 
  & 0.35 & 0.65 & -2.00 & 2.00 &-0.21 & -1.12 & -2.85 & 4.62 & NO &-0.15 & 0.065\\ 
  & 0.30 & 0.70 & -2.00 & 2.00 & -0.28 & NO & & & &-0.21 & 0.298\\ 
  & 0.25 & 0.75 & -2.00 & 2.00 & -0.37 & NO & & & &-0.28 & 0.248\\ 
  & 0.20 & 0.80 & -2.00 & 2.00 & -0.47 & NO & & & &-0.35 & 0.198\\ 
  & 0.15 & 0.85 & -2.00 & 2.00 & -0.58 & NO & & & &-0.43 & 0.148\\ 
  & 0.10 & 0.90 & -2.00 & 2.00 & -0.74 & NO & & &  &-0.55 & 0.097\\ 
 \hline
 \hline
  \multirow{9}{*}{$\bm{|\mu_2-\mu_1| = 3}$}
  & 0.50 & 0.50 & -1.50 & 1.50  & 0.00 & 0.00 & -2.68 & 2.68 & NO & 0.00 & 0.000\\ 
  & 0.45 & 0.55 & -1.50 & 1.50  & -0.12 & -0.50 & -2.29 & 2.97 & NO & -0.07 & 0.057 \\ 
  & 0.40 & 0.60 & -1.50 & 1.50  & -0.24& NO & & & & -0.14 & 0.396\\ 
  & 0.35 & 0.65 & -1.50 & 1.50 & -0.38 & NO & & & & -0.21 & 0.344\\ 
  & 0.30 & 0.70 & -1.50 & 1.50 & -0.53 & NO & & & & -0.28 & 0.293\\ 
  & 0.25 & 0.75 & -1.50 & 1.50 & -0.71 & NO & & & & -0.37 & 0.241\\ 
  & 0.20 & 0.80 & -1.50 & 1.50 &-1.50 & NO & & & & -0.46 & 0.190\\ 
  & 0.15 & 0.85 & -1.50 & 1.50 & -1.50 & NO & & & & -0.58 & 0.139\\ 
  & 0.10 & 0.90 & -1.50 & 1.50 & -1.50 & NO & & & & -0.73 & 0.090\\ 
  \hline
  \hline
\end{tabular}
}
\caption{\small{Finding the population splits for $2$-normal mixtures: $p_1 \, N(\mu_1,1) +p_2\, N(\mu_2,1)$.}}\label{table1}
\end{table}

%% file: appendix-3.tex
\subsection{Further Details on the Simulation Study \& Real Data Analysis}\label{append:sec4}

For the numerical experiments in Section~\ref{sec.sim} of the main paper,  we implemented the BMT with threshold $\alpha$ uniformly set at $10\%$, and  with the adjustment of at most $50\%$ truncation for the first split. The details are provided in Algorithm~\ref{al.complete}.

\begin{algorithm}
\caption{$\alpha$-thresholded BMT algorithm with truncation adjustment} \label{al.complete}
\vspace*{-12pt}
{
\begin{tabbing}
   \enspace INITIALIZE: \\[1ex]
    \qquad $K= \text{number of clusters } =n.$\\[1ex]
    \qquad Sort data in ascending order and store them as: $\xx=\{x_1,\ldots,x_n\}$.\\[1ex]
    \qquad Assign cluster mean $\{a_1,a_2,\ldots,a_n\}$ to them: $a_i=x_i \text{ for } i=1,\ldots,n$.\\[1ex]
    \qquad Cluster size: $s_i=1,\, i=1,\ldots,n$.\\[1ex]
    \qquad Cluster Membership Indices of $\xx$: $I(\xx)=\{1,\ldots,n\}$.\\[1.5ex]
   \enspace WHILE $K> 1$: \\[1ex]
   \qquad Find the consecutive adjacent centroid distance standardized by cluster sizes:\\[1ex]
   \qquad \qquad  $d(j,j+1)  \leftarrow (a_{j+1} - a _j)/(s_j+s_{j+1})$ \\[1ex]
   \qquad Find the clusters with minimum merging distance: \\[1ex]
   \qquad \qquad $ j^{\star} \leftarrow \argmin_{1\leq j \leq K-1} d(j,j+1)$\\[1ex]
   \qquad Check if it is a Big Merge: $\min \{s_{j^{\star}},s_{j^{\star}+1}\} > \lceil n\alpha \rceil $\\[1ex]
   \qquad \qquad IF Big Merge: Find and Store \textit{Mass after merge} $= (s_{j^{\star}}+s_{j^{\star}+1})/n$ \text{ and }\\ [1ex]
   \qquad \qquad New Split = \small{$\big \{ \max \{ \xx [I(\xx) \text{ is } j^{\star} \} \, s_{j^{\star}} +  \min \{ \xx [ I(\xx)\text{ is } (j^{\star}+1)]\} \, s_{j^{\star}+1}  \big \}\big/ (s_{j^{\star}}+s_{j^{\star}+1});$} \\[1ex]
   \qquad Merge the $j^{\star}$ and $(j^{\star}+1)$ clusters and update the centroid and size of the new cluster:\\[1ex]
   \qquad \qquad $a_{j^{\star}} \leftarrow (s_{j^{\star}} a_{j^{\star}} + s_{{j^{\star}}+1} a_{{j^{\star}}+1})/ (s_{j^{\star}}+s_{{j^{\star}}+1})$ \\[1ex]
   \qquad \qquad  $s_{j^{\star}}\leftarrow (s_{j^{\star}}+s_{{j^{\star}}+1})$\\[1ex]
   \qquad Reduce the number of clusters in path    $K \leftarrow (K-1)$ \\[1ex]
   \qquad Change cluster indices \& cluster member indices of data according to the above reduction:\\[1ex]
   \qquad \qquad FOR $ k \text{ in }  ({j^{\star}}+1):K,\; s_k \leftarrow s_{k+1}; a_k \leftarrow a_{k+1}$\\[1ex]
   \qquad \qquad FOR ALL $\text{I}(\xx)>{j^{\star}}$: reduce index by 1 ,i.e., $I(x) = I(x)-1$\\[1.5ex]
   \qquad ADJUSTMENT: IF \textit{Mass after merge} in the TOP SPLIT $< 50 \%,$ Stored.Splits = NULL;\\[1.5ex]
\enspace OUTPUT \texttt{Stored Splits}.
\end{tabbing}
}
\end{algorithm}
\subsubsection{Modality Assessment}\label{mode.literature}
We compare the performance of the BMT with the following two popular modality assessment procedures:
\begin{description}
\item [Silverman Test] is based on a kernel density estimate.  It uses the idea that if the population density is non-unimodal, a large value of the bandwidth will be required to smooth the data to a unimodal density estimate \citep{Silverman-81} .  The test uses the minimum bandwidth that produces a unimodal kernel estimator. Large values of the minimum bandwidth based test-statistic provide evidence to support the alternative hypothesis of multi-modality. To conduct the Silverman test, we use the R-package referenced in \cite{Vollmer-13}. It is based on Gaussian kernels and incorporates \cite{Hall-01} adjustment for calculating the p-value.
\item [The Dip Test] proposed by \cite{Hartigan-85} is a histogram based method, which does not require estimating the density. The Dip-statistic is the minimum Kolmogorov-Smirnov distance between the empirical distribution and the class of unimodal distributions. Larger values of the Dip-statistic signify departure from the null hypothesis of unimodality. P-values are calculated using the  R-package of \cite{Diptest}. The p-value of this test is quite conservative.
\end{description}

\subsubsection{Estimating the number of clusters}\label{number.of.clusters.literature}
We compare the performance of the BMT with eight  statistical methods that are  popularly used for estimating the number of clusters in a dataset.  A comparison study of~$30$ different approaches in \cite{Milligan-85}  reports the approach in \cite{Calinski-74} as being one of the best performing global methods. It prescribes maximizing the following index over $k$:
\begin{align*}
\text{CH}(k)=\frac{B(k)/(k-1)}{W(k)/(n-k)},
\end{align*}
where $B(k)$ and $W(k)$ are respectively the between and the within clusters sum of squares for $k$ clusters.
Another popular approach, due to \cite{Krzanowski-88}, is based on the  changes in the within clusters sum of squares as new clusters are formed, and seeks to maximize the following ratio over k:
\begin{align*}
\text{KL}(k)=\bigg\vert \frac{\text{DIFF}(k)}{\text{DIFF}(k+1)}\bigg\vert \text{ where } \text{DIFF}(k)=(k-1)^{2/p}W_{k-1}-k^{2/p}W_k.
\end{align*}
Both these approaches are not defined for $k=1$ and can not be used for testing population unimodality.
\cite{Hartigan-75} proposed using the smallest $k$ for which the following ratio of the within cluster sum of squares is greater than $10$:
\begin{align*}
\text{H}(k)=\bigg\{ \frac{W(k)}{W(k+1)}-1\bigg\}\bigg/(n-k-1).
\end{align*}
\begin{figure}[h]
	\includegraphics[width=\textwidth]{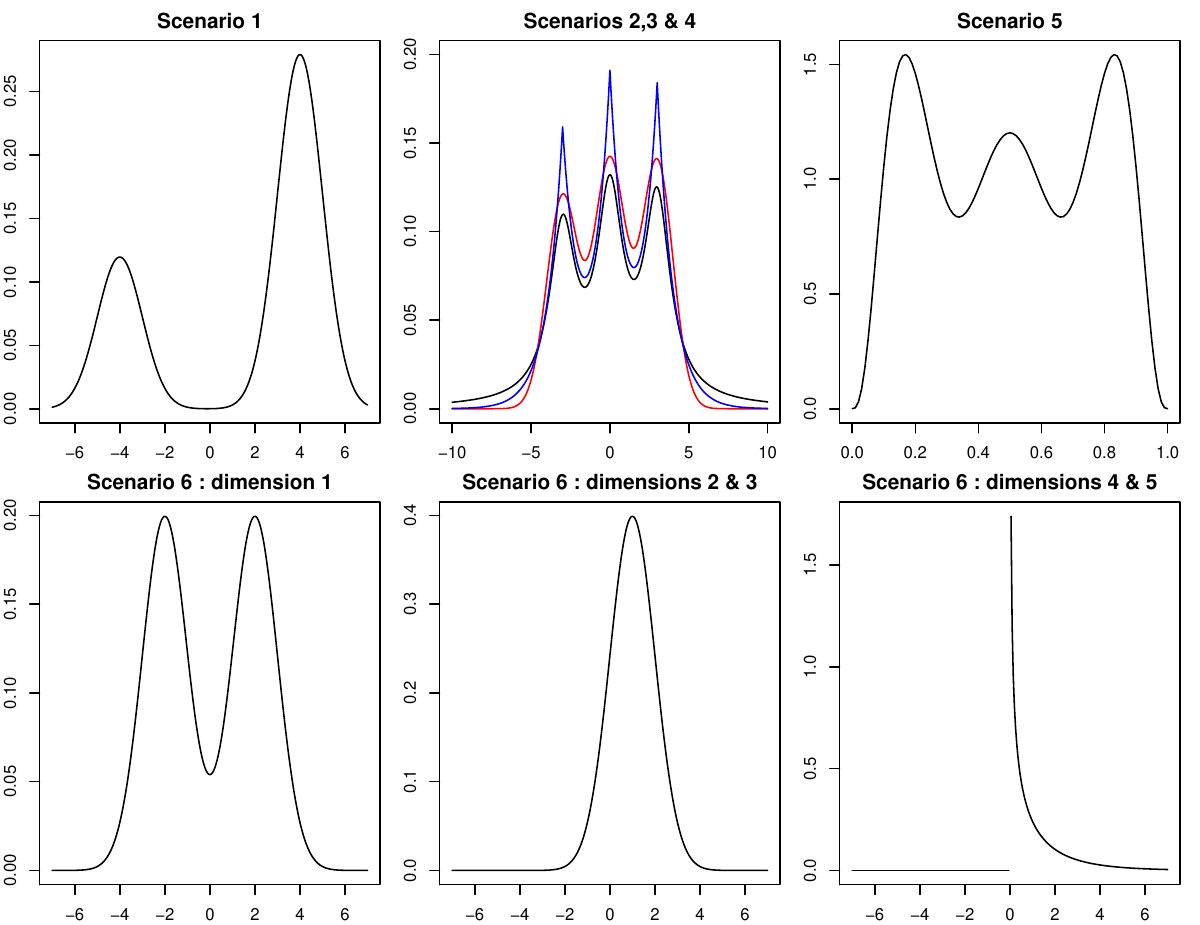}
	\caption{Plot of the univariate densities used in the different numerical experiments of Table~\ref{table3} of the main paper.}\label{fig-table3}
\end{figure}
It can be used for testing presence of only one cluster.  Theoretical thresholds  based on the $F$ distribution can also be used. \citet{Gordon-96} further sub-divides these approaches into local and global methods.
Local methods consider individual pairs of clusters and check whether they should be merged. On the other hand, global methods incorporate the entire data in evaluating measures that are subsequently optimized as a function of the number of clusters.
Note that the BMT is a local method. In addition to the above three methods, we compare BMT with the following five methods.
\par
Given $k$ clusters, for each data-point $\xx_i$ the silhouette statistic of \cite{Kaufman-09} uses:
$\bm{a(i)}$ - the average distance of $\xx_i$  to other points in its cluster, and
$\bm{b(i)}$ - the average distance of $\xx_i$ from points in its nearest neighboring cluster.
It is given by $\text{sh}_k(i)=(b(i)-a(i))/{\max \{ a(i),b(i)\}}$. Large values of $\text{sh}_k(i)$ signify good clustering. A popular estimate of the optimal number of clusters is based on maximizing the average silhouette statistic,
$$\text{KR}(k)=n^{-1} \sum_{i=1}^n \text{sh}_k(i) \quad \text{ over } k \geq 2.$$
\par
The Gap statistic of \cite{Tibshirani-01} uses the `elbow phenomenon' \citep{Thorndike-53} by estimating the number of clusters at the transition point, where the decline in the within cluster dispersion first slackens. The goodness of clustering for $k$ clusters is defined as:
\begin{align*}
\text{Gap}_n(k)=\ex_n^{\star}\{\log (\tilde{W}(k))\} - \log (\tilde{W}(k)),
\end{align*}
where $\tilde{W}(k)$ is the size-normalized intra-cluster sums of squares. The expectation is over reference datasets and can be estimated  by the mean of $\log \tilde{W}^*(k)$ over $B$ i.i.d. datasets that are generated by sampling uniformly from the original dataset's range. The standard deviation,  $\text{std}(k)$ of  $\log \tilde{W}^*(k)$,  is also recorded, and an estimate of the optimal number of clusters in the datasets is given by the smallest $k$ for which the following holds:
$$ \text{Gap}(k) \geq \text{Gap}(k+1) - (1+B^{-1})^{-1/2} \text{std}(k+1).$$
\par
Another method that uses the  `elbow phenomenon' is the Jump statistic of \citet{Sugar-03}. It again is a global method and involves computation of the mean square error (\textrm{mse}) associated with $k$ clusters for different choices of $k$. The `jumps'  in the $p/2$ moments of the mean square errors are subsequently calculated: $J_k =   \{\textrm{mse}(k)\}^{-p/2}-\{\textrm{mse}(k-1)\}^{-p/2}$ for $k=1,\ldots,n$. The estimate of the number of clusters is the value of $k$ that maximizes the jump $J_k$.
\par
The prediction strength criterion of \citet{Tibshirani-05} is computed by repeatedly dividing the data-set into two halves: train and test. For each value of $k\geq 1$, we
cluster the test and training data into $k$ groups and measure how well the training set cluster centroids predict co-memberships in the test set. For each of the $k$ test clusters, the proportion of observation pairs that are also assigned to the same cluster by the training set centroids is computed. The prediction strength is the minimum of this value over the $k$ test clusters. We have used the R-package of \citet{fpc} for computing the prediction strength criterion. $50$ divisions of the data were used and the estimate of the number of clusters is the maximum the value of $k$ which has prediction strength above $0.8$.
\par
The number of clusters estimate by \cite{fang2012selection} involves drawing $2$ bootstrap samples from the data several times. For each of these bootstrap sample pairs and for each value of $k$, a global clustering instability measure \citep{wang-10}, which reflects the distance between clusterings in paired samples, is computed. The number of clusters estimate minimizes the clustering instability aggregated over the bootstrap samples.
\par
In Tables~\ref{table3-1}, \ref{table3-1-a} and \ref{table3-2} we report the performance of the  aforementioned seven number of clusters estimation approaches (all except for the BMT and Jump), across the six simulation experiments, for the following clustering algorithms:
\begin{itemize}
\item Ward's algorithm \citep{ward1963hierarchical}, which was implemented through the \texttt{NbCLust} package.
\item Centroid-based clustering algorithm \citep{Kaufman-09}.
\item PAM (Partitioning Around Medoids,) which is the most common realization of $k$-medoid clustering  algorithm \citep{kaufman1990partitioning}. It uses a greedy search, which may not find the optimum solution but is faster than exhaustive search.
\item CLARA (Clustering LARge Applications) algorithm of \citet{kaufman1986clustering}, which extends the $k$-medoids approach for large sample sizes. It first conducts a down-sampling and then clusters the down-sampled observations. Thereafter, it assigns all objects in the dataset to the clusters that were formed.
\item GMM-merge, which conducts clustering by merging Gaussian mixture components from an initial mclust clustering \citep{hennig2010methods}. The mixture components are merged in a hierarchical fashion. The merging criterion is computed for all pairs of current clusters and the two clusters with the extremal criterion value are merged. Then, criterion values are recomputed for the merged cluster and the merging process is continued up to the length dictated by number of clusters selection methods.
\item GMM-BIC, which is based on \citet{fraley2002model}. It fits a parameterized Gaussian mixture models by EM algorithm which is initialized by model-based hierarchical clustering.
\end{itemize}
\input{table3-1}
\input{table3-1-a}
\input{table3-2}
Not all the above clustering algorithms were compatible with all of the number of clusters methods that we consider. Some of the combinations returned frequent computational errors due to non-convergence and related issues. We restricted ourselves to those pairs which were implementable in all our simulation set-ups. Also, as most of these clustering algorithms were very computationally expensive, we could not conduct $100$ repetitions of the simulation scenarios as we had done for Table~\ref{table3} in the main paper with the much faster $k$-means algorithm. We limited ourselves to~$20$ independent trails here, and most of the performance patterns were clearly visible by then. Figure~\ref{fig-table3}, provided in this subsection, contains the plots of the densities used in our numerical experiments. In the second sub-plot of the top row we have the true population densities for the second, third and fourth  simulation set-ups. In red, black and blue we have mixture of normals, $t$ and double-exponential densities, respectively. Bootstrap stability with CLARA or with the GMM-merge algorithm and Prediction Strength with CLARA produced performances that are similar to those seen with the $k$-means clustering algorithm.  The other combinations were not as effective.

\subsubsection{Performance on Large Data Sets}\label{append.large.data}
In Table~\ref{table4} we report the performance of the BMT across $4$ different simulation examples involving large samples generated from Gaussian mixtures.  For each example, $100$ independent data sets are generated from the population density and the distribution of the number of clusters detected by the BMT is reported (the frequencies are in parenthesis).  We also report the average and the standard deviation of the Mean Square Error (MSE) over the cases where the the true number of clusters is correctly detected.  The oracle MSE is calculated based on the partition that uses the minima of the true population density.
\input{table-4}
\par
Overall, the BMT correctly detects the true number of clusters with high certainly.  Also, the average MSE is observed to be very close to the Oracle one.
To demonstrate the scalability of the proposed method, we report the average elapsed time (in seconds) per replication.
The numerical experiments were performed at the Center for High Performance Computing (\url{http://hpcc.usc.edu}) of the University of Southern California.
The computations were done in R version 3.1.1 on Dual Quadcore Intel Xeon 2.33 GHz, 16GB Memory nodes. We used the \texttt{Snowfall} package of \cite{Snowfall} to distribute computations over~$100$ CPUs.
\par
Most of the competitor methods discussed above fail to accommodate the large sample sizes of the data sets in Table~\ref{table4}. To get a better understanding of the relative scalibility of the BMT, we implemented the HMAC algorithm of \citealt{li2007nonparametric},  which can handle large sample sizes. HMAC is a  non-parametric mode identification based clustering approach, which produces clusters in hierarchical levels.
Compared to HMAC, the BMT was found to be substantially faster in producing the entire hierarchical path. On an iMac desktop with 2.9 GHz Intel Core i5 processor and 8 GB memory,   for grids of 20 hierarchical levels, the R-code associated with the HMAC paper required approximate run times of 3 minutes, 70 minutes and 4 hours for the sample sizes of 10000, 50000 and 100000, respectively.  On the other hand, the corresponding average run times for the BMT were 10 seconds, 5 minutes and 20 minutes, respectively.

\subsubsection{Performance for smaller sample sizes and the choice of~$\alpha$}\label{append.small.sample}
For the numerical experiments in Section~\ref{sec.sim} of the main paper we used relatively large sample sizes and implemented the BMT with the threshold~$\alpha$ uniformly set at $10\%$.
The choice of the threshold~$\alpha$ is, however, important in smaller sample sizes.  Our theoretical results suggest that~$\alpha$ should, in general, be increased when the sample size decreases.  For this purpose, we have conducted an extensive simulation study, where, for a variety of sample sizes (100, 500, 1000, 2000, 5000 and 10000), and for a number of different threshold values (1\%, 2\%, 5\%, 7.5\%, 10\%, 15\%, 20\% and 25\%), we record the percentage of cases where the BMT approach produces splits, despite the fact that the true population density is unimodal.  We would like the threshold~$\alpha$ to be sufficiently high, so that the instances of splits in unimodal cases (which represent  false discoveries) are well controlled.
We tried~$4$ different population densities: (a) standard normal, (b) t with 1 df, (c) exponential with rate parameter 1, and (d) cauchy.  The results are reported in Table~\ref{table-th-1}.  We found that the Gaussian case is the most difficult, requiring the highest threshold.  Based on the simulation results reported in Table~\ref{table-th-1}, we suggest threshold sizes of $20\%$ and~$15\%$ for sample sizes $100$ and $500$, respectively, and $10\%$ or, possibly, lower for sample sizes $1000$ and larger.

We also report the results for modality detection and number of cluster estimation for smaller sample sizes of~$500$ and~$1000$. The comparisons with modality detection methods are reported in tables~\ref{table-modality-1} and~\ref{table-modality-2}, and the results on estimating the number of clusters by BMT are presented in tables~\ref{table-clus-bmt-1}, \ref{table-clus-bmt-2}, \ref{table-clus-1} and \ref{table-clus-2}.  The underlying distributions in these tables exactly match those reported in the simulation study in the main paper. We report the performance of our method for a number of different threshold sizes.  In the modality detection experiments, our proposed BMT approach performed better than the Dip and Silverman tests in detecting modes, though it made a few more false discoveries.   We found that even in low sample sizes, BMT generally outperformed other number of clusters approaches for finding the number of Beta and~$t$ mixture components (cases III and V).
\input{table-th-size-w}

\input{table-modality}
\input{table-clus-bmt}

\input{table-clus-others}

\subsubsection{BMT \& Sub-population Analysis in Single Cell Virology}\label{append.real.data}
\begin{figure}[!htbp]
\includegraphics[height=42pc,width=35pc]{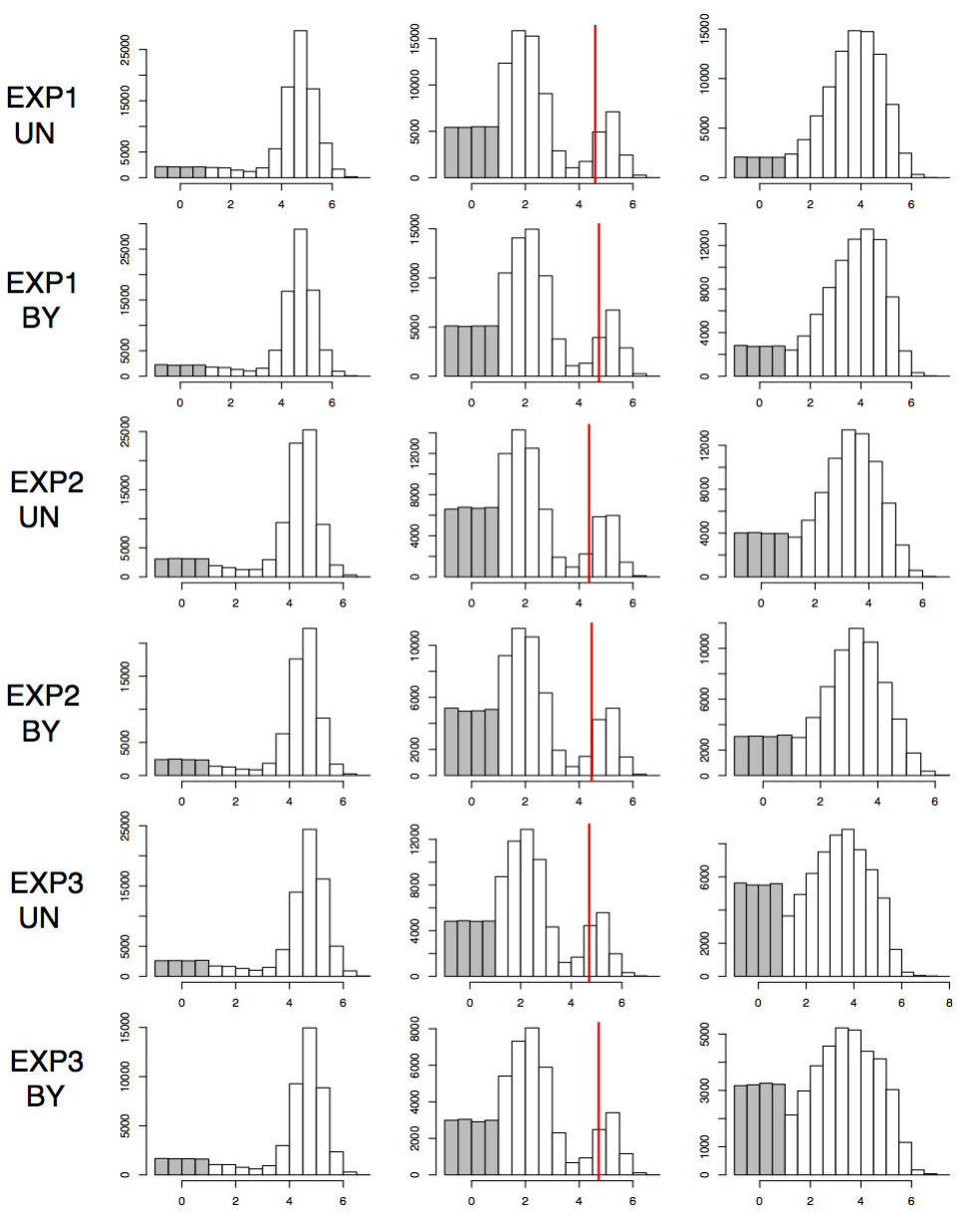}
\caption{Across columns we have histograms of the expression values of the proteins CD4, CD8 and CD45RA, respectively. Along rows, from top to bottom, we have the histograms of the Uninfected and  Bystander population, respectively, for the independent experiments I-III. The shaded gray region denotes unexpressed values.  Splits in the expression values (if any) detected by BMT are shown by vertical red lines.}\label{fig-data-1}
\end{figure}

\begin{figure}[!htbp]
\includegraphics[height=23pc,width=35pc]{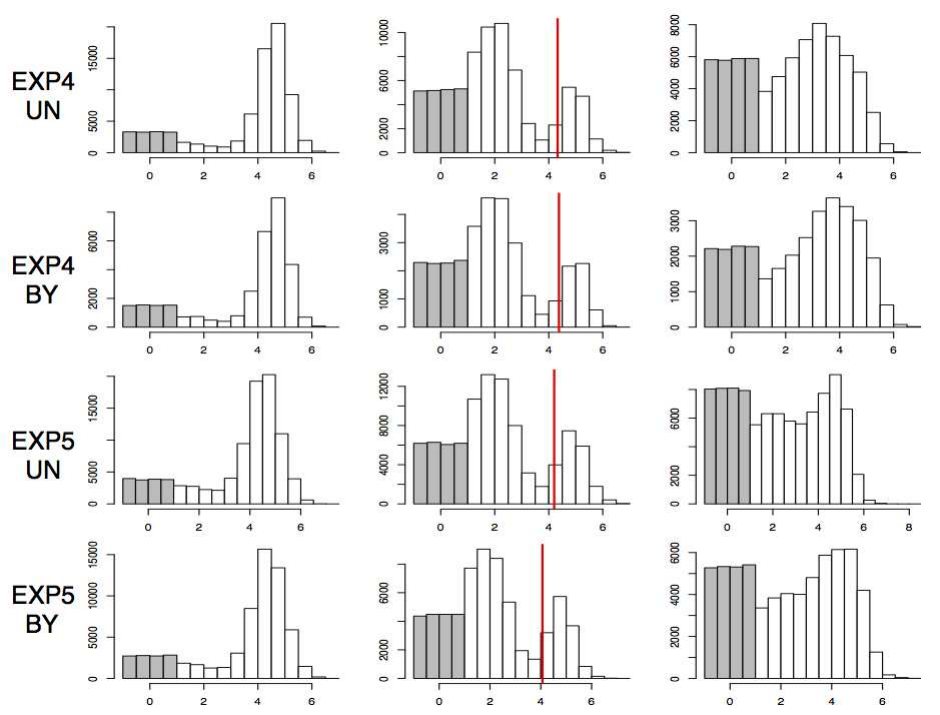}
\caption{Across columns we have histograms of the expression values of CD4, CD8 and CD45RA. Across rows are the histograms of UN and BY populations for Experiments IV-V. Splits detected by BMT (if any) are shown by red lines.}\label{fig-data-2}
\end{figure}

\begin{figure}[!htbp]
\begin{center}
\includegraphics[height=13pc,width=27pc]{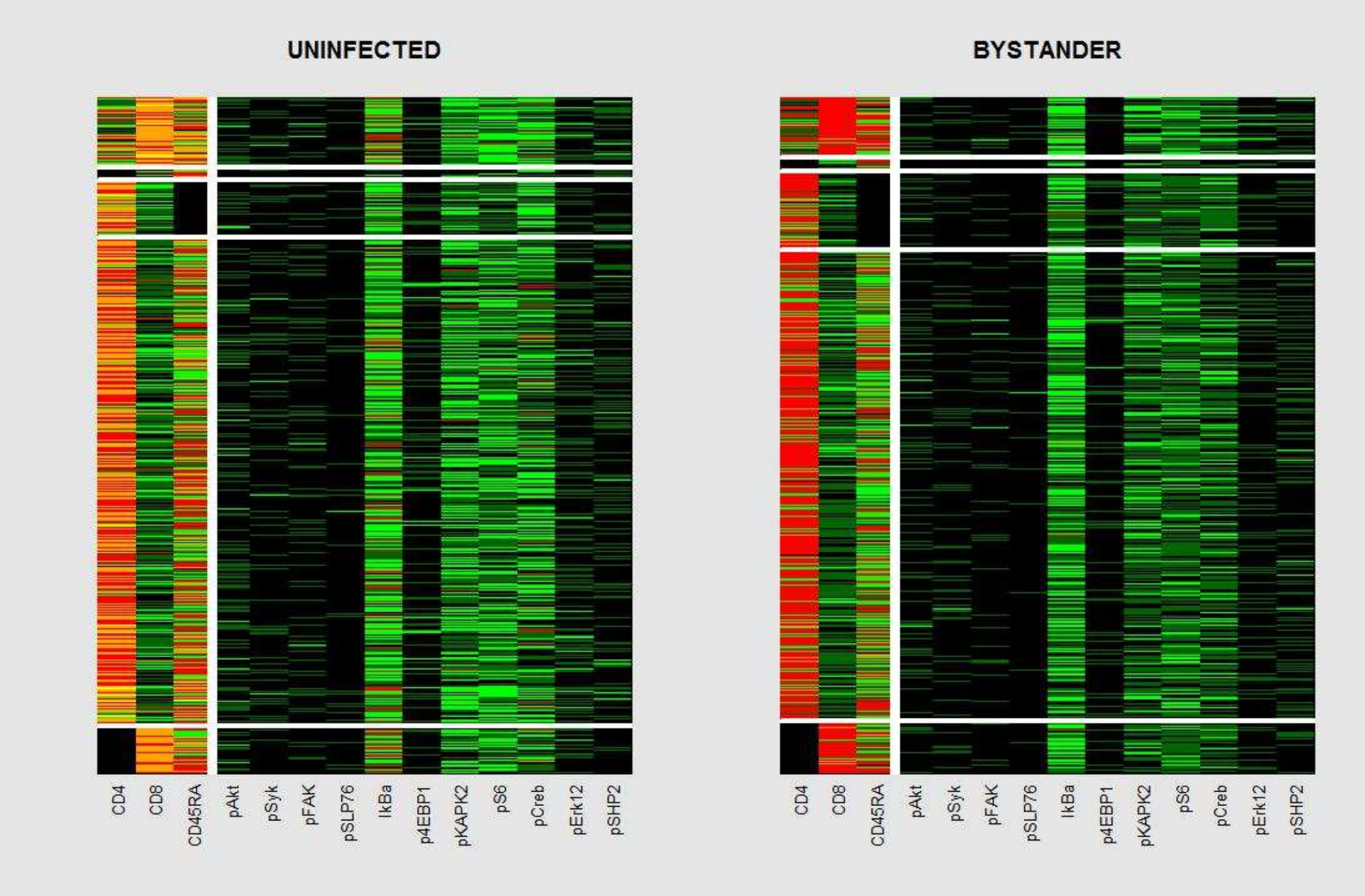}
\caption{The above plot shows the heatmaps of the protein expression values (in order of decreasing intensity: Red, Yellow, Green and Black) of the Uninfected and Bystander populations in Experiment I.
The horizontal white lines demarcate the five major sub-population detected by BMT algorithm, based on the expression of the three surface markers on the left of the vertical white line. The proteins on the right of the vertical line are associated with cell-signaling. The heatmaps are standardized separately for the two populations. }\label{fig-data-3}
\end{center}
\end{figure}

We demonstrate an application of our clustering method in an immunology study conducted at single cell level. Emerging technologies \citep{Wang11} have recently enabled us to collect proteomic data sets at single cell resolution. These data sets reflect the variations of protein expressions across cells and need clustering techniques for detection of cellular sub-populations.  Typically, sub-populations are detected by core-protein expressions based cluster analysis of the samples, and the signaling expressions of the resultant sub-populations are subsequently studied. In Figures~\ref{fig-data-1} and \ref{fig-data-2} we display the results of the BMT induced clustering on the virology datasets of \citet{Sen14}.
Figure~\ref{fig-data-3} shows the post-clustering, sub-population level signaling expressions.


%% file: table3-1.tex
\begin{table}
\centering
\scalebox{0.8}{
\begin{tabular}{|c|r|r|r|r|r|r|r|r|r|r|r|}
\cline{1-12}
True Population Density & Methods &\multicolumn{10}{ |c| }{\;\;\;\;Number of Clusters} \\ \cline{3-12}
&   &  1 & 2 & 3 & 4 & 5 & 6 & 7 & 8 & 9 & 10+\\ \cline{1-12}
\multicolumn{1}{ |c  }{\multirow{5}{*}{$0.3\, N(-4,1) + 0.7 \,N(4,1) $} } &
\multicolumn{1}{ |c| }{CH} & 0 & \textbf{3} & 1 &   1  & 0 &   2 &  1 &  0 &  5 &  7  \\ \cline{2-12}
\multicolumn{1}{ |c  }{}       &
\multicolumn{1}{ |c| }{KL} & 0 & \textbf{11} & 1 &  0 &  1  & 1 &  2 &  2 &  1 &  1  \\ \cline{2-12}
\multicolumn{1}{ |c  }{}       &
\multicolumn{1}{ |c| }{Hartigan} & 0 & \textbf{0} & 7 & 4 & 6 & 3 & 0 &  0 & 0  & 0  \\ \cline{2-12}
\multicolumn{1}{ |c  }{}       &
\multicolumn{1}{ |c| }{Silhouette} & 0 & \textbf{20} & 0 & 0 & 0 & 0 & 0 & 0 & 0 & 0  \\ \cline{2-12}
\multicolumn{1}{ |c  }{}       &
\multicolumn{1}{ |c| }{Gap} & 0 & \textbf{20} & 0 & 0 & 0 & 0 & 0 & 0 & 0 & 0    \\ \cline{2-12}
\hline
\hline
\multicolumn{1}{ |c  }{\multirow{5}{*}{$0.3\, N(-3,1) + 0.35 \,N(0,1) + 0.35 \,N(3,1)$} } &
\multicolumn{1}{ |c| }{CH} &   0 &   1 &  \textbf{ 0} &   0 &   0 &   2 &   0 &   1 &   5 &  11     \\ \cline{2-12}
\multicolumn{1}{ |c  }{}       &
\multicolumn{1}{ |c| }{KL} &   0 &   2 &  \textbf{ 4} &   1 &   1 &   2 &   1 &   6 &   1 &   2       \\ \cline{2-12}
\multicolumn{1}{ |c  }{}       &
\multicolumn{1}{ |c| }{Hartigan}  &   0 &  0  &  \textbf{11} &   1 &   3 &   1 &   2 &   1 &   0 &   1  \\ \cline{2-12}
\multicolumn{1}{ |c  }{}       &
\multicolumn{1}{ |c| }{Silhouette} & 0 &   9 &  \textbf{11} &   0 &   0 &   0 &   0 &   0 &   0 &   0  \\ \cline{2-12}
\multicolumn{1}{ |c  }{}       &
\multicolumn{1}{ |c| }{Gap}   & 0 &  20 &   \textbf{0} &   0 &   0 &   0 &   0 &   0 &   0 &   0    \\ \cline{2-12}
\hline
\hline
\multicolumn{1}{ |c  }{\multirow{5}{*}{$0.3\, t_1(-3) + 0.35 \,t_1(0) + 0.35 \,t_1(3)$} } &
\multicolumn{1}{ |c| }{CH}    &   0 &   0 &   \textbf{0} &   0 &   1 &   0 &   1 &   1 &   3 &  14 \\ \cline{2-12}
\multicolumn{1}{ |c  }{}       &
\multicolumn{1}{ |c| }{KL} &   0 &   2 &   \textbf{5} &   2 &   0 &   0 &   2 &   5 &   2 &   2  \\ \cline{2-12}
\multicolumn{1}{ |c  }{}       &
\multicolumn{1}{ |c| }{Hartigan} &   0 &   0 &   \textbf{7} &   4 &   1 &   0 &   3 &   2 &   1 &   2  \\ \cline{2-12}
\multicolumn{1}{ |c  }{}       &
\multicolumn{1}{ |c| }{Silhouette} &   0 &  16 &   \textbf{4} &   0 &   0 &   0 &   0 &   0 &   0 &   0 \\ \cline{2-12}
\multicolumn{1}{ |c  }{}       &
\multicolumn{1}{ |c| }{Gap} &   0 &  10 &  \textbf{10} &   0 &   0 &   0 &   0 &   0 &   0 &   0    \\ \cline{2-12}
\hline
\hline
\multicolumn{1}{ |c  }{\multirow{5}{*}{$0.3\, \text{dexp}(-3) + 0.35 \,\text{dexp}(0) + 0.35 \,\text{dexp}(3)$ }} &
\multicolumn{1}{ |c| }{CH} &  0 &   0 &  \textbf{ 3} &   0 &   1 &   0 &   3 &   1 &   6 &   6  \\ \cline{2-12}
\multicolumn{1}{ |c  }{}       &
\multicolumn{1}{ |c| }{KL} &  0 &   2 &   \textbf{3} &   3 &   4 &   1 &   6 &   0 &   0 &   1  \\ \cline{2-12}
\multicolumn{1}{ |c  }{}       &
\multicolumn{1}{ |c| }{Hartigan} & 0 &   0 & \textbf{ 14} &   3 &   0 &   0 &   1 &   1 &   0 &   1  \\ \cline{2-12}
\multicolumn{1}{ |c  }{}       &
\multicolumn{1}{ |c| }{Silhouette} &   0 &   7 & \textbf{ 13} &   0 &   0 &   0 &   0 &   0 &   0 &   0 \\ \cline{2-12}
\multicolumn{1}{ |c  }{}       &
\multicolumn{1}{ |c| }{Gap} &  0 &  19 &  \textbf{ 1} &   0 &   0 &   0 &   0 &   0 &   0 &   0     \\ \cline{2-12}
\hline
\hline
\multicolumn{1}{ |c  }{\multirow{5}{*}{$ \big \{ Beta(8,2)+Beta(5,5)+Beta(2,8) \big\}\big /3$ }} &
\multicolumn{1}{ |c| }{CH}  &   0 &   0 &   \textbf{0} &   0 &   0 &   0 &   2 &   2 &   6 &  10  \\ \cline{2-12}
\multicolumn{1}{ |c  }{}       &
\multicolumn{1}{ |c| }{KL} &   0 &   4 &  \textbf{ 1 }&   5 &   4 &   3 &   0 &   1 &   0 &   2  \\ \cline{2-12}
\multicolumn{1}{ |c  }{}       &
\multicolumn{1}{ |c| }{Hartigan} &   0 &   0 &  \textbf{11 }&   5 &   0 &   1 &   0 &   0 &   2 &   1\\ \cline{2-12}
\multicolumn{1}{ |c  }{}       &
\multicolumn{1}{ |c| }{Silhouette} & 0 &  11 &   \textbf{8} &   1 &   0 &   0 &   0 &   0 &   0 &   0  \\ \cline{2-12}
\multicolumn{1}{ |c  }{}       &
\multicolumn{1}{ |c| }{Gap} & 0 &  17 &   \textbf{3} &   0 &   0 &   0 &   0 &   0 &   0 &   0   \\ \cline{2-12}
\hline
\hline
\multicolumn{1}{ |c  }{$\{ 0.5\, N(-2,1) + 0.5\,N(2,1) \}$}       &
\multicolumn{1}{ |c| }{CH} &   0 &   \textbf{19} &   0 &   1 &   0 &   0 &   0 &   0 &   0 &   0 \\ \cline{2-12}
\multicolumn{1}{ |c  }{$\quad \otimes \; N(0,1) $}       &
\multicolumn{1}{ |c| }{KL} &   0 &    \textbf{5} &   3 &   3 &   1 &   2 &   2 &   0 &   3 &   1  \\ \cline{2-12}
\multicolumn{1}{ |c  }{$ \quad \otimes \; N(0,1) $}       &
\multicolumn{1}{ |c| }{Hartigan}&   0 &    \textbf{0} &   5 &  13 &   0 &   2 &   0 &   0 &   0 &   0   \\ \cline{2-12}
\multicolumn{1}{ |c  }{$ \quad \otimes \; \chi^2_1 $}       &
\multicolumn{1}{ |c| }{Silhouette} &   0 &    \textbf{9} &   5 &   6 &   0 &   0 &   0 &   0 &   0 &   0  \\ \cline{2-12}
\multicolumn{1}{ |c  }{$ \quad \otimes \; \chi^2_1 $}       &
\multicolumn{1}{ |c| }{Gap}  &   0 &   \textbf{20} &   0 &   0 &   0 &   0 &   0 &   0 &   0 &   0    \\ \cline{2-12}
\hline
\hline

\end{tabular}
}
\caption{Number of clusters detected in $20$ trials for the Ward's clustering algorithm}\label{table3-1}
\end{table}

%% file: table3-1-a.tex
\begin{table}
\centering
\scalebox{0.8}{
\begin{tabular}{|c|r|r|r|r|r|r|r|r|r|r|r|}
\cline{1-12}
True Population Density & Methods &\multicolumn{10}{ |c| }{\;\;\;\;Number of Clusters} \\ \cline{3-12}
&   &  1 & 2 & 3 & 4 & 5 & 6 & 7 & 8 & 9 & 10+\\ \cline{1-12}
\multicolumn{1}{ |c  }{\multirow{5}{*}{$0.3\, N(-4,1) + 0.7 \,N(4,1) $} } &
\multicolumn{1}{ |c| }{CH} & 0 & \textbf{17} & 0 &   0  & 0 &   0 &  0 &  1 &  1 &  1  \\ \cline{2-12}
\multicolumn{1}{ |c  }{}       &
\multicolumn{1}{ |c| }{KL} & 0 & \textbf{1} & 0 &  1 &  2  & 1 &  1 &  1 &  9 &  4  \\ \cline{2-12}
\multicolumn{1}{ |c  }{}       &
\multicolumn{1}{ |c| }{Hartigan} & 0 & \textbf{0} & 0 & 1 & 2 & 4 & 0 &  4 & 7  & 2  \\ \cline{2-12}
\multicolumn{1}{ |c  }{}       &
\multicolumn{1}{ |c| }{Silhouette} & 0 & \textbf{20} & 0 & 0 & 0 & 0 & 0 & 0 & 0 & 0  \\ \cline{2-12}
\multicolumn{1}{ |c  }{}       &
\multicolumn{1}{ |c| }{Gap} & 0 & \textbf{20} & 0 & 0 & 0 & 0 & 0 & 0 & 0 & 0    \\ \cline{2-12}
\hline
\hline
\multicolumn{1}{ |c  }{\multirow{5}{*}{$0.3\, N(-3,1) + 0.35 \,N(0,1) + 0.35 \,N(3,1)$} } &
\multicolumn{1}{ |c| }{CH} &   0 &   1 &   \textbf{1} &   0 &   0 &   0 &   1 &   3 &   6 &   8    \\ \cline{2-12}
\multicolumn{1}{ |c  }{}       &
\multicolumn{1}{ |c| }{KL}  &   0 &   2 &   \textbf{1} &   4 &   2 &   2 &   0 &   4 &   3 &   2    \\ \cline{2-12}
\multicolumn{1}{ |c  }{}       &
\multicolumn{1}{ |c| }{Hartigan}  &   0 &   0 &  \textbf{12} &   2 &   2 &   1 &   2 &   1 &   0 &   0 \\ \cline{2-12}
\multicolumn{1}{ |c  }{}       &
\multicolumn{1}{ |c| }{Silhouette}  &   0 &   5 &   \textbf{3} &   8 &   3 &   1 &   0 &   0 &   0 &   0  \\ \cline{2-12}
\multicolumn{1}{ |c  }{}       &
\multicolumn{1}{ |c| }{Gap} & 0 &  17 &   \textbf{3} &   0 &   0 &   0 &   0 &   0 &   0 &   0  \\ \cline{2-12}
\hline
\hline
\multicolumn{1}{ |c  }{\multirow{5}{*}{$0.3\, t_1(-3) + 0.35 \,t_1(0) + 0.35 \,t_1(3)$} } &
\multicolumn{1}{ |c| }{CH}   &   0 &   1 &   \textbf{1} &   0 &   0 &   5 &   3 &   1 &   2 &   7\\ \cline{2-12}
\multicolumn{1}{ |c  }{}       &
\multicolumn{1}{ |c| }{KL} &   0 &   3 &  \textbf{ 4} &   3 &   5 &   2 &   1 &   2 &   0 &   0    \\ \cline{2-12}
\multicolumn{1}{ |c  }{}       &
\multicolumn{1}{ |c| }{Hartigan}&   0 &   0 &  \textbf{12} &   4 &   2 &   0 &   0 &   0 &   0 &   2 \\ \cline{2-12}
\multicolumn{1}{ |c  }{}       &
\multicolumn{1}{ |c| }{Silhouette} &   0 &  20 &   \textbf{0} &   0 &   0 &   0 &   0 &   0 &   0 &   0  \\ \cline{2-12}
\multicolumn{1}{ |c  }{}       &
\multicolumn{1}{ |c| }{Gap}  &   0 &   6 &  \textbf{14} &   0 &   0 &   0 &   0 &   0 &   0 &   0     \\ \cline{2-12}
\hline
\hline
\multicolumn{1}{ |c  }{\multirow{5}{*}{$0.3\, \text{dexp}(-3) + 0.35 \,\text{dexp}(0) + 0.35 \,\text{dexp}(3)$ }} &
\multicolumn{1}{ |c| }{CH} &   0 &   0 &   \textbf{0} &   0 &   1 &   2 &   1 &   2 &   5 &   9   \\ \cline{2-12}
\multicolumn{1}{ |c  }{}       &
\multicolumn{1}{ |c| }{KL}  &   0 &   1 &   \textbf{4} &   2 &   2 &   3 &   1 &   4 &   2 &   1  \\ \cline{2-12}
\multicolumn{1}{ |c  }{}       &
\multicolumn{1}{ |c| }{Hartigan}&   0 &   0 &  \textbf{ 8} &   2 &   5 &   4 &   0 &   0 &   1 &   0 \\ \cline{2-12}
\multicolumn{1}{ |c  }{}       &
\multicolumn{1}{ |c| }{Silhouette}  &   0 &  11 &   \textbf{0} &   0 &   3 &   3 &   3 &   0 &   0 &   0 \\ \cline{2-12}
\multicolumn{1}{ |c  }{}       &
\multicolumn{1}{ |c| }{Gap} & 0 &  19 &   \textbf{1} &   0 &   0 &   0 &   0 &   0 &   0 &   0  \\ \cline{2-12}
\hline
\hline
\multicolumn{1}{ |c  }{\multirow{5}{*}{$ \big \{ Beta(8,2)+Beta(5,5)+Beta(2,8) \big\}\big /3$ }} &
\multicolumn{1}{ |c| }{CH} & 0 &   0 &  \textbf{ 1} &   0 &   0 &   0 &   2 &   1 &   5 &  11   \\ \cline{2-12}
\multicolumn{1}{ |c  }{}       &
\multicolumn{1}{ |c| }{KL}&   0 &   2 &  \textbf{ 1} &   5 &   1 &   1 &   3 &   0 &   5 &   2   \\ \cline{2-12}
\multicolumn{1}{ |c  }{}       &
\multicolumn{1}{ |c| }{Hartigan} &   0 &   0 & \textbf{ 11 }&   6 &   2 &   1 &   0 &   0 &   0 &   0\\ \cline{2-12}
\multicolumn{1}{ |c  }{}       &
\multicolumn{1}{ |c| }{Silhouette} &   0 &   5 & \textbf{ 12} &   3 &   0 &   0 &   0 &   0 &   0 &   0  \\ \cline{2-12}
\multicolumn{1}{ |c  }{}       &
\multicolumn{1}{ |c| }{Gap} &   0 &  19 &  \textbf{ 1} &   0 &   0 &   0 &   0 &   0 &   0 &   0    \\ \cline{2-12}
\hline
\hline
\multicolumn{1}{ |c  }{$\{ 0.5\, N(-2,1) + 0.5\,N(2,1) \}$}       &
\multicolumn{1}{ |c| }{CH}&   0 &  \textbf{ 7} &   4 &   3 &   3 &   1 &   0 &   1 &   1 &   0  \\ \cline{2-12}
\multicolumn{1}{ |c  }{$\quad \otimes \; N(0,1) $}       &
\multicolumn{1}{ |c| }{KL} &   0 &  \textbf{ 5} &   4 &   2 &   1 &   0 &   3 &   2 &   1 &   2  \\ \cline{2-12}
\multicolumn{1}{ |c  }{$ \quad \otimes \; N(0,1) $}       &
\multicolumn{1}{ |c| }{Hartigan} &   0 &   \textbf{0} &   4 &   5 &   2 &   2 &   3 &   0 &   2 &   2 \\ \cline{2-12}
\multicolumn{1}{ |c  }{$ \quad \otimes \; \chi^2_1 $}       &
\multicolumn{1}{ |c| }{Silhouette} &   0 & \textbf{ 20} &   0 &   0 &   0 &   0 &   0 &   0 &   0 &   0  \\ \cline{2-12}
\multicolumn{1}{ |c  }{$ \quad \otimes \; \chi^2_1 $}       &
\multicolumn{1}{ |c| }{Gap}&   0 &  \textbf{20} &   0 &   0 &   0 &   0 &   0 &   0 &   0 &   0    \\ \cline{2-12}
\hline
\hline

\end{tabular}
}
\caption{Number of clusters detected in $20$ trials for the centroid clustering algorithm}\label{table3-1-a}
\end{table}

%% file: table3-2.tex
\begin{table}
\centering
\scalebox{0.8}{
\begin{tabular}{|c|r|r|r|r|r|r|r|r|r|r|r|}
\cline{1-12}
True Population Density & Methods &\multicolumn{10}{ |c| }{\;\;\;\;Number of Clusters} \\ \cline{3-12}
&   &  1 & 2 & 3 & 4 & 5 & 6 & 7 & 8 & 9 & 10+  \\ \cline{1-12}
\multicolumn{1}{ |c  }{\multirow{6}{*}{$0.3\, N(-4,1) + 0.7 \,N(4,1) $} } &
\multicolumn{1}{ |c| }{BS and GMM-BIC} & 0 & \textbf{20} & 0 &   0  & 0 &   0 & 0 &   0 & 0 &   0    \\ \cline{2-12}
\multicolumn{1}{ |c  }{}       &
\multicolumn{1}{ |c| }{BS and GMM-merge}& 0 & \textbf{20} & 0 &   0  & 0 &   0 & 0 &   0 & 0 &   0   \\ \cline{2-12}
\multicolumn{1}{ |c  }{}       &
\multicolumn{1}{ |c| }{BS and CLARA} & 0 & \textbf{20} & 0 &   0  & 0 &   0 & 0 &   0 & 0 &   0    \\ \cline{2-12}
\multicolumn{1}{ |c  }{}       &
\multicolumn{1}{ |c| }{PS and CLARA} & 0 & \textbf{20} & 0 &   0  & 0 &   0 & 0 &   0 & 0 &   0   \\ \cline{2-12}
\multicolumn{1}{ |c  }{}       &
\multicolumn{1}{ |c| }{BS and PAM} & 0 & \textbf{20} & 0 &   0  & 0 &   0 & 0 &   0 & 0 &   0   \\ \cline{2-12}
\multicolumn{1}{ |c  }{}       &
\multicolumn{1}{ |c| }{PS and PAM} & 0 & \textbf{6} & 14 &   0  & 0 &   0 & 0 &   0 & 0 &   0    \\ \cline{2-12}
\hline
\hline
\multicolumn{1}{ |c  }{\multirow{6}{*}{$0.3\, N(-3,1) + 0.35 \,N(0,1) + 0.35 \,N(3,1)$} } &
\multicolumn{1}{ |c| }{BS and GMM-BIC}& 0  & 0 & \textbf{19} & 0 &   0  & 0 &   0 & 0 &   0 & 1     \\ \cline{2-12}
\multicolumn{1}{ |c  }{}       &
\multicolumn{1}{ |c| }{BS and GMM-merge}& 0 & 0& \textbf{20} & 0 &   0  & 0 &   0 & 0 &   0 & 0     \\ \cline{2-12}
\multicolumn{1}{ |c  }{}       &
\multicolumn{1}{ |c| }{BS and CLARA} & 0& 1 & \textbf{19} & 0 &   0  & 0 &   0 & 0 &   0 & 0    \\ \cline{2-12}
\multicolumn{1}{ |c  }{}       &
\multicolumn{1}{ |c| }{PS and CLARA}& 0& 0 & \textbf{20} & 0 &   0  & 0 &   0 & 0 &   0 & 0    \\ \cline{2-12}
\multicolumn{1}{ |c  }{}       &
\multicolumn{1}{ |c| }{BS and PAM}& 0& 2 & \textbf{18} & 0 &   0  & 0 &   0 & 0 &   0 & 0     \\ \cline{2-12}
\multicolumn{1}{ |c  }{}       &
\multicolumn{1}{ |c| }{PS and PAM} & 5 & 15 & \textbf{0} & 0 &   0  & 0 &   0 & 0 &   0 & 0    \\ \cline{2-12}
\hline
\hline
\multicolumn{1}{ |c  }{\multirow{6}{*}{$0.3\, t_1(-3) + 0.35 \,t_1(0) + 0.35 \,t_1(3)$} } &
\multicolumn{1}{ |c| }{BS and GMM-BIC} & 0 & 17 & \textbf{3} & 0 &   0  & 0 &   0 & 0 &   0 & 0   \\ \cline{2-12}
\multicolumn{1}{ |c  }{}       &
\multicolumn{1}{ |c| }{BS and GMM-merge} & 0 & 20 & \textbf{0} & 0 &   0  & 0 &   0 & 0 &   0 & 0    \\ \cline{2-12}
\multicolumn{1}{ |c  }{}       &
\multicolumn{1}{ |c| }{BS and CLARA} & 0 & 1 & \textbf{0} & 3 &   8  & 3 &   4 & 1 &   0 & 0   \\ \cline{2-12}
\multicolumn{1}{ |c  }{}       &
\multicolumn{1}{ |c| }{PS and CLARA} & 6 & 7 & \textbf{2} & 4 &   1  & 0 &   0 & 0 &   0 & 0     \\ \cline{2-12}
\multicolumn{1}{ |c  }{}       &
\multicolumn{1}{ |c| }{BS and PAM} & 0 & 3 & \textbf{0} & 0 &   0  &   2 & 3 &   6 & 6 & 0  \\ \cline{2-12}
\multicolumn{1}{ |c  }{}       &
\multicolumn{1}{ |c| }{PS and PAM} & 20 & 0 & \textbf{0} & 0 &   0  & 0 &   0 & 0 &   0 & 0   \\ \cline{2-12}
\hline
\hline
\multicolumn{1}{ |c  }{\multirow{6}{*}{$0.3\, \text{dexp}(-3) + 0.35 \,\text{dexp}(0) + 0.35 \,\text{dexp}(3)$ }} &
\multicolumn{1}{ |c| }{BS and GMM-BIC}& 0 & 1 & \textbf{5} & 0 &  4 & 5 &   2 & 0 &   3  & 0  \\ \cline{2-12}
\multicolumn{1}{ |c  }{}       &
\multicolumn{1}{ |c| }{BS and GMM-merge}& 0 & 0 & \textbf{20} & 0 &   0  & 0 &   0 & 0 &   0 & 0 \\ \cline{2-12}
\multicolumn{1}{ |c  }{}       &
\multicolumn{1}{ |c| }{BS and CLARA} & 0 & 0& \textbf{20} & 0 &   0  & 0 &   0 & 0 &   0 & 0 \\ \cline{2-12}
\multicolumn{1}{ |c  }{}       &
\multicolumn{1}{ |c| }{PS and CLARA} & 0 & 0& \textbf{20} & 0 &   0  & 0 &   0 & 0 &   0 & 0  \\ \cline{2-12}
\multicolumn{1}{ |c  }{}       &
\multicolumn{1}{ |c| }{BS and PAM} & 0 & 0& \textbf{20} & 0 &   0  & 0 &   0 & 0 &   0 & 0  \\ \cline{2-12}
\multicolumn{1}{ |c  }{}       &
\multicolumn{1}{ |c| }{PS and PAM} & 11 & 9 & \textbf{0} & 0 &   0  & 0 &   0 & 0 &   0 & 0  \\ \cline{2-12}
\hline
\hline
\multicolumn{1}{ |c  }{\multirow{6}{*}{$ \big \{ Beta(8,2)+Beta(5,5)+Beta(2,8) \big\}\big /3$ }} &
\multicolumn{1}{ |c| }{BS and GMM-BIC} & 0 & 0 & \textbf{1} & 2 & 1 & 5 & 2 & 3 & 1 & 5  \\ \cline{2-12}
\multicolumn{1}{ |c  }{}       &
\multicolumn{1}{ |c| }{BS and GMM-merge} & 0 & 20 &\textbf{0} & 0 & 0 & 0 & 0 & 0 & 0 & 0   \\ \cline{2-12}
\multicolumn{1}{ |c  }{}       &
\multicolumn{1}{ |c| }{BS and CLARA} & 0 & 12 & \textbf{5} & 0 & 1 & 0 & 0 & 0 & 2 & 0  \\ \cline{2-12}
\multicolumn{1}{ |c  }{}       &
\multicolumn{1}{ |c| }{PS and CLARA} & 1 & 9 & \textbf{9} & 1 & 0 & 0 & 0 & 0 & 0 & 0   \\ \cline{2-12}
\multicolumn{1}{ |c  }{}       &
\multicolumn{1}{ |c| }{BS and PAM} & 0 & 7 & \textbf{12}  & 0 & 1 & 0 & 0 & 0 & 0 & 0  \\ \cline{2-12}
\multicolumn{1}{ |c  }{}       &
\multicolumn{1}{ |c| }{PS and PAM} & 20 & 0 & \textbf{0} & 0 & 0 & 0 & 0 & 0 & 0 & 0  \\ \cline{2-12}
\hline
\hline
\multicolumn{1}{ |c  }{}       &
\multicolumn{1}{ |c| }{BS and GMM-BIC} & 0 & \textbf{0} & 0 &   0  & 0 &   0 & 0 &   0 & 0 &   20    \\ \cline{2-12}
\multicolumn{1}{ |c  }{$\{ 0.5\, N(-2,1) + 0.5\,N(2,1) \}$}       &
\multicolumn{1}{ |c| }{BS and GMM-merge} & 0 & \textbf{20} & 0 &   0  & 0 &   0 & 0 &   0 & 0 &   0    \\ \cline{2-12}
\multicolumn{1}{ |c  }{$\quad \otimes \; N(0,1) \;  \otimes \; N(0,1) $}       &
\multicolumn{1}{ |c| }{BS and CLARA} & 0 & \textbf{20} & 0 &   0  & 0 &   0 & 0 &   0 & 0 &   0    \\ \cline{2-12}
\multicolumn{1}{ |c  }{$ \quad \otimes \; \chi^2_1 \; \otimes \; \chi^2_1 $}       &
\multicolumn{1}{ |c| }{PS and CLARA} & 0 & \textbf{20} & 0 &   0  & 0 &   0 & 0 &   0 & 0 &   0    \\ \cline{2-12}
\multicolumn{1}{ |c  }{}       &
\multicolumn{1}{ |c| }{BS and PAM} & 0 & \textbf{5} & 0 &   0  & 0 &   0 & 0 &   1 & 0 &   14   \\ \cline{2-12}
\multicolumn{1}{ |c  }{}       &
\multicolumn{1}{ |c| }{PS and PAM} & 15 & \textbf{5} & 0 &   0  & 0 &   0 & 0 &   0 & 0 &   0   \\ \cline{2-12}
\hline
\hline
\end{tabular}
}
\caption{Number of clusters detected in $20$ trials by the prediction strength (PS) and bootstrap stability (BS) methods for selecting the number of clusters.  Four clustering algorithms, CLARA, PAM, GMM-BIC and GMM-merge, were used in combination with each of the two aforementioned number of clusters methods whenever the two parts were compatible  in the R-package \textit{fpc}.}\label{table3-2}
\end{table}





%% file: table-4.tex
\begin{table}[h]
\centering
\scalebox{0.74}{
\begin{tabular}{|c|c|c|c|c|c|c|c|}
 \hline
 Population Density  & Sample  &Time in Sec & Number of Clusters & \multicolumn{2}{c|}{MSE} & Oracle\\
  \cline{5-6}
& Size & per Replicate& & Mean & SD & MSE \\
  \hline & & & & &  & \\
 $\big\{N(-2.5,1) + N(0,1) + N(2.5,1)\big\}/3$ & $10^4$  & 0.70  & 2 (3),  \textbf{3 (96)}, 4 (1)  & 0.6817 & 0.0405 & 0.6564\\[1ex]
  \hline
 \hline & & & & &  & \\
 $0.5\,N(-5,1) + 0.25\,N(0,1) + 0.25\,N(5,1)$ & $5 \times 10^4$ & 11.58  & \textbf{3 (100)} & 1.1152 & 0.0125 & 0.9769\\[1ex]
 \hline
  \hline & & & & &  & \\
 $\big\{N(-1.1,1)+N(1.1,1)\big\}/2$ &  $10^5$  & 44.07  & 1 (25),  \textbf{2 (70)}, 3 (5)  & 0.6905 & 0.0258 & 0.6789\\[1ex]
 \hline
  \hline & & & & &  & \\
$\big  \{ N(0,1) \pm  N(4,1) \pm N(8,1)\big\}/5$ & $10^5$ & 38.98   & \textbf{5 (100)}  &  0.8909 & 0.0036 & 0.8909
\\[1ex]
   \hline
  \hline
\end{tabular}
}
\caption{Performance of the BMT on simulated datasets of large sample sizes}\label{table4}
\end{table} 

%% file: table-th-size-w.tex
\begin{table}[hb]
	\centering
	\caption{The table reports the percentage of cases where splits were detected by the BMT algorithm, as the sample size, $n$, and the threshold size, $\alpha$, are varied.  We consider $4$ unimodal densities: standard normal, $t$ with $1$ degrees of freedom, exponential with a unit rate parameter and Cauchy.}
	\scalebox{0.8}{
		\begin{tabular}{|r|c|r|r|r|r|r|r|r|r|}
			\hline
			\multicolumn{1}{|c|}{Density} & \multicolumn{1}{l|}{sample size} & \multicolumn{8}{c|}{Threshold size (in \%)} \bigstrut\\
			\cline{3-10}          &       & 1     & 2     & 5     & 7.5   & 10    & 15    & 20    & 25 \bigstrut\\
			\hline
			& 100   & 100   & 99    & 97    & 81    & 42    & 16    & 5     & 5 \bigstrut\\
			\cline{2-10}          & 500   & 100   & 75    & 44    & 28    & 14    & 6     & 1     & 1 \bigstrut\\
			\cline{2-10}    \multicolumn{1}{|c|}{NORMAL} & 1000  & 99    & 59    & 36    & 28    & 16    & 6     & 2     & 2 \bigstrut\\
			\cline{2-10}          & 2000  & 80    & 20    & 9     & 5     & 2     & 1     & 1     & 1 \bigstrut\\
			\cline{2-10}          & 5000  & 28    & 2     & 1     & 1     & 1     & 1     & 0     & 0 \bigstrut\\
			\cline{2-10}          & 10000 & 3     & 0     & 0     & 0     & 0     & 0     & 0     & 0 \bigstrut\\
			\hline
			& 100   & 100   & 64    & 39    & 23    & 14    & 5     & 2     & 2 \bigstrut\\
			\cline{2-10}          & 500   & 54    & 5     & 3     & 0     & 0     & 0     & 0     & 0 \bigstrut\\
			\cline{2-10}          & 1000  & 19    & 1     & 0     & 0     & 0     & 0     & 0     & 0 \bigstrut\\
			\cline{2-10}    \multicolumn{1}{|c|}{t with 1 df} & 2000  & 0     & 0     & 0     & 0     & 0     & 0     & 0     & 0 \bigstrut\\
			\cline{2-10}          & 5000  & 0     & 0     & 0     & 0     & 0     & 0     & 0     & 0 \bigstrut\\
			\cline{2-10}          & 10000 & 0     & 0     & 0     & 0     & 0     & 0     & 0     & 0 \bigstrut\\
			\hline
			& 100   & 100   & 97    & 82    & 63    & 38    & 23    & 10    & 10 \bigstrut\\
			\cline{2-10}          & 500   & 99    & 42    & 25    & 9     & 3     & 0     & 0     & 0 \bigstrut\\
			\cline{2-10}          & 1000  & 78    & 17    & 3     & 0     & 0     & 0     & 0     & 0 \bigstrut\\
			\cline{2-10}    \multicolumn{1}{|c|}{EXP} & 2000  & 30    & 2     & 0     & 0     & 0     & 0     & 0     & 0 \bigstrut\\
			\cline{2-10}          & 5000  & 6     & 0     & 0     & 0     & 0     & 0     & 0     & 0 \bigstrut\\
			\cline{2-10}          & 10000 & 0     & 0     & 0     & 0     & 0     & 0     & 0     & 0 \bigstrut\\
			\hline
			& 100   & 100   & 64    & 37    & 17    & 7     & 4     & 1     & 1 \bigstrut\\
			\cline{2-10}          & 500   & 59    & 7     & 3     & 0     & 0     & 0     & 0     & 0 \bigstrut\\
			\cline{2-10}          & 1000  & 15    & 0     & 0     & 0     & 0     & 0     & 0     & 0 \bigstrut\\
			\cline{2-10}    \multicolumn{1}{|c|}{CAUCHY} & 2000  & 1     & 0     & 0     & 0     & 0     & 0     & 0     & 0 \bigstrut\\
			\cline{2-10}          & 5000  & 0     & 0     & 0     & 0     & 0     & 0     & 0     & 0 \bigstrut\\
			\cline{2-10}          & 10000 & 0     & 0     & 0     & 0     & 0     & 0     & 0     & 0 \bigstrut\\
			\hline
		\end{tabular}\label{table-th-1}%
	}
	\end{table}%

%% file: table-modality.tex

\begin{table}[htbp]
	\centering
	\caption{Results comparing various multi-modality detection methods, across the five different simulation scenarios considered in Table 1, are reported for the sample size of $500$. The results for the BMT algorithm with threshold sizes 5\%, 10\%, 15\% and 20\% are reported. }
	  \scalebox{0.8}{
	\begin{tabular}{|l|r|r|r|r|r|r|r|r|r|r|}
	 \hline
	 & \multicolumn{3}{c|}{Dip Test P-value (D)} & \multicolumn{3}{c|}{Silverman Test P-value (S)} & \multicolumn{4}{c|}{BMT with threshold size } \bigstrut\\
	 & \multicolumn{3}{c|}{} & \multicolumn{3}{c|}{} & \multicolumn{4}{c|}{ \% multi-mode } \bigstrut\\
	 \cline{2-11}          & \multicolumn{1}{l|}{Mean (D)} & \multicolumn{1}{l|}{Std (D)} & \multicolumn{1}{l|}{\% multi-mode} & \multicolumn{1}{l|}{Mean (S)} & \multicolumn{1}{l|}{Std (S)} & \multicolumn{1}{l|}{\% multi-mode} & 5\%   & 10\%  & 15\%  & 20\% \bigstrut\\
		\hline
		Case I & 0.9   & 0.16  & 0     & 0.51  & 0.25  & 0     & 71    & 35    & 12    & 7 \bigstrut\\
		\hline
		Case II  & 0.86  & 0.18  & 0     & 0.59  & 0.26  & 1     & 85    & 46    & 27    & 15 \bigstrut\\
		\hline
		Case III   & 0.8   & 0.24  & 0     & 0.39  & 0.27  & 7     & 97    & 74    & 55    & 40 \bigstrut\\
		\hline
		Case IV  & 0.72  & 0.26  & 1     & 0.32  & 0.24  & 19    & 100   & 89    & 65    & 50 \bigstrut\\
		\hline
		Case V  & 0.6   & 0.28  & 0     & 0.23  & 0.19  & 14    & 100   & 96    & 90    & 68 \bigstrut\\
		\hline
	\end{tabular}%
}
	\label{table-modality-1}%
\end{table}%

\begin{table}[htbp]
  \centering
	\caption{Results comparing various multi-modality detection methods, across the five different simulation scenarios considered in Table 1, are reported for the sample size of~$1000$. The results for the BMT algorithm with threshold sizes 5\%, 10\%, 15\% and 20\% are reported. }
  \scalebox{0.8}{
    \begin{tabular}{|l|r|r|r|r|r|r|r|r|r|r|}
    \hline
          & \multicolumn{3}{c|}{Dip Test P-value (D)} & \multicolumn{3}{c|}{Silverman Test P-value (S)} & \multicolumn{4}{c|}{BMT with threshold size } \bigstrut\\
          & \multicolumn{3}{c|}{} & \multicolumn{3}{c|}{} & \multicolumn{4}{c|}{ \% multi-mode } \bigstrut\\
\cline{2-11}          & \multicolumn{1}{l|}{Mean (D)} & \multicolumn{1}{l|}{Std (D)} & \multicolumn{1}{l|}{\% multi-mode} & \multicolumn{1}{l|}{Mean (S)} & \multicolumn{1}{l|}{Std (S)} & \multicolumn{1}{l|}{\% multi-mode} & 5\%   & 10\%  & 15\%  & 20\% \bigstrut\\
    \hline
    Case I & 0.94  & 0.11  & 0     & 0.48  & 0.22  & 0     & 45    & 12    & 9     & 4 \bigstrut\\
    \hline
    Case II  & 0.92  & 0.12  & 0     & 0.61  & 0.25  & 0     & 65    & 26    & 14    & 4 \bigstrut\\
    \hline
    Case III    & 0.8   & 0.22  & 0     & 0.33  & 0.28  & 20    & 97    & 70    & 57    & 47 \bigstrut\\
    \hline
    Case IV    & 0.76  & 0.25  & 1     & 0.35  & 0.26  & 16    & 99    & 83    & 57    & 41 \bigstrut\\
    \hline
    Case V    & 0.5   & 0.31  & 2     & 0.2   & 0.19  & 19    & 100   & 98    & 92    & 74 \bigstrut\\
    \hline
    \end{tabular}%
}
  \label{table-modality-2}%
\end{table}%

%% file: table-clus-bmt.tex
\begin{table}[htbp]
	\centering
	\caption{\small The table reports the number of clusters detected by the BMT algorithm, with~$3$ different prefixed threshold choices, in $100$ trials for six simulation scenarios of Table 2, with sample sizes being kept fixed at $500$ for all the experiments.}
	\scalebox{1.2}{\begin{tabular}{|c|c|r|r|r|r|r|r|r|r|r|r|}
		\hline
		\textbf{Threshold} & \textbf{Simulation} & \multicolumn{10}{c|}{\textbf{Number of clusters}} \bigstrut\\
		\cline{3-12}    \textbf{size} & \textbf{Scenario} & \textbf{1} & \textbf{2} & \textbf{3} & \textbf{4} & \textbf{5} & \textbf{6} & \textbf{7} & \textbf{8} & \textbf{9} & \multicolumn{1}{c|}{\textbf{10+}} \bigstrut\\
		\hline
		& \textbf{I} & 0     & \textbf{53} & 41    & 6     & 0     & 0     & 0     & 0     & 0     & 0 \bigstrut\\
		\cline{2-12}          & \textbf{II} & 0     & 9     & \textbf{65} & 26    & 0     & 0     & 0     & 0     & 0     & 0 \bigstrut\\
		\cline{2-12}          & \textbf{III} & 0     & 12    & \textbf{84} & 4     & 0     & 0     & 0     & 0     & 0     & 0 \bigstrut\\
		\cline{2-12}    \textbf{10\%} & \textbf{IV} & 0     & 0     & \textbf{100} & 0     & 0     & 0     & 0     & 0     & 0     & 0 \bigstrut\\
		\cline{2-12}          & \textbf{V} & 0     & 3     & \textbf{77} & 20    & 0     & 0     & 0     & 0     & 0     & 0 \bigstrut\\
		\cline{2-12}          & \textbf{VI} & 0     & \textbf{27} & 15    & 20    & 0     & 19    & 0     & 6     & 6     & 7 \bigstrut\\
		\hline
		& \textbf{I} & 0     & \textbf{87} & 13    & 0     & 0     & 0     & 0     & 0     & 0     & 0 \bigstrut\\
		\cline{2-12}          & \textbf{II} & 1     & 35    & \textbf{64} & 0     & 0     & 0     & 0     & 0     & 0     & 0 \bigstrut\\
		\cline{2-12}          & \textbf{III} & 1     & 36    & \textbf{63} & 0     & 0     & 0     & 0     & 0     & 0     & 0 \bigstrut\\
		\cline{2-12}    \textbf{15\%} & \textbf{IV} & 0     & 8     & \textbf{92} & 0     & 0     & 0     & 0     & 0     & 0     & 0 \bigstrut\\
		\cline{2-12}          & \textbf{V} & 0     & 17    & \textbf{83} & 0     & 0     & 0     & 0     & 0     & 0     & 0 \bigstrut\\
		\cline{2-12}          & \textbf{VI} & 0     & \textbf{67} & 1     & 26    & 0     & 1     & 0     & 5     & 0     & 0 \bigstrut\\
		\hline
		& \textbf{I} & 0     & \textbf{100} & 0     & 0     & 0     & 0     & 0     & 0     & 0     & 0 \bigstrut\\
		\cline{2-12}          & \textbf{II} & 7     & 61    & \textbf{32} & 0     & 0     & 0     & 0     & 0     & 0     & 0 \bigstrut\\
		\cline{2-12}          & \textbf{III} & 18    & 63    & \textbf{19} & 0     & 0     & 0     & 0     & 0     & 0     & 0 \bigstrut\\
		\cline{2-12}    \textbf{20\%} & \textbf{IV} & 2     & 41    & \textbf{57} & 0     & 0     & 0     & 0     & 0     & 0     & 0 \bigstrut\\
		\cline{2-12}          & \textbf{V} & 0     & 34    & \textbf{66} & 0     & 0     & 0     & 0     & 0     & 0     & 0 \bigstrut\\
		\cline{2-12}          & \textbf{VI} & 0     & \textbf{83} & 16    & 0     & 0     & 0     & 0     & 1     & 0     & 0 \bigstrut\\
		\hline
	\end{tabular}
	}
	\label{table-clus-bmt-1}%
\end{table}%
\begin{table}[htbp]
	\centering
	\caption{\small The table reports the number of clusters detected by the BMT algorithm, with 3 different prefixed threshold choices, in $100$ trials for six simulation scenarios of Table 2, with sample sizes being kept fixed at $1000$ for all the experiments.}
	\scalebox{1.2}{
	\begin{tabular}{|c|c|r|r|r|r|r|r|r|r|r|r|}
		\hline
		\textbf{Threshold} & \textbf{Simulation} & \multicolumn{10}{c|}{\textbf{Number of clusters}} \bigstrut\\
		\cline{3-12}    \textbf{Size} & \textbf{Scenario} & \textbf{1} & \textbf{2} & \textbf{3} & \textbf{4} & \textbf{5} & \textbf{6} & \textbf{7} & \textbf{8} & \textbf{9} & \multicolumn{1}{c|}{\textbf{10+}} \bigstrut\\
		\hline
		& \textbf{I} & 0     & \textbf{68} & 32    & 0     & 0     & 0     & 0     & 0     & 0     & 0 \bigstrut\\
		\cline{2-12}          & \textbf{II} & 0     & 9     & \textbf{83} & 8     & 0     & 0     & 0     & 0     & 0     & 0 \bigstrut\\
		\cline{2-12}          & \textbf{III} & 0     & 8     & \textbf{88} & 4     & 0     & 0     & 0     & 0     & 0     & 0 \bigstrut\\
		\cline{2-12}    \textbf{10\%} & \textbf{IV} & 0     & 1     & \textbf{99} & 0     & 0     & 0     & 0     & 0     & 0     & 0 \bigstrut\\
		\cline{2-12}          & \textbf{V} & 0     & 6     & \textbf{78} & 16    & 0     & 0     & 0     & 0     & 0     & 0 \bigstrut\\
		\cline{2-12}          & \textbf{VI} & 0     & \textbf{51} & 14    & 15    & 0     & 14    & 0     & 1     & 1     & 4 \bigstrut\\
		\hline
		& \textbf{I} & 0     & \textbf{94} & 6     & 0     & 0     & 0     & 0     & 0     & 0     & 0 \bigstrut\\
		\cline{2-12}          & \textbf{II} & 0     & 34    & \textbf{65} & 1     & 0     & 0     & 0     & 0     & 0     & 0 \bigstrut\\
		\cline{2-12}          & \textbf{III} & 1     & 36    & \textbf{63} & 0     & 0     & 0     & 0     & 0     & 0     & 0 \bigstrut\\
		\cline{2-12}    \textbf{15\%} & \textbf{IV} & 0     & 5     & \textbf{95} & 0     & 0     & 0     & 0     & 0     & 0     & 0 \bigstrut\\
		\cline{2-12}          & \textbf{V} & 0     & 11    & \textbf{89} & 0     & 0     & 0     & 0     & 0     & 0     & 0 \bigstrut\\
		\cline{2-12}          & \textbf{VI} & 0     & \textbf{82} & 0     & 14    & 0     & 2     & 0     & 2     & 0     & 0 \bigstrut\\
		\hline
		& \textbf{I} & 0     & \textbf{100} & 0     & 0     & 0     & 0     & 0     & 0     & 0     & 0 \bigstrut\\
		\cline{2-12}          & \textbf{II} & 0     & 62    & \textbf{38} & 0     & 0     & 0     & 0     & 0     & 0     & 0 \bigstrut\\
		\cline{2-12}          & \textbf{III} & 15    & 67    & \textbf{18} & 0     & 0     & 0     & 0     & 0     & 0     & 0 \bigstrut\\
		\cline{2-12}    \textbf{20\%} & \textbf{IV} & 0     & 36    & \textbf{64} & 0     & 0     & 0     & 0     & 0     & 0     & 0 \bigstrut\\
		\cline{2-12}          & \textbf{V} & 0     & 27    & \textbf{73} & 0     & 0     & 0     & 0     & 0     & 0     & 0 \bigstrut\\
		\cline{2-12}          & \textbf{VI} & 0     & \textbf{94} & 6     & 0     & 0     & 0     & 0     & 0     & 0     & 0 \bigstrut\\
		\hline
	\end{tabular}%
	\label{table-clus-bmt-2}%
}
\end{table}%

%% file: table-clus-others.tex
\begin{table}[htbp]
  \centering
  	\caption{Number of clusters detected in 100 trials for simulation scenarios of Table 2, with the sample size of $500$.}
  \scalebox{0.6}{
    \begin{tabular}{|c|l|r|r|r|r|r|r|r|r|r|r|}
    \hline
    \textbf{Simulation } & \textbf{Methods} & \multicolumn{10}{c|}{\textbf{Number of clusters}} \bigstrut\\
\cline{3-12}    \textbf{Scenario} &       & \textbf{1} & \textbf{2} & \textbf{3} & \textbf{4} & \textbf{5} & \textbf{6} & \textbf{7} & \textbf{8} & \textbf{9} & \multicolumn{1}{c|}{\textbf{10+}} \bigstrut\\
    \hline
          & CH    & 0     & \textbf{33} & 9     & 3     & 6     & 4     & 4     & 8     & 9     & 24 \bigstrut\\
\cline{2-12}          & KL    & 0     & \textbf{58} & 4     & 8     & 3     & 3     & 4     & 11    & 4     & 5 \bigstrut\\
\cline{2-12}          & Hartigan & 0     & \textbf{0} & 38    & 10    & 12    & 9     & 8     & 8     & 5     & 10 \bigstrut\\
\cline{2-12}    \textbf{I} & Silhouette & 0     & \textbf{100} & 0     & 0     & 0     & 0     & 0     & 0     & 0     & 0 \bigstrut\\
\cline{2-12}          & Gap   & 0     & \textbf{100} & 0     & 0     & 0     & 0     & 0     & 0     & 0     & 0 \bigstrut\\
\cline{2-12}          & Jump  & 0     & \textbf{100} & 0     & 0     & 0     & 0     & 0     & 0     & 0     & 0 \bigstrut\\
\cline{2-12}          & Pred Str. & 0     & \textbf{100} & 0     & 0     & 0     & 0     & 0     & 0     & 0     & 0 \bigstrut\\
\cline{2-12}          & Stability & 0     & \textbf{100} & 0     & 0     & 0     & 0     & 0     & 0     & 0     & 0 \bigstrut\\
    \hline
          & CH    & 0     & 0     & \textbf{4} & 1     & 1     & 4     & 6     & 14    & 28    & 42 \bigstrut\\
\cline{2-12}          & KL    & 0     & 10    & \textbf{18} & 13    & 17    & 10    & 9     & 8     & 8     & 7 \bigstrut\\
\cline{2-12}          & Hartigan & 0     & 0     & \textbf{58} & 17    & 9     & 5     & 2     & 4     & 2     & 3 \bigstrut\\
\cline{2-12}    \textbf{II} & Silhouette & 0     & 37    & \textbf{63} & 0     & 0     & 0     & 0     & 0     & 0     & 0 \bigstrut\\
\cline{2-12}          & Gap   & 0     & 100   & \textbf{0} & 0     & 0     & 0     & 0     & 0     & 0     & 0 \bigstrut\\
\cline{2-12}          & Jump  & 6     & 0     & \textbf{76} & 0     & 0     & 6     & 3     & 7     & 1     & 1 \bigstrut\\
\cline{2-12}          & Pred Str. & 0     & 2     & \textbf{98} & 0     & 0     & 0     & 0     & 0     & 0     & 0 \bigstrut\\
\cline{2-12}          & Stability & 0     & 19    & \textbf{81} & 0     & 0     & 0     & 0     & 0     & 0     & 0 \bigstrut\\
    \hline
          & CH    & 0     & 8     & \textbf{7} & 1     & 3     & 5     & 6     & 14    & 20    & 36 \bigstrut\\
\cline{2-12}          & KL    & 0     & 26    & \textbf{13} & 12    & 11    & 6     & 8     & 10    & 9     & 5 \bigstrut\\
\cline{2-12}          & Hartigan & 0     & 0     & \textbf{39} & 12    & 11    & 13    & 8     & 6     & 7     & 4 \bigstrut\\
\cline{2-12}    \textbf{III} & Silhouette & 0     & 72    & \textbf{26} & 0     & 0     & 0     & 1     & 0     & 1     & 0 \bigstrut\\
\cline{2-12}          & Gap   & 0     & 60    & \textbf{40} & 0     & 0     & 0     & 0     & 0     & 0     & 0 \bigstrut\\
\cline{2-12}          & Jump  & 18    & 12    & \textbf{6} & 5     & 4     & 5     & 9     & 13    & 17    & 11 \bigstrut\\
\cline{2-12}          & Pred Str. & 10    & 53    & \textbf{36} & 1     & 0     & 0     & 0     & 0     & 0     & 0 \bigstrut\\
\cline{2-12}          & Stability & 0     & 47    & \textbf{9} & 0     & 0     & 0     & 6     & 5     & 21    & 12 \bigstrut\\
    \hline
          & CH    & 0     & 2     & \textbf{4} & 5     & 5     & 5     & 7     & 8     & 20    & 44 \bigstrut\\
\cline{2-12}          & KL    & 0     & 17    & \textbf{6} & 9     & 8     & 11    & 13    & 11    & 13    & 12 \bigstrut\\
\cline{2-12}          & Hartigan & 0     & 0     & \textbf{54} & 18    & 12    & 6     & 4     & 4     & 1     & 1 \bigstrut\\
\cline{2-12}    \textbf{IV} & Silhouette & 0     & 8     & \textbf{92} & 0     & 0     & 0     & 0     & 0     & 0     & 0 \bigstrut\\
\cline{2-12}          & Gap   & 0     & 100   & \textbf{0} & 0     & 0     & 0     & 0     & 0     & 0     & 0 \bigstrut\\
\cline{2-12}          & Jump  & 67    & 0     & \textbf{27} & 0     & 0     & 0     & 1     & 0     & 2     & 3 \bigstrut\\
\cline{2-12}          & Pred Str. & 0     & 0     & \textbf{100} & 0     & 0     & 0     & 0     & 0     & 0     & 0 \bigstrut\\
\cline{2-12}          & Stability & 0     & 4     & \textbf{96} & 0     & 0     & 0     & 0     & 0     & 0     & 0 \bigstrut\\
    \hline
          & CH    & 0     & 0     & \textbf{2} & 3     & 2     & 4     & 5     & 15    & 28    & 41 \bigstrut\\
\cline{2-12}          & KL    & 0     & 13    & \textbf{18} & 10    & 14    & 13    & 9     & 6     & 8     & 9 \bigstrut\\
\cline{2-12}          & Hartigan & 0     & 0     & \textbf{57} & 17    & 10    & 7     & 2     & 3     & 2     & 2 \bigstrut\\
\cline{2-12}    \textbf{V} & Silhouette & 0     & 54    & \textbf{46} & 0     & 0     & 0     & 0     & 0     & 0     & 0 \bigstrut\\
\cline{2-12}          & Gap   & 0     & 100   & \textbf{0} & 0     & 0     & 0     & 0     & 0     & 0     & 0 \bigstrut\\
\cline{2-12}          & Jump  & 0     & 0     & \textbf{62} & 2     & 0     & 6     & 8     & 8     & 4     & 10 \bigstrut\\
\cline{2-12}          & Pred Str. & 13    & 77    & \textbf{10} & 0     & 0     & 0     & 0     & 0     & 0     & 0 \bigstrut\\
\cline{2-12}          & Stability & 0     & 43    & \textbf{57} & 0     & 0     & 0     & 0     & 0     & 0     & 0 \bigstrut\\
    \hline
          & CH    & 0     & \textbf{1} &       & 4     & 1     & 3     & 9     & 16    & 24    & 42 \bigstrut\\
\cline{2-12}          & KL    & 0     & \textbf{13} & 14    & 14    & 14    & 10    & 8     & 7     & 14    & 6 \bigstrut\\
\cline{2-12}          & Hartigan & 0     & \textbf{0} & 41    & 24    & 13    & 9     & 4     & 1     & 7     & 1 \bigstrut\\
\cline{2-12}    \textbf{VI} & Silhouette & 0     & \textbf{5} & 53    & 19    & 0     & 9     & 4     & 5     & 3     & 2 \bigstrut\\
\cline{2-12}          & Gap   & 0     & \textbf{100} & 0     & 0     & 0     & 0     & 0     & 0     & 0     & 0 \bigstrut\\
\cline{2-12}          & Jump  & 0     & \textbf{2} & 0     & 13    & 0     & 0     & 2     & 9     & 24    & 50 \bigstrut\\
\cline{2-12}          & Pred Str. & 0     & \textbf{100} & 0     & 0     & 0     & 0     & 0     & 0     & 0     & 0 \bigstrut\\
\cline{2-12}          & Stability & 0     & \textbf{68} & 27    & 5     & 0     & 0     & 0     & 0     & 0     & 0 \bigstrut\\
    \hline
    \end{tabular}%
  \label{table-clus-1}%
}
\end{table}%

\begin{table}[htbp]
	\centering
	\caption{Number of clusters detected in 100 trials for simulation scenarios of Table 2, with the sample size of $1000$.}
	\scalebox{0.6}{
	\begin{tabular}{|c|l|r|r|r|r|r|r|r|r|r|r|}
		\hline
		\textbf{Simulation } & \textbf{Methods} & \multicolumn{10}{c|}{\textbf{Number of clusters}} \bigstrut\\
		\cline{3-12}    \textbf{Scenario} &       & \textbf{1} & \textbf{2} & \textbf{3} & \textbf{4} & \textbf{5} & \textbf{6} & \textbf{7} & \textbf{8} & \textbf{9} & \multicolumn{1}{c|}{\textbf{10+}} \bigstrut\\
		\hline
		& CH    & 0     & \textbf{31} & 4     & 3     & 4     & 4     & 6     & 5     & 15    & 28 \bigstrut\\
		\cline{2-12}          & KL    & 0     & \textbf{46} & 11    & 6     & 4     & 5     & 3     & 4     & 12    & 9 \bigstrut\\
		\cline{2-12}          & Hartigan & 0     & \textbf{0} & 25    & 15    & 14    & 10    & 2     & 15    & 7     & 12 \bigstrut\\
		\cline{2-12}    \textbf{I} & Silhouette & 0     & \textbf{100} & 0     & 0     & 0     & 0     & 0     & 0     & 0     & 0 \bigstrut\\
		\cline{2-12}          & Gap   & 0     & \textbf{100} & 0     & 0     & 0     & 0     & 0     & 0     & 0     & 0 \bigstrut\\
		\cline{2-12}          & Jump  & 0     & \textbf{100} & 0     & 0     & 0     & 0     & 0     & 0     & 0     & 0 \bigstrut\\
		\cline{2-12}          & Pred Str. & 0     & \textbf{100} & 0     & 0     & 0     & 0     & 0     & 0     & 0     & 0 \bigstrut\\
		\cline{2-12}          & Stability & 0     & \textbf{100} & 0     & 0     & 0     & 0     & 0     & 0     & 0     & 0 \bigstrut\\
		\hline
		& CH    & 0     & 0     & \textbf{2} & 0     & 3     & 5     & 4     & 14    & 25    & 47 \bigstrut\\
		\cline{2-12}          & KL    & 0     & 11    & \textbf{3} & 12    & 10    & 11    & 19    & 14    & 13    & 7 \bigstrut\\
		\cline{2-12}          & Hartigan & 0     & 0     & \textbf{55} & 15    & 15    & 7     & 1     & 4     & 3     & 0 \bigstrut\\
		\cline{2-12}    \textbf{II} & Silhouette & 0     & 27    & \textbf{73} & 0     & 0     & 0     & 0     & 0     & 0     & 0 \bigstrut\\
		\cline{2-12}          & Gap   & 0     & 100   & \textbf{0} & 0     & 0     & 0     & 0     & 0     & 0     & 0 \bigstrut\\
		\cline{2-12}          & Jump  & 4     & 0     & \textbf{96} & 0     & 0     & 0     & 0     & 0     & 0     & 0 \bigstrut\\
		\cline{2-12}          & Pred Str. & 0     & 0     & \textbf{100} & 0     & 0     & 0     & 0     & 0     & 0     & 0 \bigstrut\\
		\cline{2-12}          & Stability & 0     & 14    & \textbf{86} & 0     & 0     & 0     & 0     & 0     & 0     & 0 \bigstrut\\
		\hline
		& CH    & 0     & 10    & \textbf{10} & 6     & 0     & 5     & 12    & 11    & 12    & 34 \bigstrut\\
		\cline{2-12}          & KL    & 0     & 19    & \textbf{14} & 11    & 11    & 9     & 12    & 8     & 6     & 10 \bigstrut\\
		\cline{2-12}          & Hartigan & 0     & 0     & \textbf{39} & 20    & 9     & 10    & 8     & 5     & 4     & 5 \bigstrut\\
		\cline{2-12}    \textbf{III} & Silhouette & 0     & 72    & \textbf{28} & 0     & 0     & 0     & 0     & 0     & 0     & 0 \bigstrut\\
		\cline{2-12}          & Gap   & 0     & 60    & \textbf{40} & 0     & 0     & 0     & 0     & 0     & 0     & 0 \bigstrut\\
		\cline{2-12}          & Jump  & 22    & 12    & \textbf{11} & 4     & 6     & 2     & 19    & 4     & 10    & 10 \bigstrut\\
		\cline{2-12}          & Pred Str. & 3     & 40    & \textbf{51} & 6     & 0     & 0     & 0     & 0     & 0     & 0 \bigstrut\\
		\cline{2-12}          & Stability & 0     & 64    & \textbf{14} & 0     & 0     & 0     & 0     & 0     & 9     & 13 \bigstrut\\
		\hline
		& CH    & 0     & 2     & \textbf{4} & 0     & 1     & 3     & 5     & 16    & 20    & 49 \bigstrut\\
		\cline{2-12}          & KL    & 0     & 6     & \textbf{10} & 6     & 13    & 13    & 11    & 10    & 14    & 17 \bigstrut\\
		\cline{2-12}          & Hartigan & 0     & 0     & \textbf{53} & 9     & 15    & 6     & 8     & 3     & 3     & 3 \bigstrut\\
		\cline{2-12}    \textbf{IV} & Silhouette & 0     & 0     & \textbf{100} & 0     & 0     & 0     & 0     & 0     & 0     & 0 \bigstrut\\
		\cline{2-12}          & Gap   & 0     & 100   & \textbf{0} & 0     & 0     & 0     & 0     & 0     & 0     & 0 \bigstrut\\
		\cline{2-12}          & Jump  & 80    & 0     & \textbf{20} & 0     & 0     & 0     & 0     & 0     & 0     & 0 \bigstrut\\
		\cline{2-12}          & Pred Str. & 0     & 0     & \textbf{100} & 0     & 0     & 0     & 0     & 0     & 0     & 0 \bigstrut\\
		\cline{2-12}          & Stability & 0     & 2     & \textbf{98} & 0     & 0     & 0     & 0     & 0     & 0     & 0 \bigstrut\\
		\hline
		& CH    & 0     & 0     & \textbf{1} & 1     & 1     & 5     & 12    & 22    & 16    & 42 \bigstrut\\
		\cline{2-12}          & KL    & 0     & 12    & \textbf{18} & 14    & 12    & 12    & 14    & 3     & 8     & 7 \bigstrut\\
		\cline{2-12}          & Hartigan & 0     & 0     & \textbf{70} & 10    & 4     & 5     & 5     & 3     & 2     & 1 \bigstrut\\
		\cline{2-12}    \textbf{V} & Silhouette & 0     & 53    & \textbf{47} & 0     & 0     & 0     & 0     & 0     & 0     & 0 \bigstrut\\
		\cline{2-12}          & Gap   & 0     & 100   & \textbf{0} & 0     & 0     & 0     & 0     & 0     & 0     & 0 \bigstrut\\
		\cline{2-12}          & Jump  & 0     & 0     & \textbf{90} & 0     & 0     & 0     & 2     & 1     & 4     & 3 \bigstrut\\
		\cline{2-12}          & Pred Str. & 0     & 67    & \textbf{32} & 1     & 0     & 0     & 0     & 0     & 0     & 0 \bigstrut\\
		\cline{2-12}          & Stability & 0     & 38    & \textbf{62} & 0     & 0     & 0     & 0     & 0     & 0     & 0 \bigstrut\\
		\hline
		& CH    & 0     & \textbf{0} & 3     & 2     & 0     & 7     & 8     & 15    & 21    & 44 \bigstrut\\
		\cline{2-12}          & KL    & 0     & \textbf{13} & 15    & 7     & 11    & 13    & 13    & 9     & 11    & 8 \bigstrut\\
		\cline{2-12}          & Hartigan & 0     & \textbf{0} & 42    & 26    & 8     & 9     & 9     & 2     & 1     & 3 \bigstrut\\
		\cline{2-12}    \textbf{VI} & Silhouette & 0     & \textbf{1} & 69    & 13    & 1     & 13    & 2     & 0     & 1     & 0 \bigstrut\\
		\cline{2-12}          & Gap   & 0     & \textbf{100} & 0     & 0     & 0     & 0     & 0     & 0     & 0     & 0 \bigstrut\\
		\cline{2-12}          & Jump  & 0     & \textbf{8} & 0     & 14    & 0     & 0     & 0     & 9     & 21    & 48 \bigstrut\\
		\cline{2-12}          & Pred Str. & 0     & \textbf{100} & 0     & 0     & 0     & 0     & 0     & 0     & 0     & 0 \bigstrut\\
		\cline{2-12}          & Stability & 0     & \textbf{79} & 10    & 11    & 0     & 0     & 0     & 0     & 0     & 0 \bigstrut\\
		\hline
	\end{tabular}%
	\label{table-clus-2}%
}
\end{table}%